\documentclass[
superscriptaddress,
amsmath,amssymb,
aps,
twocolumn,
prl
]{revtex4-2}

\usepackage[utf8]{inputenc}
\usepackage{graphicx}
\usepackage{dcolumn}
\usepackage{xcolor}
\usepackage{bm}
\usepackage{soul}
\usepackage{braket}
\usepackage{cleveref}

\newcommand{\iSWAP}{\mathrm{\lowercase{i}SWAP}}
\newcommand{\CNOT}{\mathrm{CNOT}}
\newcommand{\BGate}{\mathrm{B}}
\newcommand{\CZ}{\mathrm{CZ}}

\newcommand{\SWAP}{\mathrm{SWAP}}

\newcommand{\SU}[1]{\mathrm{SU}( #1 )}

\begin{document}


\title{Efficient Implementation of Arbitrary Two-Qubit Gates via Unified Control}

\newcommand{\BAQIS}{\affiliation{1}{Beijing Academy of Quantum Information Sciences, Beijing 100193, China}}
\newcommand{\IOP}{\affiliation{2}{Institute of Physics, Chinese Academy of Science, Beijing 100190, China}}
\newcommand{\UCAS}{\affiliation{3}{University of Chinese Academy of Sciences, Beijing 101408, China}}
\newcommand{\THUCS}{\affiliation{4}{Department of Computer Science and Technology, Tsinghua University, Beijing 100084, China}}

\author{Zhen Chen}
\thanks{These authors have contributed equally to this work.}
\affiliation{\BAQIS}

\author{Weiyang Liu}
\thanks{These authors have contributed equally to this work.}
\affiliation{\BAQIS}

\author{Yanjun Ma}
\affiliation{\BAQIS}
\author{Weijie Sun}
\affiliation{\BAQIS}
\author{Ruixia Wang}
\affiliation{\BAQIS}
\author{He Wang}
\affiliation{\BAQIS}
\author{Huikai Xu}
\affiliation{\BAQIS}
\author{Guangming Xue}
\affiliation{\BAQIS}
\author{Haisheng Yan}
\affiliation{\BAQIS}
\author{Zhen Yang}
\affiliation{\BAQIS}

\author{Jiayu Ding}
\affiliation{\BAQIS}
\author{Yang Gao}
\affiliation{\BAQIS}
\affiliation{\IOP}
\affiliation{\UCAS}
\author{Feiyu Li}
\affiliation{\BAQIS}
\affiliation{\IOP}
\affiliation{\UCAS}
\author{Yujia Zhang}
\affiliation{\BAQIS}
\affiliation{\IOP}
\affiliation{\UCAS}
\author{Zikang Zhang}
\affiliation{\THUCS}
\author{Yirong Jin}
\affiliation{\BAQIS}
\author{Haifeng Yu}
\affiliation{\BAQIS}

\author{Jianxin Chen}
\email{jianxinchen@acm.org}
\affiliation{\THUCS}

\author{Fei Yan}
\email{yanfei@baqis.ac.cn}
\affiliation{\BAQIS}




\begin{abstract}
The native gate set is fundamental to the performance of quantum devices, as it governs the accuracy of basic quantum operations and dictates the complexity of implementing quantum algorithms. Traditional approaches to extending gate sets often require accessing multiple transitions within an extended Hilbert space, leading to increased control complexity while offering only a limited set of gates. Here, we experimentally demonstrate a unified and highly versatile gate scheme capable of natively generating arbitrary two-qubit gates using only exchange interaction and qubit driving on a superconducting quantum processor, achieving maximum expressivity. Using a state-of-the-art transmon-coupler-transmon architecture, we achieve high fidelities averaging $99.37 \pm 0.07\%$ across a wide range of commonly used two-qubit unitaries. This outstanding performance, combined with reduced complexity, enables precise multipartite entangled state preparation, as demonstrated. To further enhance its applicability, we also show the high-fidelity realization of the unique $\BGate$ gate, which efficiently synthesizes the entire family of two-qubit gates. Our results highlight that fully exploiting the capabilities of a single interaction can yield a comprehensive and highly accurate gate set. With maximum expressivity, gate-time optimality, demonstrated high fidelity, and easy adaptability to other quantum platforms, our unified control scheme paves the way for optimal performance in quantum devices, offering exciting prospects for advancing quantum hardware and algorithm development.
\end{abstract}
\maketitle



\section{Introduction}

Quantum computing stands at the forefront of technological innovation, offering the potential to solve complex problems beyond the reach of classical computers.
At the heart of quantum computation are quantum algorithms, which are executed using circuits typically composed of hardware-native single-qubit and two-qubit gates. Many existing quantum algorithms are inspired by classical computing paradigms and predominantly rely on gate sets formed from the two-qubit Controlled-NOT (CNOT) gate or its variants. While this approach has been instrumental in a wide range of developments, it does not fully exploit the inherent capabilities of quantum hardware~\cite{vidal2002interaction}. In addition to improving gate fidelity through better coherence and faster operation times, another promising avenue to enhance performance is to expand the repertoire of available quantum gates, enabling more efficient circuit construction with reduced depth and gate count~\cite{chong2017programming}.

Implementing different types of native gates on the same device often requires accessing various transitions within the Hilbert space, complicating hardware design and device calibration~\cite{sung2021realization, PhysRevApplied.10.034050, foxen2020demonstrating}. In superconducting qubits, for example, the $\ket{01}$–$\ket{10}$ transition is used to implement iSWAP or $\sqrt{\mathrm{iSWAP}}$ gates~\cite{sung2021realization, huang2023quantum}, while the $\ket{11}$–$\ket{20}$ transition supports the Controlled-Z (CZ) gate, equivalent to the CNOT up to single-qubit rotations~\cite{dicarlo2009demonstration}. However, utilizing non-computational states like $\ket{2}$ can lead to faster decoherence and, more critically, correlated errors that undermine quantum error correction protocols~\cite{terhal2015quantum}.

In addition to these discrete gates, researchers have explored continuous two-qubit gate sets, such as the $f\text{Sim}$~\cite{foxen2020demonstrating} or XXZ family~\cite{nguyen2024programmable}, the XY family~\cite{abrams2020implementation}, and recently introduced fractional gates in both superconducting~\cite{ibm_fractional_gates} and trapped-ion systems~\cite{decross2024computational}---offering better expressivity than traditional discrete sets. Despite these advances, such gates still represent only a tiny fraction (technically, a measure-zero subset) of all possible two-qubit operations within the entire special unitary group $\SU{4}$. As a result, compiling arbitrary quantum operations into sequences of these gates remains nontrivial and typically provides limited practical benefits for general-purpose applications. Further expanding the gate alphabet often requires more complex device architectures or intricate control protocols, adding overhead and potential error sources. This balance, between the desire for a comprehensive and expressive gate set and the constraints of practical hardware, is a central challenge in quantum computing. Although it is theoretically possible to access the entire $\SU{4}$ group in different systems~\cite{sastry2013nonlinear,liu2006controllable,vandersypen2004nmr}, a native two-qubit gate scheme that provides full operational capability yet remains straightforward to implement on state-of-the-art hardware has not yet been demonstrated.

In this work, we experimentally demonstrate the AshN gate scheme~\cite{chen2024one}, capable of natively generating arbitrary two-qubit gates. By addressing the exchange interaction between qubits with near-resonant driving on a superconducting quantum processor, we implement the time-optimal generation of a diverse mixture of commonly used two-qubit gates in one step, achieving an average fidelity of $99.37 \pm 0.07\%$. As a practical application, we generate multi-qubit Dicke states with single and double excitations, achieving notably high accuracy, by leveraging high-fidelity AshN gates and a substantial reduction in the two-qubit gate count compared to traditional $\CNOT$-based implementations. To further enhance its practicality, particularly in terms of calibration costs, we demonstrate native generation of the $\BGate$ gate, the unique operation that enables the synthesis of any two-qubit operation using only two applications of the gate combined with single-qubit rotations~\cite{PhysRevLett.93.020502, wei2024native}. This approach not only minimizes the calibration cost but also facilitates more uniform performance throughout the $\SU{4}$ manifold, enhancing practicality for general-purpose applications.  

The implications of this scheme are multifaceted.  
First, it relies on a single type of interaction---the exchange interaction between $\ket{10}$ and $\ket{01}$---to realize various two-qubit gates, thereby unifying the control strategy. Compared to conventional approaches that rely on different transitions to produce different gate types, our scheme simplifies the frequency allocation strategy in large-scale processors and avoids leakage into non-computational states.
Second, while synthesizing arbitrary two-qubit gates typically requires three applications of $\CZ$\ or $\iSWAP$ gates~\cite{PhysRevA.69.010301}, our approach enables these operations to be implemented directly, with evolution times and fidelities comparable to those of a single $\CZ$\ or $\iSWAP$. By combining the full $\SU{4}$ expressivity with time-optimal implementations of target unitaries supported by the available Hamiltonian, the AshN control scheme is able not only to deliver optimal performance with current superconducting processor architectures,
but also to encourage hardware designers to tailor their systems and algorithm developers to rethink quantum circuits under the AshN framework.
Finally, our scheme only requires qubits to be brought into resonance for exchange interaction and individually driven --- a capability available in many physical platforms.

\section{Implementing the AshN Gate with Superconducting Qubits}

\begin{figure*}[hbtp]
\centering
\includegraphics[scale=1]{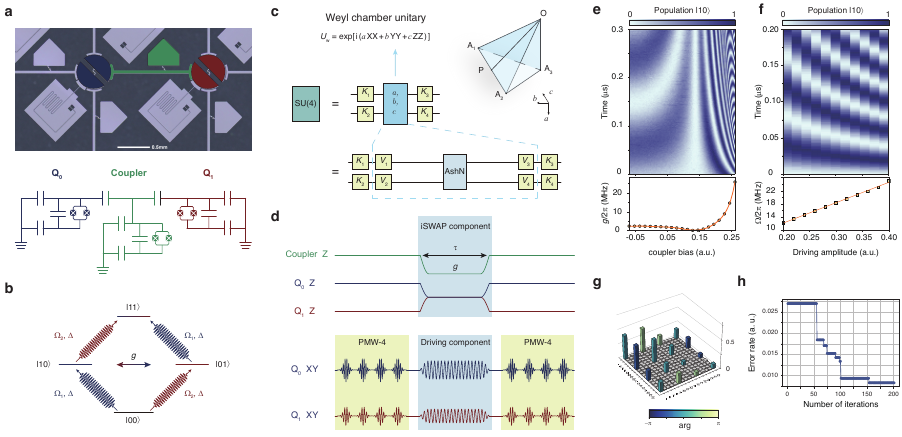}
\caption{
\textbf{Implementation of the AshN scheme with superconducting qubits}.
\textbf{a}, Top: microscope image of the superconducting quantum processor focusing on two transmon qubits (blue and red) and the tunable coupler (green). The meandering wires are quarter-wave transmission-line resonators for qubit readout. Bottom: the circuit diagram.  
\textbf{b}, Energy diagram of a two-qubit subsystem showing the exchange interaction (strength $g$) and qubit drivings (amplitudes $\Omega_{1,2}$ and detuning $\Delta$) used in an AshN gate. 
\textbf{c}, Synthesizing an arbitrary $\SU{4}$ unitary using the AshN gate. According to the KAK decomposition, any $\SU{4}$ operation is equivalent to a Weyl chamber unitary (the part in the dashed box) up to single-qubit operations $K_i$ ($i = 1, 2, 3, 4$). A calibrated AshN gate is also equivalent to the same Weyl chamber unitary up to a different set of single-qubit operations $V_i$ ($i = 1, 2, 3, 4$). The Weyl chamber unitary, derived from the Heisenberg Hamiltonian, can be visualized within a tetrahedral region $\mathrm{OA_{1}A_{2}A_{3}}$ known as the Weyl chamber and parameterized by the three coordinates $(a, b, c)$. The coordinates of the Vertices are $\mathrm{O:(0,0,0)}$, $\mathrm{A_{1}:(\pi/4,\pi/4,-\pi/4)}$, $\mathrm{A_{2}:(\pi/4,\pi/4,\pi/4)}$, $\mathrm{A_{3}:(\pi/4,0,0)}$. 
\textbf{d}, Compiled pulse sequences for an arbitrary $\SU{4}$ operation. The AshN gate control has two parts: the $\iSWAP$ component and the driving component. They are implemented with flux pulses via the Z control lines and microwave pulses via the XY control lines, respectively. The typical duration of flux pulses is 40~ns, including a 5~ns rise and fall time. Successive single-qubit operations ($V_i$ and $K_i$) are merged into a single $\mathrm{SU}(2)$ operation, which is further compiled into four $\pi/2$ pulses using the PMW-4 method.
\textbf{e}, Top panel: measured population swapping at different coupler biases by initializing one of the qubits at its excited state and bringing the two qubits into resonance. Bottom panel: extracted coupling strength versus coupler bias.
\textbf{f}, Top panel: measured Rabi oscillation of one of the qubits at different driving amplitudes. Bottom panel: extracted Rabi frequency versus drive amplitude.
\textbf{g}, Quantum process tomography for KAK decomposition. After combining the $\iSWAP$ component and the driving component simultaneously, we use QPT to infer the compensatory single-qubit gates ($V_i$).
\textbf{h}, Closed-loop optimization of the cross-entropy benchmarking fidelity at a fixed cycle number, typically 50-100, using a Bayesian optimizer.
}
\label{fig1}
\end{figure*}

Our experiment utilizes qubits embedded in a 72-qubit superconducting quantum processor arranged in a square lattice. The qubits are made of tantalum on sapphire~\cite{wang2022towards}, yielding an average $T_1$ relaxation times of 68.8 $\mathrm{\mu s}$. The basic building block consists of two transmon qubits (with frequencies $\omega_1$ and $\omega_2$) and a tunable coupler (frequency $\omega_\mathrm{c}$). The coupler, also by design a transmon qubit, is used to dynamically adjust the qubit-qubit coupling strength for fast gate operations and low crosstalk (Fig.~\ref{fig1}a). All components are frequency-tunable and have a floating design~\cite{sete2016functional}. Individual control lines are directly wired to the corresponding qubits and couplers, delivering diplexed signals for both frequency modulation (DC--500~MHz) and microwave driving (around 4~GHz)~\cite{chu_scalable_2023}. Detailed device information and experimental setup can be found in the Supplementary Materials.

The energy level diagram in Fig.~\ref{fig1}b depicts the computational subspace of two qubits set to resonance with $\omega_1 = \omega_2 = \omega$ and the inter-level couplings, including tunable exchange coupling ($XX+YY$) with strength $g(\omega_\mathrm{c})$ and local driving transitions with amplitudes $\Omega_{1,2}$ and detuning $\Delta = \omega - \omega_\mathrm{d}$ between the qubit frequency and the drive frequency $\omega_\mathrm{d}$. 
These transitions span the entire subspace, and their intensities are all programmable in our system, opening up more possibilities for the system dynamics when acting together.
In the rotating frame, the two-qubit Hamiltonian can be expressed as
\begin{eqnarray}
H  &= \Delta (ZI + IZ)/2 + g (XX+YY)/2 \nonumber \\
&+ \Omega_{1} XI/2 + \Omega_{2} IX/2 \;,
\label{eq:ashnH}
\end{eqnarray}
where $X, Y, Z, I$ are the Pauli operators. It suffices to verify that the operators $XX+YY$, $ZI+IZ$, $XI$, and $IX$ generate the Lie algebra $\mathfrak{su}(4)$ by iteratively applying the Lie bracket operation, enabling the implementation of any unitary operation through a sequence of exponentials of the control Hamiltonians~\cite{sastry2013nonlinear,vandersypen2004nmr}. 

The AshN gate scheme, proposed in Ref.~\cite{chen2024one}, represents a stronger form of the well-established controllability results and provides a straightforward method for utilizing the independent control of the parameters in Eq.~\ref{eq:ashnH} together with the evolution time $\tau$ to generate the local equivalent of an arbitrary two-qubit operation in $\SU{4}$ via a single pulse and, for most cases, within an optimal time evolution. Here, time optimality means the theoretical lower bound of $\tau$ required to achieve a certain unitary given a fixed coupling $g$. In practice, it is crucial to minimize the exposure of the qubit to decoherence. Below, we begin with a brief overview of the AshN scheme and describe our protocol for implementing it with superconducting qubits. 

According to the KAK decomposition or the more general Cartan's decomposition~\cite{tucci2005introduction}, any two-qubit unitary $U \in \mathrm{SU}(4)$ can be expressed as  
\begin{eqnarray}
U = \lambda \cdot (K_1 \otimes K_2) \, U_\mathrm{w} \, (K_3 \otimes K_4) \;,
\label{eq:kak}
\end{eqnarray}
where $\lambda\in\{1,i\}$ and
\begin{eqnarray}
U_\mathrm{w}(a,b,c) = \exp[\mathrm{i} (a XX + b YY + c ZZ)] \;.
\label{eq:uw}
\end{eqnarray}
Here, $K_1, K_2, K_3, K_4 \in \mathrm{SU}(2)$ are single-qubit unitaries and $a,b,c\in \mathbb{R}$. The operators $U$ and $U_\mathrm{w}$ are said to be locally equivalent because they differ only by global phase and single-qubit operations. Due to symmetry reasons, the geometric structure of $U_\mathrm{w}$ can be visualized within a tetrahedral region known as the \emph{Weyl chamber}~\cite{zhang2003geometric}, parameterized by the coordinates $(a, b, c)$ as shown in Fig.~\ref{fig1}c. 
Unitaries that have the same coordinates are then locally equivalent and belong to the same class. For example, some hardware-native gates, such as the Controlled-Z (CZ) gate~\cite{dicarlo2009demonstration, levine2019parallel, xue_quantum_2022} and the Cross-Resonance (CR) gate~\cite{PhysRevLett.107.080502} are all locally equivalent to the $\CNOT$ gate.

For any unitary represented by its local equivalence class $U_\mathrm{w}(a,b,c)$, the AshN scheme provides a convenient algorithm that determines the values of the corresponding control parameters $\Omega_{1}$, $\Omega_{2}$, $\Delta$, and $\tau$ for a given $g$. By applying these control parameters to the Hamiltonian in Eq.~\ref{eq:ashnH} and allowing it to evolve for a duration $\tau$, an AshN gate $U_\mathrm{AshN}$ is generated, which is locally equivalent to the target $U_\mathrm{w}$. Therefore, following the local equivalence relations $U \to U_\mathrm{w} \to U_\mathrm{AshN}$, any two-qubit operation in $\mathrm{SU}(4)$ can be decomposed into an AshN gate sandwiched between single-qubit operations, as illustrated in Fig.~\ref{fig1}c. The single-qubit operations on either side of $U_\mathrm{AshN}$ can be seamlessly merged with adjacent single-qubit operations when implementing quantum algorithms with $U_\mathrm{AshN}$, ensuring that implementing a locally equivalent two-qubit operation does not introduce any additional single-qubit overhead. The AshN scheme is by far the only approach we know that enables native generation of the entire Weyl chamber, thereby achieving maximum expressivity, particularly at the current stage where multi-qubit control remains underdeveloped in most platforms. In comparison, the regions correspond to the $f\text{Sim}$ family can be identified as the three faces of the tetrahedron, i.e.\ $\mathrm{OA_{1}A_{2}}$, $\mathrm{OA_{2}A_{3}}$ and $\mathrm{OA_{3}A_{1}}$ in  Fig.\ref{fig1}c; the line $\mathrm{OP}$ represents the $\text{XY}$ family. 

In our superconducting processor, the AshN gate is realized by synchronizing the time windows of the $XX+YY$ interaction and the qubit driving signals (Fig.~\ref{fig1}d). The former involves adjusting the two qubits to be resonant and turning on the coupling by simultaneously modulating the frequencies of the qubits and coupler via Z pulses. This part alone generates an iSWAP-type operation with a swap angle $\theta = g\tau$ (with $\theta = \pi/2$ corresponding to an exact iSWAP gate); hence we refer to it as the iSWAP component. The latter part involves qubit driving using XY pulses, similar to those in a standard Rabi experiment, and is referred to as the driving component.

Since single-qubit phase gates cannot propagate through a general two-qubit gate, software solutions such as virtual-Z gates or phase-swapping techniques are not directly applicable~\cite{mckay_efficient_2017}. Meanwhile, applying physical phase gates with various phases adds complexity during practical implementation. To address this, we adopt a generalization of the virtual-Z gate that is compatible with arbitrary two-qubit operations: the PMW-4 scheme~\cite{chen2023compiling}. This approach decomposes any $\SU{2}$ operation into four $\pi/2$ microwave pulses (which can be reduced to three if a $\pi$ pulse is used) with analytically calculated phases. In addition to these advantages, maintaining a uniform pulse pattern causes the signal that cross-talks to nearby qubits to impose a fixed Stark effect and thereby simplifying the correction procedure~\cite{PhysRevResearch.6.013015}. Further details on the PMW-4 scheme and the calculation of compensatory gates are provided in the Supplementary Materials.

\begin{figure*}[hbtp]
\centering
\includegraphics[scale=1]{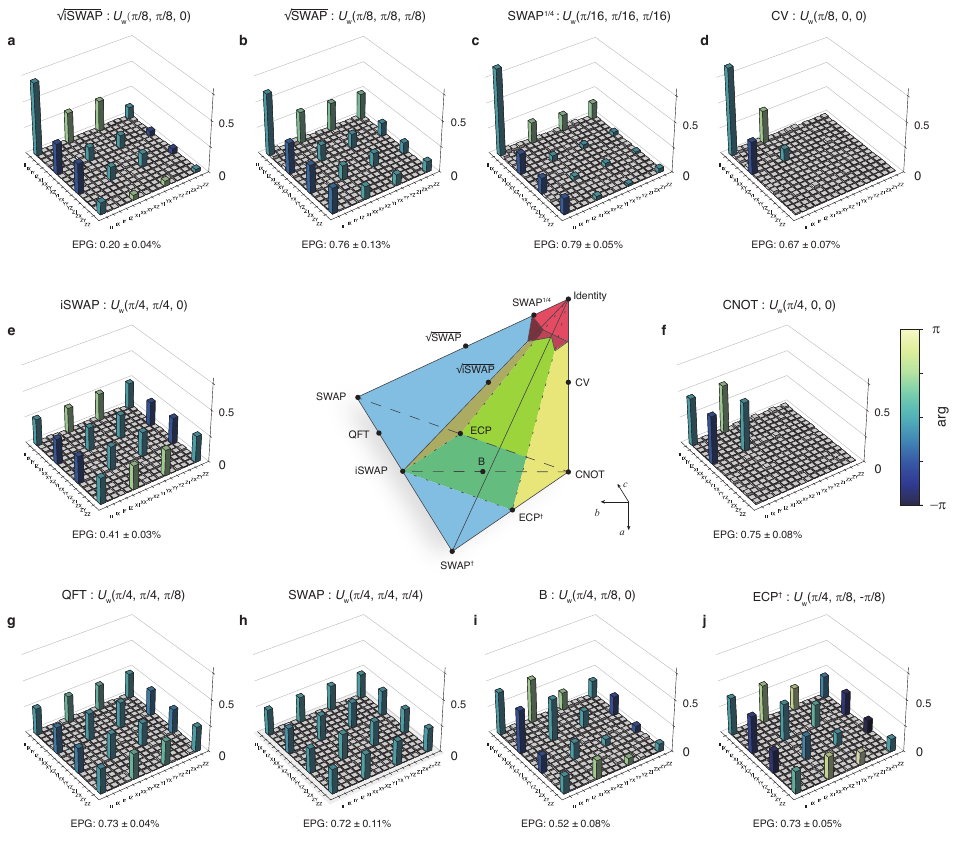}
\caption{
\textbf{Quantum process tomography (QPT) of the AshN gates for several commonly used $\mathrm{SU}(4)$ unitaries.} The error per gate (EPG) is obtained from cross-entropy benchmarking (XEB) experiments to avoid state preparation and measurement errors. The uncertainties represent the standard deviation over five repetitions. See Supplementary Materials for details of the benchmarking results.
Note that throughout this manuscript, we define the Weyl chamber unitary according to Eq.~\ref{eq:uw}. In the otherwise convention with a negative sign on the exponent, the gate we demonstrate here would be the corresponding Hermitian conjugate. Here, CV refers to the Controlled-V gate, i.e., square root of $\CNOT$ gate~\cite{crooks2020gates}. ECP refers to the peak of the pyramid of gates in the Weyl chamber that can be created with a $\sqrt{\mathrm{iSWAP}}$ sandwich~\cite{crooks2020gates,peterson2020fixed}. QFT refers to the quantum Fourier transform, which acts on two qubits and occupies a position halfway between $\mathrm{iSWAP}$ and $\mathrm{SWAP}$~\cite{crooks2020gates}.
}
\label{fig2}
\end{figure*}

\begin{table*}[hbtp]
    \centering
    \def\arraystretch{1.5}
    \begin{tabular}{cccccccc}
    \hline
    \hline
        \textrm{SU}{(4)} & (a,b,c)& $g/2\pi$ (MHz) & $\tau$ (ns) & $\Omega_\mathrm{1}/2\pi$ (MHz) & $\Omega_\mathrm{2}/2\pi$ (MHz)  &$\Delta/2\pi$ (MHz) 
         & $\epsilon $ (\%) \\ 
       & &  & & exp. $|$ th. & exp. $|$ th. & exp. $|$ th. & \\ \hline
       $\sqrt{\mathrm{iSWAP}}$ & $(\frac{\pi}{8},\frac{\pi}{8},0)$ & 6.25 & 20& 0 $ | $ 0& 0 $ | $ 0 & 0 $ | $ 0 &  0.20 \\    
       iSWAP & $(\frac{\pi}{4},\frac{\pi}{4},0)$ & 6.25 &  40& 0 $|$ 0 & 0 $|$ 0 & 0 $|$ 0 &  0.41   \\ 
        $\mathrm{SWAP}^{1/4}$ & $(\frac{\pi}{16},\frac{\pi}{16},\frac{\pi}{16})$ & 6.25 &  70& 18.13 $|$ 10.92& 4.58 $|$ 1.93& 0.02 $|$ 0& 0.79   \\
      $\sqrt{\mathrm{SWAP}}$ & $(\frac{\pi}{8},\frac{\pi}{8},\frac{\pi}{8})$ & 6.25 &  60& 13.66 $|$ 4.62& 17.60 $|$ 10.83& 1.41 $|$ 0&  0.76   \\ 
       SWAP & $(\frac{\pi}{4},\frac{\pi}{4},\frac{\pi}{4})$ & 6.25 & 60& 13.66 $|$ 13.18 & 17.60 $|$ 13.18& 10.00 $|$ 9.55&  0.72   \\
      CV & $(\frac{\pi}{8},0,0)$ & 6.25 & 60& 21.00 $|$ 15.45& 1.80 $|$ 0& 0.41 $|$ 0&  0.67  \\ 
      CNOT & $(\frac{\pi}{4},0,0)$ & 6.25 &  40& 32.65 $|$ 24.21& 1.45 $|$ 0& 1.21 $|$ 0 & 0.75  \\
       B & $(\frac{\pi}{4},\frac{\pi}{8},0)$ & 6.25 &  40& 18.60 $|$ 13.99& 1.00 $|$ 0& 1.94 $|$ 0 & 0.52   \\ 
     $\mathrm{ECP}^\dagger$ & $(\frac{\pi}{4},\frac{\pi}{8},-\frac{\pi}{8})$ &6.25 &  40& 19.86 $|$ 12.10& 11.95$|$ 12.10& 2.18 $|$ 0 & 0.73   \\ 
      QFT & $(\frac{\pi}{4},\frac{\pi}{4},\frac{\pi}{8})$ & 6.25 & 50& 9.42 $|$  12.42& 20.30 $|$ 12.42& 15.35 $|$ 15.98 & 0.73   \\ 
         \hline
         \hline
    \end{tabular}
    \caption{\label{table:ashn} Parameters of ten AshN-generated gates and comparison between experiment and theory. 
    In the experiment, the pulse total time $\tau$ includes rising and falling edges (each 2.5~ns). 
    $\epsilon$ is the average gate error from the XEB experiments. 
    For gates in the $\mathrm{iSWAP}$ family, we fix the $\Omega_\mathrm{1}$, $\Omega_\mathrm{2}$ and $\Delta$ to zero.
   We set $\tau$ to the theoretical optimal time for all gates, except for $\mathrm{SWAP}^{1/4}$, $\sqrt{\mathrm{SWAP}}$, and CV, where the optimal times are $15$ ns, $30$ ns, and $20$ ns, respectively. In these cases, the driving amplitudes required for the optimal time exceeded the range of our setup. Therefore we chose to implement another variant of the AshN scheme with an alternative evolution time.}
\end{table*}

The calibration of AshN gates follows a systematic procedure. First, we separately calibrate the iSWAP and driving components to determine the dependencies of the control parameters $g$ and $\Omega$, as shown in Fig.\ref{fig1}e and Fig.\ref{fig1}f, providing an initial estimate of the control parameters for targeting $U_\mathrm{w}(a,b,c)$. Next, we combine the iSWAP and driving components by diplexing Z pulses and XY pulses to evaluate the gate’s performance and proceed with multiple stages of optimization. In the first stage, quantum process tomography (QPT) is performed to extract the actual values of $(a, b, c)$ and minimize the distance from the target coordinates (Fig.\ref{fig1}g). Following this, single-qubit compensatory gates are fixed, and the second optimization stage fine-tunes the control parameters by minimizing the error rate through cross-entropy benchmarking (XEB), as shown in Fig.\ref{fig1}h. This stepwise calibration process ensures precise control and optimal performance of the AshN gates.

We demonstrate the expressivity of the AshN scheme by implementing a diverse set of commonly encountered $\mathrm{SU}(4)$ unitaries, covering a wide range of points across the Weyl chamber (Fig.~\ref{fig2}). Our selection also covers three distinct regions within the tetrahedron, highlighted in blue, yellow, and red, each corresponding to a specific protocol variant for converting the Weyl chamber coordinates $(a, b, c)$ into the control parameters ($g$, $\Omega_{1}$, $\Omega_{2}$, $\Delta$ and $\tau$). In this experiment, we first assume a uniform coupling strength $g$ and estimate the gate time for each AshN gate. We then fix the gate time and optimize the coupling strength and other parameters. Gate information is listed in Table~\ref{table:ashn} for comparison. 
We benchmark all these AshN gates using both QPT and XEB, with the average XEB fidelity reaches $99.37 \pm 0.07\%$.
The standard deviation of error within these gates is 0.18\%, yielding a relative standard deviation of 29\%. Such a strong variance is unsurprising, given that the control parameters cover a wide range. For example, the gate time ranges from 20~ns for $\sqrt{\mathrm{iSWAP}}$ to 70~ns for $\mathrm{SWAP^{1/4}}$, leading to drastically different susceptibility to decoherence and other non-idealities.
In general, we find that the gate error rate increases with gate time, which can be explained by the measured decoherence during interaction (see Supplementary Materials for details).

\section{Multipartite Entangled State Generation Using AshN Gates}

\begin{figure*}[hbtp]
\centering
\includegraphics[scale=1]{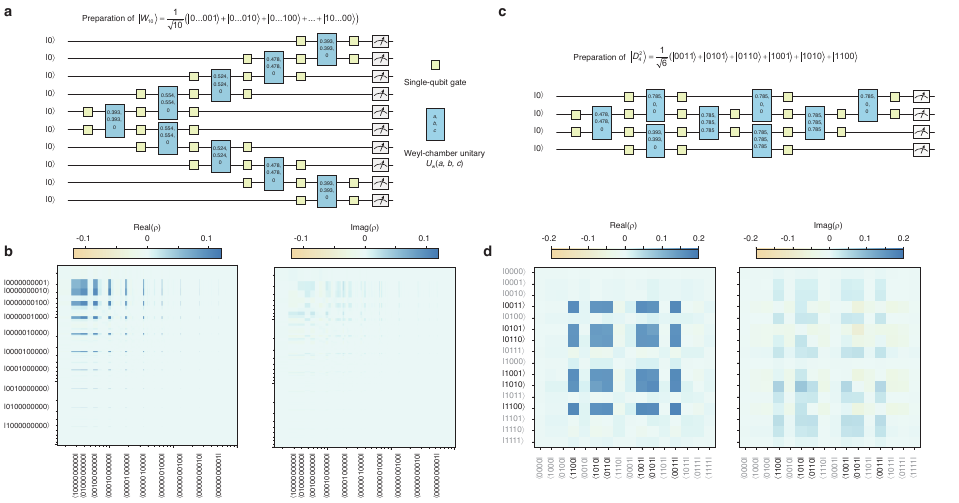}
\caption{
\textbf{Synthesizing multipartite entangled states using AshN gates.}
\textbf{a}, Circuit to prepare the 10-qubit W state $W_{10}$.
\textbf{b}, Tomography result of the $W_{10}$ state with an estimated fidelity of $\mathrm{0.913 \pm 0.012 }$.
\textbf{c}, Circuit to prepare the 4-qubit two-excitation Dicke state $D_{4}^{2}$.
\textbf{d}, Quantum state tomography of the $D_{4}^{2}$ state with an estimated fidelity of $\mathrm{0.926 \pm 0.001}$.
}
\label{fig4}
\end{figure*}

To show the advantages of the AshN gates, we demonstrate the efficient preparation of Dicke states with single and double excitations. These states form a family of highly entangled states with important applications in quantum computation, quantum networking, and quantum metrology~\cite{luo2017deterministic,prevedel2009experimental,PhysRevLett.98.063604}.

The single-excitation Dicke state is also known as the $W$ state. Here, we implement a $10$-qubit $W$ state, $\ket{W_{10}}$, using the circuit shown in Fig.~\ref{fig4}a which consists of $9$ two-qubit gates. This protocol also enables the generation of an $N$-qubit $W$ state using only $(N-1)$ two-qubit gates, which represents the theoretical lower bound under the assumption of no multi-qubit gates or ancillary qubits. In comparison, achieving the same goal with $\CNOT$ gates has a provable lower bound of $\frac{15N-3}{14}$ gates~\cite{huang2023quantum}, which already exceeds $N-1$. Moreover, numerical evidence suggests that the actual lower bound is $(2N-3)$ (see Supplementary Materials for details). We select $N=10$ to balance the cost of quantum state tomography, and this choice adequately demonstrates the advantage. We repeat standard quantum state tomography seven times and estimate the fidelity to be $\mathrm{0.913 \pm 0.012 }$ (Fig.~\ref{fig4}b), surpassing previous demonstrations~\cite{Häffner2005, PhysRevLett.124.013601, PhysRevLett.133.160801}. With our method, it is also straightforward to generate $W$ states with arbitrary phases.

Adding more excitations to the Dicke state significantly increases the circuit complexity, posing substantial challenges for solid-state qubits, which often suffer from limited connectivity. 
By utilizing the full $\SU{4}$ expressivity of the AshN gates, we implement a circuit that generates the 4-qubit double-excitation Dicke state $\ket{D_{4}^{2}}$ with eight $\SU{4}$ operations (Fig.~\ref{fig4}c); for comparison, an existing scheme for exact synthesis of such a state requires fourteen $\CNOT$ gates.~\cite{bartschi2019deterministic}.
Our final state yields a fidelity of $\mathrm{0.926 \pm 0.001} $(Fig.~\ref{fig4}d), demonstrating the advantage of the digital synthesis of such a complex entangled state compared to the analog approach~\cite{PhysRevLett.98.063604,PhysRevLett.130.223601} (see Supplementary Materials for details).

\section{Uniform Synthesis of Arbitrary $\mathrm{SU}(4)$ Operations Using B Gates}
\label{sec:synthsis_b}

\begin{figure}[hbtp]
\centering
\includegraphics[scale=1]{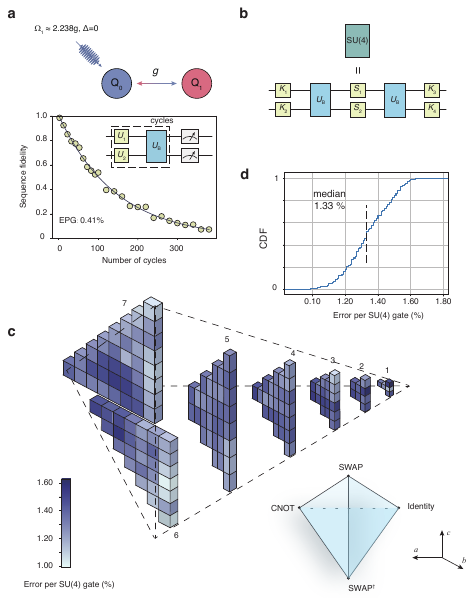}
\caption{
\textbf{Synthesizing arbitrary $\mathrm{SU}(4)$ operations with $\BGate$ gates.} \textbf{a}, Cross-entropy benchmarking (XEB) of the $\BGate$ gate. Under the AshN framework, the $\BGate$ gate is achieved by driving one of the qubits at a Rabi frequency of $\Omega_1 \approx 2.238g$ (top panel). \textbf{b}, Decomposition of an arbitrary $\mathrm{SU}(4)$ into two $\BGate$ gates plus single-qubit operations. \textbf{c}, Error per gate for each $\BGate$-composed Weyl chamber unitary $U_\mathrm{w}$. We sample 152 different unitaries, uniformly selecting 7 plane cuts in the Weyl chamber (planes 6 and 7 are very close to each other). For better visualization, we changed the Weyl chamber orientation as indicated in the lower right. \textbf{d}, Cumulative distribution of the gate errors.
}
\label{fig3}
\end{figure}

While we have demonstrated the capability to implement arbitrary two-qubit operations in a single step, realizing the full $\SU{4}$ --- there are an infinite number of possibilities --- remains practically infeasible due to the cost of calibration. We will explain in the Supplementary Materials how this issue can be resolved at the software level. Here we introduce an alternative approach for efficiently implementing arbitrary two-qubit operation. The $\BGate$ gate, located at $(\pi/4,\, \pi/8,\, 0)$ --- halfway between the iSWAP and CNOT in the Weyl chamber --- is known as a unique two-qubit gate that can synthesize any two-qubit operation with only two applications~\cite{PhysRevLett.93.020502}, compared to the three applications needed for CNOT or iSWAP (see Supplementary Materials for details). This property makes the $\BGate$ gate a promising building block for efficient circuit compilation.

The native $\BGate$ gate can be conveniently generated using the AshN scheme, which requires driving only one of the qubits ($g = \pi/2\tau$, $\Omega_1 \approx 2.238g$, $\Omega_2=0$, $\Delta = 0$). In Fig.~\ref{fig3}a, we show the XEB result of an AshN-generated B gate; the gate error is 0.41\% with a 40-ns pulse.

Using this calibrated $\BGate$ gate and the decomposition rule (Fig.~\ref{fig3}b), we reconstruct the Weyl chamber and plot the error map in Fig.~\ref{fig3}c.
The average error per $\mathrm{SU}(4)$ and the standard deviation are 1.34\% and 0.14\%, respectively, which gives a substantially improved relative standard deviation of 10\% compared to the native AshN case. This is due to the fact that all $\mathrm{SU}(4)$ gates are generated in a uniform fashion with two applications of $\BGate$ gates. This enhanced uniformity leads to more stable and predictable circuit performance, which can be highly beneficial in scenarios like error analysis and error mitigation that rely on a stable noise model~\cite{kim2023evidence}. 
In addition to the uniformity improvement, we observe smaller errors near the boundary between SWAP and SWAP$^\dagger$ within the Weyl chamber. This can be attributed to the faster dephasing of one qubit, as well as the reduced sensitivity of the single-qubit operation on this qubit to dephasing noise when $a = b = \pi/4$ (see Supplementary Materials for details).

Another advantage of using the $\BGate$ gate is the convenience in synthesizing unitary operations near the identity region, the red pinnacle in the tetrahedron (Fig.~\ref{fig2}). Gates in this small region remain essential for Trotterization in Hamiltonian simulations~\cite{daley2022practical}. Implementing these gates with the time-optimal AshN protocol may require impractically large amplitudes, making experimental realization challenging. In contrast, the exact synthesis enabled by the $\BGate$ gate, combined with its uniform performance, is ideally suited for this situation.

\section{Discussion}

Our demonstration of AshN gates has broad implications for both experimentalists and theorists. We show that the AshN scheme is compatible with state-of-the-art hardware designs, which have been successfully employed in recent milestone demonstrations~\cite{acharya2024quantum,cao2023generation,zhang2022digital}. This suggests that the AshN scheme can significantly enhance the functionality and performance of existing systems, as evidenced by its validation in a wide range of benchmarking quantum algorithms~\cite{yang2024_reQISC} and its application to error correction codes~\cite{zhou2024_defects}.
One immediate benefit is that traditional gates, such as the CNOT or CZ gate, supported by the $\ket{11}-\ket{20}$ transition, typically require longer evolution times compared to iSWAP-like gates, leading to degraded fidelity performance~\cite{acharya2024quantum,morvan_phase_2024}. The AshN-generated CZ gate, however, shares the same gate time as the iSWAP gate, which may help eliminate this fidelity gap.
Furthermore, since the AshN scheme only requires a single resonant condition, it simplifies the frequency allocation problem and reduces the likelihood of collisions with spurious two-level systems—one of the main challenges when optimizing the overall performance of multi-qubit devices~\cite{klimov_optimizing_2024}.
Another particularly important aspect is that the exclusive use of the $\ket{10}-\ket{01}$ transition helps avoid leakage into non-computational states. This advantage in reducing correlated errors in error correction codes has been experimentally demonstrated recently~\cite{eickbusch2024demonstratingdynamicsurfacecodes}.

The AshN scheme requires only that qubits be tuned to resonance and possess exchange-type interactions (a requirement that may be relaxed in an extended AshN scheme~\cite{yang2024_reQISC}), along with the ability to drive qubits independently. Given these minimal requirements, many other promising qubit modalities, such as superconducting fluxonium qubits~\cite{nguyen2019high,bao_fluxonium_2022,ding2023high} or alternative physical platforms including semiconductor spin qubits~\cite{dijkema2024cavity}, neutral atoms~\cite{bluvstein_logical_2024}, and molecular qubits~\cite{picard2024entanglement} could immediately benefit from implementing AshN gates.

Looking ahead, this successful demonstration may inspire researchers to rethink algorithm and application design, encouraging a move beyond the CNOT-based systems that have dominated the past few decades. In addition to its demonstrated advantages in state preparation and the direct implementation of $\SWAP$ which is useful for efficient routing, the AshN scheme retains the benefits of earlier gate sets, including the XY family for variational quantum eigensolvers~\cite{abrams2020implementation} and fractional gates for simulating Ising models~\cite{ibm_fractional_gates}. The elegant structure of $\SU{4}$ also enables more refined and efficient constructions than previously studied gate sets. Furthermore, extending the scheme to multiple qubits—where more qubits can be prepared in resonance and driven independently, as in ion trap systems~\cite{PhysRevLett.102.040501}—could further enrich the capabilities of multi-qubit operations.

By demonstrating this flexible and efficient two-qubit gate scheme, our work opens new avenues for optimizing quantum hardware and algorithm design, bringing practical quantum computing a significant step closer to reality.

{\bf Acknowledgment:}
The authors thank Cheng Chen, Chunqing Deng, Xiangliang Li, Khiwan Kim, Feng Wu, Xiao Xue, Hao Zhang, Jun Zhang, Huihai Zhao for their valuable comments. This work was supported by the National Natural Science Foundation of China (Grants No. 12322413, No. 92365206, and No. 92476206, No. 12347104, No. 12404563), National Key Research and Development Program of China (Grant No. 2023YFA1009403),  Innovation Program for Quantum Science and Technology (Grants No.2021ZD0301802), Beijing Natural Science Foundation (Grant No. Z220002, Grant No. 1244065).

\bibliography{AshN}

\begin{thebibliography}{57}%
\makeatletter
\providecommand \@ifxundefined [1]{%
 \@ifx{#1\undefined}
}%
\providecommand \@ifnum [1]{%
 \ifnum #1\expandafter \@firstoftwo
 \else \expandafter \@secondoftwo
 \fi
}%
\providecommand \@ifx [1]{%
 \ifx #1\expandafter \@firstoftwo
 \else \expandafter \@secondoftwo
 \fi
}%
\providecommand \natexlab [1]{#1}%
\providecommand \enquote  [1]{``#1''}%
\providecommand \bibnamefont  [1]{#1}%
\providecommand \bibfnamefont [1]{#1}%
\providecommand \citenamefont [1]{#1}%
\providecommand \href@noop [0]{\@secondoftwo}%
\providecommand \href [0]{\begingroup \@sanitize@url \@href}%
\providecommand \@href[1]{\@@startlink{#1}\@@href}%
\providecommand \@@href[1]{\endgroup#1\@@endlink}%
\providecommand \@sanitize@url [0]{\catcode `\\12\catcode `\$12\catcode
  `\&12\catcode `\#12\catcode `\^12\catcode `\_12\catcode `\%12\relax}%
\providecommand \@@startlink[1]{}%
\providecommand \@@endlink[0]{}%
\providecommand \url  [0]{\begingroup\@sanitize@url \@url }%
\providecommand \@url [1]{\endgroup\@href {#1}{\urlprefix }}%
\providecommand \urlprefix  [0]{URL }%
\providecommand \Eprint [0]{\href }%
\providecommand \doibase [0]{https://doi.org/}%
\providecommand \selectlanguage [0]{\@gobble}%
\providecommand \bibinfo  [0]{\@secondoftwo}%
\providecommand \bibfield  [0]{\@secondoftwo}%
\providecommand \translation [1]{[#1]}%
\providecommand \BibitemOpen [0]{}%
\providecommand \bibitemStop [0]{}%
\providecommand \bibitemNoStop [0]{.\EOS\space}%
\providecommand \EOS [0]{\spacefactor3000\relax}%
\providecommand \BibitemShut  [1]{\csname bibitem#1\endcsname}%
\let\auto@bib@innerbib\@empty
\bibitem [{\citenamefont {Vidal}\ \emph {et~al.}(2002)\citenamefont {Vidal},
  \citenamefont {Hammerer},\ and\ \citenamefont
  {Cirac}}]{vidal2002interaction}%
  \BibitemOpen
  \bibfield  {author} {\bibinfo {author} {\bibfnamefont {G.}~\bibnamefont
  {Vidal}}, \bibinfo {author} {\bibfnamefont {K.}~\bibnamefont {Hammerer}},\
  and\ \bibinfo {author} {\bibfnamefont {J.~I.}\ \bibnamefont {Cirac}},\
  }\bibfield  {title} {\bibinfo {title} {Interaction cost of nonlocal gates},\
  }\href {https://doi.org/10.1103/PhysRevLett.88.237902} {\bibfield  {journal}
  {\bibinfo  {journal} {Phys. Rev. Lett.}\ }\textbf {\bibinfo {volume} {88}},\
  \bibinfo {pages} {237902} (\bibinfo {year} {2002})}\BibitemShut {NoStop}%
\bibitem [{\citenamefont {Chong}\ \emph {et~al.}(2017)\citenamefont {Chong},
  \citenamefont {Franklin},\ and\ \citenamefont
  {Martonosi}}]{chong2017programming}%
  \BibitemOpen
  \bibfield  {author} {\bibinfo {author} {\bibfnamefont {F.~T.}\ \bibnamefont
  {Chong}}, \bibinfo {author} {\bibfnamefont {D.}~\bibnamefont {Franklin}},\
  and\ \bibinfo {author} {\bibfnamefont {M.}~\bibnamefont {Martonosi}},\
  }\bibfield  {title} {\bibinfo {title} {Programming languages and compiler
  design for realistic quantum hardware},\ }\href
  {https://doi.org/10.1038/nature23459} {\bibfield  {journal} {\bibinfo
  {journal} {Nature}\ }\textbf {\bibinfo {volume} {549}},\ \bibinfo {pages}
  {180} (\bibinfo {year} {2017})}\BibitemShut {NoStop}%
\bibitem [{\citenamefont {Sung}\ \emph {et~al.}(2021)\citenamefont {Sung},
  \citenamefont {Ding}, \citenamefont {Braum\"uller}, \citenamefont
  {Veps\"al\"ainen}, \citenamefont {Kannan}, \citenamefont {Kjaergaard},
  \citenamefont {Greene}, \citenamefont {Samach}, \citenamefont {McNally},
  \citenamefont {Kim}, \citenamefont {Melville}, \citenamefont {Niedzielski},
  \citenamefont {Schwartz}, \citenamefont {Yoder}, \citenamefont {Orlando},
  \citenamefont {Gustavsson},\ and\ \citenamefont
  {Oliver}}]{sung2021realization}%
  \BibitemOpen
  \bibfield  {author} {\bibinfo {author} {\bibfnamefont {Y.}~\bibnamefont
  {Sung}}, \bibinfo {author} {\bibfnamefont {L.}~\bibnamefont {Ding}}, \bibinfo
  {author} {\bibfnamefont {J.}~\bibnamefont {Braum\"uller}}, \bibinfo {author}
  {\bibfnamefont {A.}~\bibnamefont {Veps\"al\"ainen}}, \bibinfo {author}
  {\bibfnamefont {B.}~\bibnamefont {Kannan}}, \bibinfo {author} {\bibfnamefont
  {M.}~\bibnamefont {Kjaergaard}}, \bibinfo {author} {\bibfnamefont
  {A.}~\bibnamefont {Greene}}, \bibinfo {author} {\bibfnamefont {G.~O.}\
  \bibnamefont {Samach}}, \bibinfo {author} {\bibfnamefont {C.}~\bibnamefont
  {McNally}}, \bibinfo {author} {\bibfnamefont {D.}~\bibnamefont {Kim}},
  \bibinfo {author} {\bibfnamefont {A.}~\bibnamefont {Melville}}, \bibinfo
  {author} {\bibfnamefont {B.~M.}\ \bibnamefont {Niedzielski}}, \bibinfo
  {author} {\bibfnamefont {M.~E.}\ \bibnamefont {Schwartz}}, \bibinfo {author}
  {\bibfnamefont {J.~L.}\ \bibnamefont {Yoder}}, \bibinfo {author}
  {\bibfnamefont {T.~P.}\ \bibnamefont {Orlando}}, \bibinfo {author}
  {\bibfnamefont {S.}~\bibnamefont {Gustavsson}},\ and\ \bibinfo {author}
  {\bibfnamefont {W.~D.}\ \bibnamefont {Oliver}},\ }\bibfield  {title}
  {\bibinfo {title} {Realization of high-fidelity cz and $zz$-free iswap gates
  with a tunable coupler},\ }\href {https://doi.org/10.1103/PhysRevX.11.021058}
  {\bibfield  {journal} {\bibinfo  {journal} {Phys. Rev. X}\ }\textbf {\bibinfo
  {volume} {11}},\ \bibinfo {pages} {021058} (\bibinfo {year}
  {2021})}\BibitemShut {NoStop}%
\bibitem [{\citenamefont {Caldwell}\ \emph {et~al.}(2018)\citenamefont
  {Caldwell}, \citenamefont {Didier}, \citenamefont {Ryan}, \citenamefont
  {Sete}, \citenamefont {Hudson}, \citenamefont {Karalekas}, \citenamefont
  {Manenti}, \citenamefont {da~Silva}, \citenamefont {Sinclair}, \citenamefont
  {Acala}, \citenamefont {Alidoust}, \citenamefont {Angeles}, \citenamefont
  {Bestwick}, \citenamefont {Block}, \citenamefont {Bloom}, \citenamefont
  {Bradley}, \citenamefont {Bui}, \citenamefont {Capelluto}, \citenamefont
  {Chilcott}, \citenamefont {Cordova}, \citenamefont {Crossman}, \citenamefont
  {Curtis}, \citenamefont {Deshpande}, \citenamefont {Bouayadi}, \citenamefont
  {Girshovich}, \citenamefont {Hong}, \citenamefont {Kuang}, \citenamefont
  {Lenihan}, \citenamefont {Manning}, \citenamefont {Marchenkov}, \citenamefont
  {Marshall}, \citenamefont {Maydra}, \citenamefont {Mohan}, \citenamefont
  {O'Brien}, \citenamefont {Osborn}, \citenamefont {Otterbach}, \citenamefont
  {Papageorge}, \citenamefont {Paquette}, \citenamefont {Pelstring},
  \citenamefont {Polloreno}, \citenamefont {Prawiroatmodjo}, \citenamefont
  {Rawat}, \citenamefont {Reagor}, \citenamefont {Renzas}, \citenamefont
  {Rubin}, \citenamefont {Russell}, \citenamefont {Rust}, \citenamefont
  {Scarabelli}, \citenamefont {Scheer}, \citenamefont {Selvanayagam},
  \citenamefont {Smith}, \citenamefont {Staley}, \citenamefont {Suska},
  \citenamefont {Tezak}, \citenamefont {Thompson}, \citenamefont {To},
  \citenamefont {Vahidpour}, \citenamefont {Vodrahalli}, \citenamefont
  {Whyland}, \citenamefont {Yadav}, \citenamefont {Zeng},\ and\ \citenamefont
  {Rigetti}}]{PhysRevApplied.10.034050}%
  \BibitemOpen
  \bibfield  {author} {\bibinfo {author} {\bibfnamefont {S.~A.}\ \bibnamefont
  {Caldwell}}, \bibinfo {author} {\bibfnamefont {N.}~\bibnamefont {Didier}},
  \bibinfo {author} {\bibfnamefont {C.~A.}\ \bibnamefont {Ryan}}, \bibinfo
  {author} {\bibfnamefont {E.~A.}\ \bibnamefont {Sete}}, \bibinfo {author}
  {\bibfnamefont {A.}~\bibnamefont {Hudson}}, \bibinfo {author} {\bibfnamefont
  {P.}~\bibnamefont {Karalekas}}, \bibinfo {author} {\bibfnamefont
  {R.}~\bibnamefont {Manenti}}, \bibinfo {author} {\bibfnamefont {M.~P.}\
  \bibnamefont {da~Silva}}, \bibinfo {author} {\bibfnamefont {R.}~\bibnamefont
  {Sinclair}}, \bibinfo {author} {\bibfnamefont {E.}~\bibnamefont {Acala}},
  \bibinfo {author} {\bibfnamefont {N.}~\bibnamefont {Alidoust}}, \bibinfo
  {author} {\bibfnamefont {J.}~\bibnamefont {Angeles}}, \bibinfo {author}
  {\bibfnamefont {A.}~\bibnamefont {Bestwick}}, \bibinfo {author}
  {\bibfnamefont {M.}~\bibnamefont {Block}}, \bibinfo {author} {\bibfnamefont
  {B.}~\bibnamefont {Bloom}}, \bibinfo {author} {\bibfnamefont
  {A.}~\bibnamefont {Bradley}}, \bibinfo {author} {\bibfnamefont
  {C.}~\bibnamefont {Bui}}, \bibinfo {author} {\bibfnamefont {L.}~\bibnamefont
  {Capelluto}}, \bibinfo {author} {\bibfnamefont {R.}~\bibnamefont {Chilcott}},
  \bibinfo {author} {\bibfnamefont {J.}~\bibnamefont {Cordova}}, \bibinfo
  {author} {\bibfnamefont {G.}~\bibnamefont {Crossman}}, \bibinfo {author}
  {\bibfnamefont {M.}~\bibnamefont {Curtis}}, \bibinfo {author} {\bibfnamefont
  {S.}~\bibnamefont {Deshpande}}, \bibinfo {author} {\bibfnamefont {T.~E.}\
  \bibnamefont {Bouayadi}}, \bibinfo {author} {\bibfnamefont {D.}~\bibnamefont
  {Girshovich}}, \bibinfo {author} {\bibfnamefont {S.}~\bibnamefont {Hong}},
  \bibinfo {author} {\bibfnamefont {K.}~\bibnamefont {Kuang}}, \bibinfo
  {author} {\bibfnamefont {M.}~\bibnamefont {Lenihan}}, \bibinfo {author}
  {\bibfnamefont {T.}~\bibnamefont {Manning}}, \bibinfo {author} {\bibfnamefont
  {A.}~\bibnamefont {Marchenkov}}, \bibinfo {author} {\bibfnamefont
  {J.}~\bibnamefont {Marshall}}, \bibinfo {author} {\bibfnamefont
  {R.}~\bibnamefont {Maydra}}, \bibinfo {author} {\bibfnamefont
  {Y.}~\bibnamefont {Mohan}}, \bibinfo {author} {\bibfnamefont
  {W.}~\bibnamefont {O'Brien}}, \bibinfo {author} {\bibfnamefont
  {C.}~\bibnamefont {Osborn}}, \bibinfo {author} {\bibfnamefont
  {J.}~\bibnamefont {Otterbach}}, \bibinfo {author} {\bibfnamefont
  {A.}~\bibnamefont {Papageorge}}, \bibinfo {author} {\bibfnamefont {J.-P.}\
  \bibnamefont {Paquette}}, \bibinfo {author} {\bibfnamefont {M.}~\bibnamefont
  {Pelstring}}, \bibinfo {author} {\bibfnamefont {A.}~\bibnamefont
  {Polloreno}}, \bibinfo {author} {\bibfnamefont {G.}~\bibnamefont
  {Prawiroatmodjo}}, \bibinfo {author} {\bibfnamefont {V.}~\bibnamefont
  {Rawat}}, \bibinfo {author} {\bibfnamefont {M.}~\bibnamefont {Reagor}},
  \bibinfo {author} {\bibfnamefont {R.}~\bibnamefont {Renzas}}, \bibinfo
  {author} {\bibfnamefont {N.}~\bibnamefont {Rubin}}, \bibinfo {author}
  {\bibfnamefont {D.}~\bibnamefont {Russell}}, \bibinfo {author} {\bibfnamefont
  {M.}~\bibnamefont {Rust}}, \bibinfo {author} {\bibfnamefont {D.}~\bibnamefont
  {Scarabelli}}, \bibinfo {author} {\bibfnamefont {M.}~\bibnamefont {Scheer}},
  \bibinfo {author} {\bibfnamefont {M.}~\bibnamefont {Selvanayagam}}, \bibinfo
  {author} {\bibfnamefont {R.}~\bibnamefont {Smith}}, \bibinfo {author}
  {\bibfnamefont {A.}~\bibnamefont {Staley}}, \bibinfo {author} {\bibfnamefont
  {M.}~\bibnamefont {Suska}}, \bibinfo {author} {\bibfnamefont
  {N.}~\bibnamefont {Tezak}}, \bibinfo {author} {\bibfnamefont {D.~C.}\
  \bibnamefont {Thompson}}, \bibinfo {author} {\bibfnamefont {T.-W.}\
  \bibnamefont {To}}, \bibinfo {author} {\bibfnamefont {M.}~\bibnamefont
  {Vahidpour}}, \bibinfo {author} {\bibfnamefont {N.}~\bibnamefont
  {Vodrahalli}}, \bibinfo {author} {\bibfnamefont {T.}~\bibnamefont {Whyland}},
  \bibinfo {author} {\bibfnamefont {K.}~\bibnamefont {Yadav}}, \bibinfo
  {author} {\bibfnamefont {W.}~\bibnamefont {Zeng}},\ and\ \bibinfo {author}
  {\bibfnamefont {C.}~\bibnamefont {Rigetti}},\ }\bibfield  {title} {\bibinfo
  {title} {Parametrically activated entangling gates using transmon qubits},\
  }\href {https://doi.org/10.1103/PhysRevApplied.10.034050} {\bibfield
  {journal} {\bibinfo  {journal} {Phys. Rev. Appl.}\ }\textbf {\bibinfo
  {volume} {10}},\ \bibinfo {pages} {034050} (\bibinfo {year}
  {2018})}\BibitemShut {NoStop}%
\bibitem [{\citenamefont {Foxen}\ \emph {et~al.}(2020)\citenamefont {Foxen},
  \citenamefont {Neill}, \citenamefont {Dunsworth}, \citenamefont {Roushan},
  \citenamefont {Chiaro}, \citenamefont {Megrant}, \citenamefont {Kelly},
  \citenamefont {Chen}, \citenamefont {Satzinger}, \citenamefont {Barends},
  \citenamefont {Arute}, \citenamefont {Arya}, \citenamefont {Babbush},
  \citenamefont {Bacon}, \citenamefont {Bardin}, \citenamefont {Boixo},
  \citenamefont {Buell}, \citenamefont {Burkett}, \citenamefont {Chen},
  \citenamefont {Collins}, \citenamefont {Farhi}, \citenamefont {Fowler},
  \citenamefont {Gidney}, \citenamefont {Giustina}, \citenamefont {Graff},
  \citenamefont {Harrigan}, \citenamefont {Huang}, \citenamefont {Isakov},
  \citenamefont {Jeffrey}, \citenamefont {Jiang}, \citenamefont {Kafri},
  \citenamefont {Kechedzhi}, \citenamefont {Klimov}, \citenamefont {Korotkov},
  \citenamefont {Kostritsa}, \citenamefont {Landhuis}, \citenamefont {Lucero},
  \citenamefont {McClean}, \citenamefont {McEwen}, \citenamefont {Mi},
  \citenamefont {Mohseni}, \citenamefont {Mutus}, \citenamefont {Naaman},
  \citenamefont {Neeley}, \citenamefont {Niu}, \citenamefont {Petukhov},
  \citenamefont {Quintana}, \citenamefont {Rubin}, \citenamefont {Sank},
  \citenamefont {Smelyanskiy}, \citenamefont {Vainsencher}, \citenamefont
  {White}, \citenamefont {Yao}, \citenamefont {Yeh}, \citenamefont {Zalcman},
  \citenamefont {Neven},\ and\ \citenamefont
  {Martinis}}]{foxen2020demonstrating}%
  \BibitemOpen
  \bibfield  {author} {\bibinfo {author} {\bibfnamefont {B.}~\bibnamefont
  {Foxen}}, \bibinfo {author} {\bibfnamefont {C.}~\bibnamefont {Neill}},
  \bibinfo {author} {\bibfnamefont {A.}~\bibnamefont {Dunsworth}}, \bibinfo
  {author} {\bibfnamefont {P.}~\bibnamefont {Roushan}}, \bibinfo {author}
  {\bibfnamefont {B.}~\bibnamefont {Chiaro}}, \bibinfo {author} {\bibfnamefont
  {A.}~\bibnamefont {Megrant}}, \bibinfo {author} {\bibfnamefont
  {J.}~\bibnamefont {Kelly}}, \bibinfo {author} {\bibfnamefont
  {Z.}~\bibnamefont {Chen}}, \bibinfo {author} {\bibfnamefont {K.}~\bibnamefont
  {Satzinger}}, \bibinfo {author} {\bibfnamefont {R.}~\bibnamefont {Barends}},
  \bibinfo {author} {\bibfnamefont {F.}~\bibnamefont {Arute}}, \bibinfo
  {author} {\bibfnamefont {K.}~\bibnamefont {Arya}}, \bibinfo {author}
  {\bibfnamefont {R.}~\bibnamefont {Babbush}}, \bibinfo {author} {\bibfnamefont
  {D.}~\bibnamefont {Bacon}}, \bibinfo {author} {\bibfnamefont {J.~C.}\
  \bibnamefont {Bardin}}, \bibinfo {author} {\bibfnamefont {S.}~\bibnamefont
  {Boixo}}, \bibinfo {author} {\bibfnamefont {D.}~\bibnamefont {Buell}},
  \bibinfo {author} {\bibfnamefont {B.}~\bibnamefont {Burkett}}, \bibinfo
  {author} {\bibfnamefont {Y.}~\bibnamefont {Chen}}, \bibinfo {author}
  {\bibfnamefont {R.}~\bibnamefont {Collins}}, \bibinfo {author} {\bibfnamefont
  {E.}~\bibnamefont {Farhi}}, \bibinfo {author} {\bibfnamefont
  {A.}~\bibnamefont {Fowler}}, \bibinfo {author} {\bibfnamefont
  {C.}~\bibnamefont {Gidney}}, \bibinfo {author} {\bibfnamefont
  {M.}~\bibnamefont {Giustina}}, \bibinfo {author} {\bibfnamefont
  {R.}~\bibnamefont {Graff}}, \bibinfo {author} {\bibfnamefont
  {M.}~\bibnamefont {Harrigan}}, \bibinfo {author} {\bibfnamefont
  {T.}~\bibnamefont {Huang}}, \bibinfo {author} {\bibfnamefont {S.~V.}\
  \bibnamefont {Isakov}}, \bibinfo {author} {\bibfnamefont {E.}~\bibnamefont
  {Jeffrey}}, \bibinfo {author} {\bibfnamefont {Z.}~\bibnamefont {Jiang}},
  \bibinfo {author} {\bibfnamefont {D.}~\bibnamefont {Kafri}}, \bibinfo
  {author} {\bibfnamefont {K.}~\bibnamefont {Kechedzhi}}, \bibinfo {author}
  {\bibfnamefont {P.}~\bibnamefont {Klimov}}, \bibinfo {author} {\bibfnamefont
  {A.}~\bibnamefont {Korotkov}}, \bibinfo {author} {\bibfnamefont
  {F.}~\bibnamefont {Kostritsa}}, \bibinfo {author} {\bibfnamefont
  {D.}~\bibnamefont {Landhuis}}, \bibinfo {author} {\bibfnamefont
  {E.}~\bibnamefont {Lucero}}, \bibinfo {author} {\bibfnamefont
  {J.}~\bibnamefont {McClean}}, \bibinfo {author} {\bibfnamefont
  {M.}~\bibnamefont {McEwen}}, \bibinfo {author} {\bibfnamefont
  {X.}~\bibnamefont {Mi}}, \bibinfo {author} {\bibfnamefont {M.}~\bibnamefont
  {Mohseni}}, \bibinfo {author} {\bibfnamefont {J.~Y.}\ \bibnamefont {Mutus}},
  \bibinfo {author} {\bibfnamefont {O.}~\bibnamefont {Naaman}}, \bibinfo
  {author} {\bibfnamefont {M.}~\bibnamefont {Neeley}}, \bibinfo {author}
  {\bibfnamefont {M.}~\bibnamefont {Niu}}, \bibinfo {author} {\bibfnamefont
  {A.}~\bibnamefont {Petukhov}}, \bibinfo {author} {\bibfnamefont
  {C.}~\bibnamefont {Quintana}}, \bibinfo {author} {\bibfnamefont
  {N.}~\bibnamefont {Rubin}}, \bibinfo {author} {\bibfnamefont
  {D.}~\bibnamefont {Sank}}, \bibinfo {author} {\bibfnamefont {V.}~\bibnamefont
  {Smelyanskiy}}, \bibinfo {author} {\bibfnamefont {A.}~\bibnamefont
  {Vainsencher}}, \bibinfo {author} {\bibfnamefont {T.~C.}\ \bibnamefont
  {White}}, \bibinfo {author} {\bibfnamefont {Z.}~\bibnamefont {Yao}}, \bibinfo
  {author} {\bibfnamefont {P.}~\bibnamefont {Yeh}}, \bibinfo {author}
  {\bibfnamefont {A.}~\bibnamefont {Zalcman}}, \bibinfo {author} {\bibfnamefont
  {H.}~\bibnamefont {Neven}},\ and\ \bibinfo {author} {\bibfnamefont {J.~M.}\
  \bibnamefont {Martinis}} (\bibinfo {collaboration} {Google AI Quantum}),\
  }\bibfield  {title} {\bibinfo {title} {Demonstrating a continuous set of
  two-qubit gates for near-term quantum algorithms},\ }\href
  {https://doi.org/10.1103/PhysRevLett.125.120504} {\bibfield  {journal}
  {\bibinfo  {journal} {Phys. Rev. Lett.}\ }\textbf {\bibinfo {volume} {125}},\
  \bibinfo {pages} {120504} (\bibinfo {year} {2020})}\BibitemShut {NoStop}%
\bibitem [{\citenamefont {Huang}\ \emph {et~al.}(2023)\citenamefont {Huang},
  \citenamefont {Wang}, \citenamefont {Wu}, \citenamefont {Ding}, \citenamefont
  {Ye}, \citenamefont {Kong}, \citenamefont {Zhang}, \citenamefont {Ni},
  \citenamefont {Song}, \citenamefont {Shi}, \citenamefont {Zhao},
  \citenamefont {Deng},\ and\ \citenamefont {Chen}}]{huang2023quantum}%
  \BibitemOpen
  \bibfield  {author} {\bibinfo {author} {\bibfnamefont {C.}~\bibnamefont
  {Huang}}, \bibinfo {author} {\bibfnamefont {T.}~\bibnamefont {Wang}},
  \bibinfo {author} {\bibfnamefont {F.}~\bibnamefont {Wu}}, \bibinfo {author}
  {\bibfnamefont {D.}~\bibnamefont {Ding}}, \bibinfo {author} {\bibfnamefont
  {Q.}~\bibnamefont {Ye}}, \bibinfo {author} {\bibfnamefont {L.}~\bibnamefont
  {Kong}}, \bibinfo {author} {\bibfnamefont {F.}~\bibnamefont {Zhang}},
  \bibinfo {author} {\bibfnamefont {X.}~\bibnamefont {Ni}}, \bibinfo {author}
  {\bibfnamefont {Z.}~\bibnamefont {Song}}, \bibinfo {author} {\bibfnamefont
  {Y.}~\bibnamefont {Shi}}, \bibinfo {author} {\bibfnamefont {H.-H.}\
  \bibnamefont {Zhao}}, \bibinfo {author} {\bibfnamefont {C.}~\bibnamefont
  {Deng}},\ and\ \bibinfo {author} {\bibfnamefont {J.}~\bibnamefont {Chen}},\
  }\bibfield  {title} {\bibinfo {title} {Quantum instruction set design for
  performance},\ }\href {https://doi.org/10.1103/PhysRevLett.130.070601}
  {\bibfield  {journal} {\bibinfo  {journal} {Phys. Rev. Lett.}\ }\textbf
  {\bibinfo {volume} {130}},\ \bibinfo {pages} {070601} (\bibinfo {year}
  {2023})}\BibitemShut {NoStop}%
\bibitem [{\citenamefont {DiCarlo}\ \emph {et~al.}(2009)\citenamefont
  {DiCarlo}, \citenamefont {Chow}, \citenamefont {Gambetta}, \citenamefont
  {Bishop}, \citenamefont {Johnson}, \citenamefont {Schuster}, \citenamefont
  {Majer}, \citenamefont {Blais}, \citenamefont {Frunzio}, \citenamefont
  {Girvin},\ and\ \citenamefont {Schoelkopf}}]{dicarlo2009demonstration}%
  \BibitemOpen
  \bibfield  {author} {\bibinfo {author} {\bibfnamefont {L.}~\bibnamefont
  {DiCarlo}}, \bibinfo {author} {\bibfnamefont {J.~M.}\ \bibnamefont {Chow}},
  \bibinfo {author} {\bibfnamefont {J.~M.}\ \bibnamefont {Gambetta}}, \bibinfo
  {author} {\bibfnamefont {L.~S.}\ \bibnamefont {Bishop}}, \bibinfo {author}
  {\bibfnamefont {B.~R.}\ \bibnamefont {Johnson}}, \bibinfo {author}
  {\bibfnamefont {D.~I.}\ \bibnamefont {Schuster}}, \bibinfo {author}
  {\bibfnamefont {J.}~\bibnamefont {Majer}}, \bibinfo {author} {\bibfnamefont
  {A.}~\bibnamefont {Blais}}, \bibinfo {author} {\bibfnamefont
  {L.}~\bibnamefont {Frunzio}}, \bibinfo {author} {\bibfnamefont {S.~M.}\
  \bibnamefont {Girvin}},\ and\ \bibinfo {author} {\bibfnamefont {R.~J.}\
  \bibnamefont {Schoelkopf}},\ }\bibfield  {title} {\bibinfo {title}
  {Demonstration of two-qubit algorithms with a superconducting quantum
  processor},\ }\href {https://doi.org/10.1038/nature08121} {\bibfield
  {journal} {\bibinfo  {journal} {Nature}\ }\textbf {\bibinfo {volume} {460}},\
  \bibinfo {pages} {240} (\bibinfo {year} {2009})}\BibitemShut {NoStop}%
\bibitem [{\citenamefont {Terhal}(2015)}]{terhal2015quantum}%
  \BibitemOpen
  \bibfield  {author} {\bibinfo {author} {\bibfnamefont {B.~M.}\ \bibnamefont
  {Terhal}},\ }\bibfield  {title} {\bibinfo {title} {Quantum error correction
  for quantum memories},\ }\href {https://doi.org/10.1103/RevModPhys.87.307}
  {\bibfield  {journal} {\bibinfo  {journal} {Rev. Mod. Phys.}\ }\textbf
  {\bibinfo {volume} {87}},\ \bibinfo {pages} {307} (\bibinfo {year}
  {2015})}\BibitemShut {NoStop}%
\bibitem [{\citenamefont {Nguyen}\ \emph {et~al.}(2024)\citenamefont {Nguyen},
  \citenamefont {Kim}, \citenamefont {Hashim}, \citenamefont {Goss},
  \citenamefont {Marinelli}, \citenamefont {Bhandari}, \citenamefont {Das},
  \citenamefont {Naik}, \citenamefont {Kreikebaum}, \citenamefont {Jordan},
  \citenamefont {Santiago},\ and\ \citenamefont
  {Siddiqi}}]{nguyen2024programmable}%
  \BibitemOpen
  \bibfield  {author} {\bibinfo {author} {\bibfnamefont {L.~B.}\ \bibnamefont
  {Nguyen}}, \bibinfo {author} {\bibfnamefont {Y.}~\bibnamefont {Kim}},
  \bibinfo {author} {\bibfnamefont {A.}~\bibnamefont {Hashim}}, \bibinfo
  {author} {\bibfnamefont {N.}~\bibnamefont {Goss}}, \bibinfo {author}
  {\bibfnamefont {B.}~\bibnamefont {Marinelli}}, \bibinfo {author}
  {\bibfnamefont {B.}~\bibnamefont {Bhandari}}, \bibinfo {author}
  {\bibfnamefont {D.}~\bibnamefont {Das}}, \bibinfo {author} {\bibfnamefont
  {R.~K.}\ \bibnamefont {Naik}}, \bibinfo {author} {\bibfnamefont {J.~M.}\
  \bibnamefont {Kreikebaum}}, \bibinfo {author} {\bibfnamefont {A.~N.}\
  \bibnamefont {Jordan}}, \bibinfo {author} {\bibfnamefont {D.~I.}\
  \bibnamefont {Santiago}},\ and\ \bibinfo {author} {\bibfnamefont
  {I.}~\bibnamefont {Siddiqi}},\ }\bibfield  {title} {\bibinfo {title}
  {Programmable heisenberg interactions between floquet qubits},\ }\href
  {https://doi.org/10.1038/s41567-023-02326-7} {\bibfield  {journal} {\bibinfo
  {journal} {Nature Physics}\ }\textbf {\bibinfo {volume} {20}},\ \bibinfo
  {pages} {240} (\bibinfo {year} {2024})}\BibitemShut {NoStop}%
\bibitem [{\citenamefont {Abrams}\ \emph {et~al.}(2020)\citenamefont {Abrams},
  \citenamefont {Didier}, \citenamefont {Johnson}, \citenamefont {Silva},\ and\
  \citenamefont {Ryan}}]{abrams2020implementation}%
  \BibitemOpen
  \bibfield  {author} {\bibinfo {author} {\bibfnamefont {D.~M.}\ \bibnamefont
  {Abrams}}, \bibinfo {author} {\bibfnamefont {N.}~\bibnamefont {Didier}},
  \bibinfo {author} {\bibfnamefont {B.~R.}\ \bibnamefont {Johnson}}, \bibinfo
  {author} {\bibfnamefont {M.~P.~d.}\ \bibnamefont {Silva}},\ and\ \bibinfo
  {author} {\bibfnamefont {C.~A.}\ \bibnamefont {Ryan}},\ }\bibfield  {title}
  {\bibinfo {title} {Implementation of xy entangling gates with a single
  calibrated pulse},\ }\href {https://doi.org/10.1038/s41928-020-00498-1}
  {\bibfield  {journal} {\bibinfo  {journal} {Nature Electronics}\ }\textbf
  {\bibinfo {volume} {3}},\ \bibinfo {pages} {744} (\bibinfo {year}
  {2020})}\BibitemShut {NoStop}%
\bibitem [{\citenamefont {Quantum}(2024)}]{ibm_fractional_gates}%
  \BibitemOpen
  \bibfield  {author} {\bibinfo {author} {\bibfnamefont {I.}~\bibnamefont
  {Quantum}},\ }\href {https://www.ibm.com/quantum/blog/fractional-gates}
  {\bibinfo {title} {New fractional gates reduce circuit depth for
  utility-scale workloads}} (\bibinfo {year} {2024}),\ \bibinfo {note}
  {accessed: Nov. 18, 2024}\BibitemShut {NoStop}%
\bibitem [{\citenamefont {DeCross}\ \emph {et~al.}(2024)\citenamefont
  {DeCross}, \citenamefont {Haghshenas}, \citenamefont {Liu}, \citenamefont
  {Rinaldi}, \citenamefont {Gray}, \citenamefont {Alexeev}, \citenamefont
  {Baldwin}, \citenamefont {Bartolotta}, \citenamefont {Bohn}, \citenamefont
  {Chertkov} \emph {et~al.}}]{decross2024computational}%
  \BibitemOpen
  \bibfield  {author} {\bibinfo {author} {\bibfnamefont {M.}~\bibnamefont
  {DeCross}}, \bibinfo {author} {\bibfnamefont {R.}~\bibnamefont {Haghshenas}},
  \bibinfo {author} {\bibfnamefont {M.}~\bibnamefont {Liu}}, \bibinfo {author}
  {\bibfnamefont {E.}~\bibnamefont {Rinaldi}}, \bibinfo {author} {\bibfnamefont
  {J.}~\bibnamefont {Gray}}, \bibinfo {author} {\bibfnamefont {Y.}~\bibnamefont
  {Alexeev}}, \bibinfo {author} {\bibfnamefont {C.~H.}\ \bibnamefont
  {Baldwin}}, \bibinfo {author} {\bibfnamefont {J.~P.}\ \bibnamefont
  {Bartolotta}}, \bibinfo {author} {\bibfnamefont {M.}~\bibnamefont {Bohn}},
  \bibinfo {author} {\bibfnamefont {E.}~\bibnamefont {Chertkov}}, \emph
  {et~al.},\ }\bibfield  {title} {\bibinfo {title} {The computational power of
  random quantum circuits in arbitrary geometries},\ }\href@noop {} {\bibfield
  {journal} {\bibinfo  {journal} {arXiv preprint arXiv:2406.02501}\ } (\bibinfo
  {year} {2024})}\BibitemShut {NoStop}%
\bibitem [{\citenamefont {Sastry}(2013)}]{sastry2013nonlinear}%
  \BibitemOpen
  \bibfield  {author} {\bibinfo {author} {\bibfnamefont {S.}~\bibnamefont
  {Sastry}},\ }\href@noop {} {\emph {\bibinfo {title} {Nonlinear systems:
  analysis, stability, and control}}},\ Vol.~\bibinfo {volume} {10}\ (\bibinfo
  {publisher} {Springer Science \& Business Media},\ \bibinfo {year}
  {2013})\BibitemShut {NoStop}%
\bibitem [{\citenamefont {Liu}\ \emph {et~al.}(2006)\citenamefont {Liu},
  \citenamefont {Wei}, \citenamefont {Tsai},\ and\ \citenamefont
  {Nori}}]{liu2006controllable}%
  \BibitemOpen
  \bibfield  {author} {\bibinfo {author} {\bibfnamefont {Y.-x.}\ \bibnamefont
  {Liu}}, \bibinfo {author} {\bibfnamefont {L.~F.}\ \bibnamefont {Wei}},
  \bibinfo {author} {\bibfnamefont {J.~S.}\ \bibnamefont {Tsai}},\ and\
  \bibinfo {author} {\bibfnamefont {F.}~\bibnamefont {Nori}},\ }\bibfield
  {title} {\bibinfo {title} {Controllable coupling between flux qubits},\
  }\href {https://doi.org/10.1103/PhysRevLett.96.067003} {\bibfield  {journal}
  {\bibinfo  {journal} {Phys. Rev. Lett.}\ }\textbf {\bibinfo {volume} {96}},\
  \bibinfo {pages} {067003} (\bibinfo {year} {2006})}\BibitemShut {NoStop}%
\bibitem [{\citenamefont {Vandersypen}\ and\ \citenamefont
  {Chuang}(2005)}]{vandersypen2004nmr}%
  \BibitemOpen
  \bibfield  {author} {\bibinfo {author} {\bibfnamefont {L.~M.~K.}\
  \bibnamefont {Vandersypen}}\ and\ \bibinfo {author} {\bibfnamefont {I.~L.}\
  \bibnamefont {Chuang}},\ }\bibfield  {title} {\bibinfo {title} {Nmr
  techniques for quantum control and computation},\ }\href
  {https://doi.org/10.1103/RevModPhys.76.1037} {\bibfield  {journal} {\bibinfo
  {journal} {Rev. Mod. Phys.}\ }\textbf {\bibinfo {volume} {76}},\ \bibinfo
  {pages} {1037} (\bibinfo {year} {2005})}\BibitemShut {NoStop}%
\bibitem [{\citenamefont {Chen}\ \emph {et~al.}(2024)\citenamefont {Chen},
  \citenamefont {Ding}, \citenamefont {Gong}, \citenamefont {Huang},\ and\
  \citenamefont {Ye}}]{chen2024one}%
  \BibitemOpen
  \bibfield  {author} {\bibinfo {author} {\bibfnamefont {J.}~\bibnamefont
  {Chen}}, \bibinfo {author} {\bibfnamefont {D.}~\bibnamefont {Ding}}, \bibinfo
  {author} {\bibfnamefont {W.}~\bibnamefont {Gong}}, \bibinfo {author}
  {\bibfnamefont {C.}~\bibnamefont {Huang}},\ and\ \bibinfo {author}
  {\bibfnamefont {Q.}~\bibnamefont {Ye}},\ }\bibfield  {title} {\bibinfo
  {title} {One gate scheme to rule them all: Introducing a complex yet reduced
  instruction set for quantum computing},\ }in\ \href
  {https://doi.org/10.1145/3620665.3640386} {\emph {\bibinfo {booktitle}
  {Proceedings of the 29th ACM International Conference on Architectural
  Support for Programming Languages and Operating Systems, Volume 2}}},\
  \bibinfo {series and number} {ASPLOS '24}\ (\bibinfo  {publisher}
  {Association for Computing Machinery},\ \bibinfo {address} {New York, NY,
  USA},\ \bibinfo {year} {2024})\ p.\ \bibinfo {pages} {779–796}\BibitemShut
  {NoStop}%
\bibitem [{\citenamefont {Zhang}\ \emph {et~al.}(2004)\citenamefont {Zhang},
  \citenamefont {Vala}, \citenamefont {Sastry},\ and\ \citenamefont
  {Whaley}}]{PhysRevLett.93.020502}%
  \BibitemOpen
  \bibfield  {author} {\bibinfo {author} {\bibfnamefont {J.}~\bibnamefont
  {Zhang}}, \bibinfo {author} {\bibfnamefont {J.}~\bibnamefont {Vala}},
  \bibinfo {author} {\bibfnamefont {S.}~\bibnamefont {Sastry}},\ and\ \bibinfo
  {author} {\bibfnamefont {K.~B.}\ \bibnamefont {Whaley}},\ }\bibfield  {title}
  {\bibinfo {title} {Minimum construction of two-qubit quantum operations},\
  }\href {https://doi.org/10.1103/PhysRevLett.93.020502} {\bibfield  {journal}
  {\bibinfo  {journal} {Phys. Rev. Lett.}\ }\textbf {\bibinfo {volume} {93}},\
  \bibinfo {pages} {020502} (\bibinfo {year} {2004})}\BibitemShut {NoStop}%
\bibitem [{\citenamefont {Wei}\ \emph {et~al.}(2024)\citenamefont {Wei},
  \citenamefont {Lauer}, \citenamefont {Pritchett}, \citenamefont {Shanks},
  \citenamefont {McKay},\ and\ \citenamefont {Javadi-Abhari}}]{wei2024native}%
  \BibitemOpen
  \bibfield  {author} {\bibinfo {author} {\bibfnamefont {K.~X.}\ \bibnamefont
  {Wei}}, \bibinfo {author} {\bibfnamefont {I.}~\bibnamefont {Lauer}}, \bibinfo
  {author} {\bibfnamefont {E.}~\bibnamefont {Pritchett}}, \bibinfo {author}
  {\bibfnamefont {W.}~\bibnamefont {Shanks}}, \bibinfo {author} {\bibfnamefont
  {D.~C.}\ \bibnamefont {McKay}},\ and\ \bibinfo {author} {\bibfnamefont
  {A.}~\bibnamefont {Javadi-Abhari}},\ }\bibfield  {title} {\bibinfo {title}
  {Native two-qubit gates in fixed-coupling, fixed-frequency transmons beyond
  cross-resonance interaction},\ }\href
  {https://doi.org/10.1103/PRXQuantum.5.020338} {\bibfield  {journal} {\bibinfo
   {journal} {PRX Quantum}\ }\textbf {\bibinfo {volume} {5}},\ \bibinfo {pages}
  {020338} (\bibinfo {year} {2024})}\BibitemShut {NoStop}%
\bibitem [{\citenamefont {Vidal}\ and\ \citenamefont
  {Dawson}(2004)}]{PhysRevA.69.010301}%
  \BibitemOpen
  \bibfield  {author} {\bibinfo {author} {\bibfnamefont {G.}~\bibnamefont
  {Vidal}}\ and\ \bibinfo {author} {\bibfnamefont {C.~M.}\ \bibnamefont
  {Dawson}},\ }\bibfield  {title} {\bibinfo {title} {Universal quantum circuit
  for two-qubit transformations with three controlled-not gates},\ }\href
  {https://doi.org/10.1103/PhysRevA.69.010301} {\bibfield  {journal} {\bibinfo
  {journal} {Phys. Rev. A}\ }\textbf {\bibinfo {volume} {69}},\ \bibinfo
  {pages} {010301} (\bibinfo {year} {2004})}\BibitemShut {NoStop}%
\bibitem [{\citenamefont {Wang}\ \emph {et~al.}(2022)\citenamefont {Wang},
  \citenamefont {Li}, \citenamefont {Xu}, \citenamefont {Li}, \citenamefont
  {Wang}, \citenamefont {Yang}, \citenamefont {Mi}, \citenamefont {Liang},
  \citenamefont {Su}, \citenamefont {Yang}, \citenamefont {Wang}, \citenamefont
  {Wang}, \citenamefont {Li}, \citenamefont {Chen}, \citenamefont {Li},
  \citenamefont {Linghu}, \citenamefont {Han}, \citenamefont {Zhang},
  \citenamefont {Feng}, \citenamefont {Song}, \citenamefont {Ma}, \citenamefont
  {Zhang}, \citenamefont {Wang}, \citenamefont {Zhao}, \citenamefont {Liu},
  \citenamefont {Xue}, \citenamefont {Jin},\ and\ \citenamefont
  {Yu}}]{wang2022towards}%
  \BibitemOpen
  \bibfield  {author} {\bibinfo {author} {\bibfnamefont {C.}~\bibnamefont
  {Wang}}, \bibinfo {author} {\bibfnamefont {X.}~\bibnamefont {Li}}, \bibinfo
  {author} {\bibfnamefont {H.}~\bibnamefont {Xu}}, \bibinfo {author}
  {\bibfnamefont {Z.}~\bibnamefont {Li}}, \bibinfo {author} {\bibfnamefont
  {J.}~\bibnamefont {Wang}}, \bibinfo {author} {\bibfnamefont {Z.}~\bibnamefont
  {Yang}}, \bibinfo {author} {\bibfnamefont {Z.}~\bibnamefont {Mi}}, \bibinfo
  {author} {\bibfnamefont {X.}~\bibnamefont {Liang}}, \bibinfo {author}
  {\bibfnamefont {T.}~\bibnamefont {Su}}, \bibinfo {author} {\bibfnamefont
  {C.}~\bibnamefont {Yang}}, \bibinfo {author} {\bibfnamefont {G.}~\bibnamefont
  {Wang}}, \bibinfo {author} {\bibfnamefont {W.}~\bibnamefont {Wang}}, \bibinfo
  {author} {\bibfnamefont {Y.}~\bibnamefont {Li}}, \bibinfo {author}
  {\bibfnamefont {M.}~\bibnamefont {Chen}}, \bibinfo {author} {\bibfnamefont
  {C.}~\bibnamefont {Li}}, \bibinfo {author} {\bibfnamefont {K.}~\bibnamefont
  {Linghu}}, \bibinfo {author} {\bibfnamefont {J.}~\bibnamefont {Han}},
  \bibinfo {author} {\bibfnamefont {Y.}~\bibnamefont {Zhang}}, \bibinfo
  {author} {\bibfnamefont {Y.}~\bibnamefont {Feng}}, \bibinfo {author}
  {\bibfnamefont {Y.}~\bibnamefont {Song}}, \bibinfo {author} {\bibfnamefont
  {T.}~\bibnamefont {Ma}}, \bibinfo {author} {\bibfnamefont {J.}~\bibnamefont
  {Zhang}}, \bibinfo {author} {\bibfnamefont {R.}~\bibnamefont {Wang}},
  \bibinfo {author} {\bibfnamefont {P.}~\bibnamefont {Zhao}}, \bibinfo {author}
  {\bibfnamefont {W.}~\bibnamefont {Liu}}, \bibinfo {author} {\bibfnamefont
  {G.}~\bibnamefont {Xue}}, \bibinfo {author} {\bibfnamefont {Y.}~\bibnamefont
  {Jin}},\ and\ \bibinfo {author} {\bibfnamefont {H.}~\bibnamefont {Yu}},\
  }\bibfield  {title} {\bibinfo {title} {Towards practical quantum computers:
  transmon qubit with a lifetime approaching 0.5 milliseconds},\ }\href
  {https://doi.org/10.1038/s41534-021-00510-2} {\bibfield  {journal} {\bibinfo
  {journal} {npj Quantum Information}\ }\textbf {\bibinfo {volume} {8}},\
  \bibinfo {pages} {3} (\bibinfo {year} {2022})}\BibitemShut {NoStop}%
\bibitem [{\citenamefont {Sete}\ \emph {et~al.}(2016)\citenamefont {Sete},
  \citenamefont {Zeng},\ and\ \citenamefont {Rigetti}}]{sete2016functional}%
  \BibitemOpen
  \bibfield  {author} {\bibinfo {author} {\bibfnamefont {E.~A.}\ \bibnamefont
  {Sete}}, \bibinfo {author} {\bibfnamefont {W.~J.}\ \bibnamefont {Zeng}},\
  and\ \bibinfo {author} {\bibfnamefont {C.~T.}\ \bibnamefont {Rigetti}},\
  }\bibfield  {title} {\bibinfo {title} {A functional architecture for scalable
  quantum computing},\ }in\ \href {https://doi.org/10.1109/ICRC.2016.7738703}
  {\emph {\bibinfo {booktitle} {2016 IEEE International Conference on Rebooting
  Computing (ICRC)}}}\ (\bibinfo {year} {2016})\ pp.\ \bibinfo {pages}
  {1--6}\BibitemShut {NoStop}%
\bibitem [{\citenamefont {Chu}\ \emph {et~al.}(2023)\citenamefont {Chu},
  \citenamefont {He}, \citenamefont {Zhou}, \citenamefont {Yuan}, \citenamefont
  {Zhang}, \citenamefont {Guo}, \citenamefont {Hai}, \citenamefont {Han},
  \citenamefont {Hu}, \citenamefont {Huang}, \citenamefont {Jia}, \citenamefont
  {Jiao}, \citenamefont {Li}, \citenamefont {Liu}, \citenamefont {Ni},
  \citenamefont {Nie}, \citenamefont {Pan}, \citenamefont {Qiu}, \citenamefont
  {Wei}, \citenamefont {Nuerbolati}, \citenamefont {Yang}, \citenamefont
  {Zhang}, \citenamefont {Zhang}, \citenamefont {Zou}, \citenamefont {Chen},
  \citenamefont {Deng}, \citenamefont {Deng}, \citenamefont {Hu}, \citenamefont
  {Li}, \citenamefont {Liu}, \citenamefont {Lu}, \citenamefont {Niu},
  \citenamefont {Tan}, \citenamefont {Xu}, \citenamefont {Yan}, \citenamefont
  {Zhong}, \citenamefont {Yan}, \citenamefont {Sun},\ and\ \citenamefont
  {Yu}}]{chu_scalable_2023}%
  \BibitemOpen
  \bibfield  {author} {\bibinfo {author} {\bibfnamefont {J.}~\bibnamefont
  {Chu}}, \bibinfo {author} {\bibfnamefont {X.}~\bibnamefont {He}}, \bibinfo
  {author} {\bibfnamefont {Y.}~\bibnamefont {Zhou}}, \bibinfo {author}
  {\bibfnamefont {J.}~\bibnamefont {Yuan}}, \bibinfo {author} {\bibfnamefont
  {L.}~\bibnamefont {Zhang}}, \bibinfo {author} {\bibfnamefont
  {Q.}~\bibnamefont {Guo}}, \bibinfo {author} {\bibfnamefont {Y.}~\bibnamefont
  {Hai}}, \bibinfo {author} {\bibfnamefont {Z.}~\bibnamefont {Han}}, \bibinfo
  {author} {\bibfnamefont {C.-K.}\ \bibnamefont {Hu}}, \bibinfo {author}
  {\bibfnamefont {W.}~\bibnamefont {Huang}}, \bibinfo {author} {\bibfnamefont
  {H.}~\bibnamefont {Jia}}, \bibinfo {author} {\bibfnamefont {D.}~\bibnamefont
  {Jiao}}, \bibinfo {author} {\bibfnamefont {S.}~\bibnamefont {Li}}, \bibinfo
  {author} {\bibfnamefont {Y.}~\bibnamefont {Liu}}, \bibinfo {author}
  {\bibfnamefont {Z.}~\bibnamefont {Ni}}, \bibinfo {author} {\bibfnamefont
  {L.}~\bibnamefont {Nie}}, \bibinfo {author} {\bibfnamefont {X.}~\bibnamefont
  {Pan}}, \bibinfo {author} {\bibfnamefont {J.}~\bibnamefont {Qiu}}, \bibinfo
  {author} {\bibfnamefont {W.}~\bibnamefont {Wei}}, \bibinfo {author}
  {\bibfnamefont {W.}~\bibnamefont {Nuerbolati}}, \bibinfo {author}
  {\bibfnamefont {Z.}~\bibnamefont {Yang}}, \bibinfo {author} {\bibfnamefont
  {J.}~\bibnamefont {Zhang}}, \bibinfo {author} {\bibfnamefont
  {Z.}~\bibnamefont {Zhang}}, \bibinfo {author} {\bibfnamefont
  {W.}~\bibnamefont {Zou}}, \bibinfo {author} {\bibfnamefont {Y.}~\bibnamefont
  {Chen}}, \bibinfo {author} {\bibfnamefont {X.}~\bibnamefont {Deng}}, \bibinfo
  {author} {\bibfnamefont {X.}~\bibnamefont {Deng}}, \bibinfo {author}
  {\bibfnamefont {L.}~\bibnamefont {Hu}}, \bibinfo {author} {\bibfnamefont
  {J.}~\bibnamefont {Li}}, \bibinfo {author} {\bibfnamefont {S.}~\bibnamefont
  {Liu}}, \bibinfo {author} {\bibfnamefont {Y.}~\bibnamefont {Lu}}, \bibinfo
  {author} {\bibfnamefont {J.}~\bibnamefont {Niu}}, \bibinfo {author}
  {\bibfnamefont {D.}~\bibnamefont {Tan}}, \bibinfo {author} {\bibfnamefont
  {Y.}~\bibnamefont {Xu}}, \bibinfo {author} {\bibfnamefont {T.}~\bibnamefont
  {Yan}}, \bibinfo {author} {\bibfnamefont {Y.}~\bibnamefont {Zhong}}, \bibinfo
  {author} {\bibfnamefont {F.}~\bibnamefont {Yan}}, \bibinfo {author}
  {\bibfnamefont {X.}~\bibnamefont {Sun}},\ and\ \bibinfo {author}
  {\bibfnamefont {D.}~\bibnamefont {Yu}},\ }\bibfield  {title} {\bibinfo
  {title} {Scalable algorithm simplification using quantum and logic},\ }\href
  {https://doi.org/10.1038/s41567-022-01813-7} {\bibfield  {journal} {\bibinfo
  {journal} {Nature Physics}\ }\textbf {\bibinfo {volume} {19}},\ \bibinfo
  {pages} {126} (\bibinfo {year} {2023})}\BibitemShut {NoStop}%
\bibitem [{\citenamefont {Tucci}(2005)}]{tucci2005introduction}%
  \BibitemOpen
  \bibfield  {author} {\bibinfo {author} {\bibfnamefont {R.~R.}\ \bibnamefont
  {Tucci}},\ }\bibfield  {title} {\bibinfo {title} {An introduction to cartan's
  kak decomposition for qc programmers},\ }\href@noop {} {\bibfield  {journal}
  {\bibinfo  {journal} {arXiv preprint quant-ph/0507171}\ } (\bibinfo {year}
  {2005})}\BibitemShut {NoStop}%
\bibitem [{\citenamefont {Zhang}\ \emph {et~al.}(2003)\citenamefont {Zhang},
  \citenamefont {Vala}, \citenamefont {Sastry},\ and\ \citenamefont
  {Whaley}}]{zhang2003geometric}%
  \BibitemOpen
  \bibfield  {author} {\bibinfo {author} {\bibfnamefont {J.}~\bibnamefont
  {Zhang}}, \bibinfo {author} {\bibfnamefont {J.}~\bibnamefont {Vala}},
  \bibinfo {author} {\bibfnamefont {S.}~\bibnamefont {Sastry}},\ and\ \bibinfo
  {author} {\bibfnamefont {K.~B.}\ \bibnamefont {Whaley}},\ }\bibfield  {title}
  {\bibinfo {title} {Geometric theory of nonlocal two-qubit operations},\
  }\href {https://doi.org/10.1103/PhysRevA.67.042313} {\bibfield  {journal}
  {\bibinfo  {journal} {Phys. Rev. A}\ }\textbf {\bibinfo {volume} {67}},\
  \bibinfo {pages} {042313} (\bibinfo {year} {2003})}\BibitemShut {NoStop}%
\bibitem [{\citenamefont {Levine}\ \emph {et~al.}(2019)\citenamefont {Levine},
  \citenamefont {Keesling}, \citenamefont {Semeghini}, \citenamefont {Omran},
  \citenamefont {Wang}, \citenamefont {Ebadi}, \citenamefont {Bernien},
  \citenamefont {Greiner}, \citenamefont {Vuleti\ifmmode~\acute{c}\else
  \'{c}\fi{}}, \citenamefont {Pichler},\ and\ \citenamefont
  {Lukin}}]{levine2019parallel}%
  \BibitemOpen
  \bibfield  {author} {\bibinfo {author} {\bibfnamefont {H.}~\bibnamefont
  {Levine}}, \bibinfo {author} {\bibfnamefont {A.}~\bibnamefont {Keesling}},
  \bibinfo {author} {\bibfnamefont {G.}~\bibnamefont {Semeghini}}, \bibinfo
  {author} {\bibfnamefont {A.}~\bibnamefont {Omran}}, \bibinfo {author}
  {\bibfnamefont {T.~T.}\ \bibnamefont {Wang}}, \bibinfo {author}
  {\bibfnamefont {S.}~\bibnamefont {Ebadi}}, \bibinfo {author} {\bibfnamefont
  {H.}~\bibnamefont {Bernien}}, \bibinfo {author} {\bibfnamefont
  {M.}~\bibnamefont {Greiner}}, \bibinfo {author} {\bibfnamefont
  {V.}~\bibnamefont {Vuleti\ifmmode~\acute{c}\else \'{c}\fi{}}}, \bibinfo
  {author} {\bibfnamefont {H.}~\bibnamefont {Pichler}},\ and\ \bibinfo {author}
  {\bibfnamefont {M.~D.}\ \bibnamefont {Lukin}},\ }\bibfield  {title} {\bibinfo
  {title} {Parallel implementation of high-fidelity multiqubit gates with
  neutral atoms},\ }\href {https://doi.org/10.1103/PhysRevLett.123.170503}
  {\bibfield  {journal} {\bibinfo  {journal} {Phys. Rev. Lett.}\ }\textbf
  {\bibinfo {volume} {123}},\ \bibinfo {pages} {170503} (\bibinfo {year}
  {2019})}\BibitemShut {NoStop}%
\bibitem [{\citenamefont {Xue}\ \emph {et~al.}(2022)\citenamefont {Xue},
  \citenamefont {Russ}, \citenamefont {Samkharadze}, \citenamefont {Undseth},
  \citenamefont {Sammak}, \citenamefont {Scappucci},\ and\ \citenamefont
  {Vandersypen}}]{xue_quantum_2022}%
  \BibitemOpen
  \bibfield  {author} {\bibinfo {author} {\bibfnamefont {X.}~\bibnamefont
  {Xue}}, \bibinfo {author} {\bibfnamefont {M.}~\bibnamefont {Russ}}, \bibinfo
  {author} {\bibfnamefont {N.}~\bibnamefont {Samkharadze}}, \bibinfo {author}
  {\bibfnamefont {B.}~\bibnamefont {Undseth}}, \bibinfo {author} {\bibfnamefont
  {A.}~\bibnamefont {Sammak}}, \bibinfo {author} {\bibfnamefont
  {G.}~\bibnamefont {Scappucci}},\ and\ \bibinfo {author} {\bibfnamefont
  {L.~M.~K.}\ \bibnamefont {Vandersypen}},\ }\bibfield  {title} {\bibinfo
  {title} {Quantum logic with spin qubits crossing the surface code
  threshold},\ }\href {https://doi.org/10.1038/s41586-021-04273-w} {\bibfield
  {journal} {\bibinfo  {journal} {Nature}\ }\textbf {\bibinfo {volume} {601}},\
  \bibinfo {pages} {343} (\bibinfo {year} {2022})}\BibitemShut {NoStop}%
\bibitem [{\citenamefont {Chow}\ \emph {et~al.}(2011)\citenamefont {Chow},
  \citenamefont {C\'orcoles}, \citenamefont {Gambetta}, \citenamefont
  {Rigetti}, \citenamefont {Johnson}, \citenamefont {Smolin}, \citenamefont
  {Rozen}, \citenamefont {Keefe}, \citenamefont {Rothwell}, \citenamefont
  {Ketchen},\ and\ \citenamefont {Steffen}}]{PhysRevLett.107.080502}%
  \BibitemOpen
  \bibfield  {author} {\bibinfo {author} {\bibfnamefont {J.~M.}\ \bibnamefont
  {Chow}}, \bibinfo {author} {\bibfnamefont {A.~D.}\ \bibnamefont
  {C\'orcoles}}, \bibinfo {author} {\bibfnamefont {J.~M.}\ \bibnamefont
  {Gambetta}}, \bibinfo {author} {\bibfnamefont {C.}~\bibnamefont {Rigetti}},
  \bibinfo {author} {\bibfnamefont {B.~R.}\ \bibnamefont {Johnson}}, \bibinfo
  {author} {\bibfnamefont {J.~A.}\ \bibnamefont {Smolin}}, \bibinfo {author}
  {\bibfnamefont {J.~R.}\ \bibnamefont {Rozen}}, \bibinfo {author}
  {\bibfnamefont {G.~A.}\ \bibnamefont {Keefe}}, \bibinfo {author}
  {\bibfnamefont {M.~B.}\ \bibnamefont {Rothwell}}, \bibinfo {author}
  {\bibfnamefont {M.~B.}\ \bibnamefont {Ketchen}},\ and\ \bibinfo {author}
  {\bibfnamefont {M.}~\bibnamefont {Steffen}},\ }\bibfield  {title} {\bibinfo
  {title} {Simple all-microwave entangling gate for fixed-frequency
  superconducting qubits},\ }\href
  {https://doi.org/10.1103/PhysRevLett.107.080502} {\bibfield  {journal}
  {\bibinfo  {journal} {Phys. Rev. Lett.}\ }\textbf {\bibinfo {volume} {107}},\
  \bibinfo {pages} {080502} (\bibinfo {year} {2011})}\BibitemShut {NoStop}%
\bibitem [{\citenamefont {McKay}\ \emph {et~al.}(2017)\citenamefont {McKay},
  \citenamefont {Wood}, \citenamefont {Sheldon}, \citenamefont {Chow},\ and\
  \citenamefont {Gambetta}}]{mckay_efficient_2017}%
  \BibitemOpen
  \bibfield  {author} {\bibinfo {author} {\bibfnamefont {D.~C.}\ \bibnamefont
  {McKay}}, \bibinfo {author} {\bibfnamefont {C.~J.}\ \bibnamefont {Wood}},
  \bibinfo {author} {\bibfnamefont {S.}~\bibnamefont {Sheldon}}, \bibinfo
  {author} {\bibfnamefont {J.~M.}\ \bibnamefont {Chow}},\ and\ \bibinfo
  {author} {\bibfnamefont {J.~M.}\ \bibnamefont {Gambetta}},\ }\bibfield
  {title} {\bibinfo {title} {Efficient $z$ gates for quantum computing},\
  }\href {https://doi.org/10.1103/PhysRevA.96.022330} {\bibfield  {journal}
  {\bibinfo  {journal} {Phys. Rev. A}\ }\textbf {\bibinfo {volume} {96}},\
  \bibinfo {pages} {022330} (\bibinfo {year} {2017})}\BibitemShut {NoStop}%
\bibitem [{\citenamefont {Chen}\ \emph
  {et~al.}(2023{\natexlab{a}})\citenamefont {Chen}, \citenamefont {Ding},
  \citenamefont {Huang},\ and\ \citenamefont {Ye}}]{chen2023compiling}%
  \BibitemOpen
  \bibfield  {author} {\bibinfo {author} {\bibfnamefont {J.}~\bibnamefont
  {Chen}}, \bibinfo {author} {\bibfnamefont {D.}~\bibnamefont {Ding}}, \bibinfo
  {author} {\bibfnamefont {C.}~\bibnamefont {Huang}},\ and\ \bibinfo {author}
  {\bibfnamefont {Q.}~\bibnamefont {Ye}},\ }\bibfield  {title} {\bibinfo
  {title} {Compiling arbitrary single-qubit gates via the phase shifts of
  microwave pulses},\ }\href
  {https://doi.org/10.1103/PhysRevResearch.5.L022031} {\bibfield  {journal}
  {\bibinfo  {journal} {Phys. Rev. Res.}\ }\textbf {\bibinfo {volume} {5}},\
  \bibinfo {pages} {L022031} (\bibinfo {year}
  {2023}{\natexlab{a}})}\BibitemShut {NoStop}%
\bibitem [{\citenamefont {Han}\ \emph {et~al.}(2024)\citenamefont {Han},
  \citenamefont {Lyu}, \citenamefont {Zhou}, \citenamefont {Yuan},
  \citenamefont {Chu}, \citenamefont {Nuerbolati}, \citenamefont {Jia},
  \citenamefont {Nie}, \citenamefont {Wei}, \citenamefont {Yang}, \citenamefont
  {Zhang}, \citenamefont {Zhang}, \citenamefont {Hu}, \citenamefont {Hu},
  \citenamefont {Li}, \citenamefont {Tan}, \citenamefont {Bayat}, \citenamefont
  {Liu}, \citenamefont {Yan},\ and\ \citenamefont
  {Yu}}]{PhysRevResearch.6.013015}%
  \BibitemOpen
  \bibfield  {author} {\bibinfo {author} {\bibfnamefont {Z.}~\bibnamefont
  {Han}}, \bibinfo {author} {\bibfnamefont {C.}~\bibnamefont {Lyu}}, \bibinfo
  {author} {\bibfnamefont {Y.}~\bibnamefont {Zhou}}, \bibinfo {author}
  {\bibfnamefont {J.}~\bibnamefont {Yuan}}, \bibinfo {author} {\bibfnamefont
  {J.}~\bibnamefont {Chu}}, \bibinfo {author} {\bibfnamefont {W.}~\bibnamefont
  {Nuerbolati}}, \bibinfo {author} {\bibfnamefont {H.}~\bibnamefont {Jia}},
  \bibinfo {author} {\bibfnamefont {L.}~\bibnamefont {Nie}}, \bibinfo {author}
  {\bibfnamefont {W.}~\bibnamefont {Wei}}, \bibinfo {author} {\bibfnamefont
  {Z.}~\bibnamefont {Yang}}, \bibinfo {author} {\bibfnamefont {L.}~\bibnamefont
  {Zhang}}, \bibinfo {author} {\bibfnamefont {Z.}~\bibnamefont {Zhang}},
  \bibinfo {author} {\bibfnamefont {C.-K.}\ \bibnamefont {Hu}}, \bibinfo
  {author} {\bibfnamefont {L.}~\bibnamefont {Hu}}, \bibinfo {author}
  {\bibfnamefont {J.}~\bibnamefont {Li}}, \bibinfo {author} {\bibfnamefont
  {D.}~\bibnamefont {Tan}}, \bibinfo {author} {\bibfnamefont {A.}~\bibnamefont
  {Bayat}}, \bibinfo {author} {\bibfnamefont {S.}~\bibnamefont {Liu}}, \bibinfo
  {author} {\bibfnamefont {F.}~\bibnamefont {Yan}},\ and\ \bibinfo {author}
  {\bibfnamefont {D.}~\bibnamefont {Yu}},\ }\bibfield  {title} {\bibinfo
  {title} {Multilevel variational spectroscopy using a programmable quantum
  simulator},\ }\href {https://doi.org/10.1103/PhysRevResearch.6.013015}
  {\bibfield  {journal} {\bibinfo  {journal} {Phys. Rev. Res.}\ }\textbf
  {\bibinfo {volume} {6}},\ \bibinfo {pages} {013015} (\bibinfo {year}
  {2024})}\BibitemShut {NoStop}%
\bibitem [{\citenamefont {Crooks}(2020)}]{crooks2020gates}%
  \BibitemOpen
  \bibfield  {author} {\bibinfo {author} {\bibfnamefont {G.~E.}\ \bibnamefont
  {Crooks}},\ }\href@noop {} {\bibinfo {title} {Gates, states, and circuits}}
  (\bibinfo {year} {2020}),\ \bibinfo {note} {available at
  \url{https://threeplusone.com/pubs/on-gates-v0-5/}}\BibitemShut {NoStop}%
\bibitem [{\citenamefont {Peterson}\ \emph {et~al.}(2020)\citenamefont
  {Peterson}, \citenamefont {Crooks},\ and\ \citenamefont
  {Smith}}]{peterson2020fixed}%
  \BibitemOpen
  \bibfield  {author} {\bibinfo {author} {\bibfnamefont {E.~C.}\ \bibnamefont
  {Peterson}}, \bibinfo {author} {\bibfnamefont {G.~E.}\ \bibnamefont
  {Crooks}},\ and\ \bibinfo {author} {\bibfnamefont {R.~S.}\ \bibnamefont
  {Smith}},\ }\bibfield  {title} {\bibinfo {title} {Fixed-{D}epth {T}wo-{Q}ubit
  {C}ircuits and the {M}onodromy {P}olytope},\ }\href
  {https://doi.org/10.22331/q-2020-03-26-247} {\bibfield  {journal} {\bibinfo
  {journal} {{Quantum}}\ }\textbf {\bibinfo {volume} {4}},\ \bibinfo {pages}
  {247} (\bibinfo {year} {2020})}\BibitemShut {NoStop}%
\bibitem [{\citenamefont {Luo}\ \emph {et~al.}(2017)\citenamefont {Luo},
  \citenamefont {Zou}, \citenamefont {Wu}, \citenamefont {Liu}, \citenamefont
  {Han}, \citenamefont {Tey},\ and\ \citenamefont
  {You}}]{luo2017deterministic}%
  \BibitemOpen
  \bibfield  {author} {\bibinfo {author} {\bibfnamefont {X.-Y.}\ \bibnamefont
  {Luo}}, \bibinfo {author} {\bibfnamefont {Y.-Q.}\ \bibnamefont {Zou}},
  \bibinfo {author} {\bibfnamefont {L.-N.}\ \bibnamefont {Wu}}, \bibinfo
  {author} {\bibfnamefont {Q.}~\bibnamefont {Liu}}, \bibinfo {author}
  {\bibfnamefont {M.-F.}\ \bibnamefont {Han}}, \bibinfo {author} {\bibfnamefont
  {M.~K.}\ \bibnamefont {Tey}},\ and\ \bibinfo {author} {\bibfnamefont
  {L.}~\bibnamefont {You}},\ }\bibfield  {title} {\bibinfo {title}
  {Deterministic entanglement generation from driving through quantum phase
  transitions},\ }\href {https://doi.org/10.1126/science.aag1106} {\bibfield
  {journal} {\bibinfo  {journal} {Science}\ }\textbf {\bibinfo {volume}
  {355}},\ \bibinfo {pages} {620} (\bibinfo {year} {2017})}\BibitemShut
  {NoStop}%
\bibitem [{\citenamefont {Prevedel}\ \emph {et~al.}(2009)\citenamefont
  {Prevedel}, \citenamefont {Cronenberg}, \citenamefont {Tame}, \citenamefont
  {Paternostro}, \citenamefont {Walther}, \citenamefont {Kim},\ and\
  \citenamefont {Zeilinger}}]{prevedel2009experimental}%
  \BibitemOpen
  \bibfield  {author} {\bibinfo {author} {\bibfnamefont {R.}~\bibnamefont
  {Prevedel}}, \bibinfo {author} {\bibfnamefont {G.}~\bibnamefont
  {Cronenberg}}, \bibinfo {author} {\bibfnamefont {M.~S.}\ \bibnamefont
  {Tame}}, \bibinfo {author} {\bibfnamefont {M.}~\bibnamefont {Paternostro}},
  \bibinfo {author} {\bibfnamefont {P.}~\bibnamefont {Walther}}, \bibinfo
  {author} {\bibfnamefont {M.~S.}\ \bibnamefont {Kim}},\ and\ \bibinfo {author}
  {\bibfnamefont {A.}~\bibnamefont {Zeilinger}},\ }\bibfield  {title} {\bibinfo
  {title} {Experimental realization of dicke states of up to six qubits for
  multiparty quantum networking},\ }\href
  {https://doi.org/10.1103/PhysRevLett.103.020503} {\bibfield  {journal}
  {\bibinfo  {journal} {Phys. Rev. Lett.}\ }\textbf {\bibinfo {volume} {103}},\
  \bibinfo {pages} {020503} (\bibinfo {year} {2009})}\BibitemShut {NoStop}%
\bibitem [{\citenamefont {Kiesel}\ \emph {et~al.}(2007)\citenamefont {Kiesel},
  \citenamefont {Schmid}, \citenamefont {T\'oth}, \citenamefont {Solano},\ and\
  \citenamefont {Weinfurter}}]{PhysRevLett.98.063604}%
  \BibitemOpen
  \bibfield  {author} {\bibinfo {author} {\bibfnamefont {N.}~\bibnamefont
  {Kiesel}}, \bibinfo {author} {\bibfnamefont {C.}~\bibnamefont {Schmid}},
  \bibinfo {author} {\bibfnamefont {G.}~\bibnamefont {T\'oth}}, \bibinfo
  {author} {\bibfnamefont {E.}~\bibnamefont {Solano}},\ and\ \bibinfo {author}
  {\bibfnamefont {H.}~\bibnamefont {Weinfurter}},\ }\bibfield  {title}
  {\bibinfo {title} {Experimental observation of four-photon entangled dicke
  state with high fidelity},\ }\href
  {https://doi.org/10.1103/PhysRevLett.98.063604} {\bibfield  {journal}
  {\bibinfo  {journal} {Phys. Rev. Lett.}\ }\textbf {\bibinfo {volume} {98}},\
  \bibinfo {pages} {063604} (\bibinfo {year} {2007})}\BibitemShut {NoStop}%
\bibitem [{\citenamefont {H{\"a}ffner}\ \emph {et~al.}(2005)\citenamefont
  {H{\"a}ffner}, \citenamefont {H{\"a}nsel}, \citenamefont {Roos},
  \citenamefont {Benhelm}, \citenamefont {Chek-al kar}, \citenamefont
  {Chwalla}, \citenamefont {K{\"o}rber}, \citenamefont {Rapol}, \citenamefont
  {Riebe}, \citenamefont {Schmidt}, \citenamefont {Becher}, \citenamefont
  {G{\"u}hne}, \citenamefont {D{\"u}r},\ and\ \citenamefont
  {Blatt}}]{Häffner2005}%
  \BibitemOpen
  \bibfield  {author} {\bibinfo {author} {\bibfnamefont {H.}~\bibnamefont
  {H{\"a}ffner}}, \bibinfo {author} {\bibfnamefont {W.}~\bibnamefont
  {H{\"a}nsel}}, \bibinfo {author} {\bibfnamefont {C.~F.}\ \bibnamefont
  {Roos}}, \bibinfo {author} {\bibfnamefont {J.}~\bibnamefont {Benhelm}},
  \bibinfo {author} {\bibfnamefont {D.}~\bibnamefont {Chek-al kar}}, \bibinfo
  {author} {\bibfnamefont {M.}~\bibnamefont {Chwalla}}, \bibinfo {author}
  {\bibfnamefont {T.}~\bibnamefont {K{\"o}rber}}, \bibinfo {author}
  {\bibfnamefont {U.~D.}\ \bibnamefont {Rapol}}, \bibinfo {author}
  {\bibfnamefont {M.}~\bibnamefont {Riebe}}, \bibinfo {author} {\bibfnamefont
  {P.~O.}\ \bibnamefont {Schmidt}}, \bibinfo {author} {\bibfnamefont
  {C.}~\bibnamefont {Becher}}, \bibinfo {author} {\bibfnamefont
  {O.}~\bibnamefont {G{\"u}hne}}, \bibinfo {author} {\bibfnamefont
  {W.}~\bibnamefont {D{\"u}r}},\ and\ \bibinfo {author} {\bibfnamefont
  {R.}~\bibnamefont {Blatt}},\ }\bibfield  {title} {\bibinfo {title} {Scalable
  multiparticle entanglement of trapped ions},\ }\href
  {https://doi.org/10.1038/nature04279} {\bibfield  {journal} {\bibinfo
  {journal} {Nature}\ }\textbf {\bibinfo {volume} {438}},\ \bibinfo {pages}
  {643} (\bibinfo {year} {2005})}\BibitemShut {NoStop}%
\bibitem [{\citenamefont {Wang}\ \emph {et~al.}(2020)\citenamefont {Wang},
  \citenamefont {Li}, \citenamefont {Feng}, \citenamefont {Song}, \citenamefont
  {Song}, \citenamefont {Liu}, \citenamefont {Guo}, \citenamefont {Zhang},
  \citenamefont {Dong}, \citenamefont {Zheng}, \citenamefont {Wang},\ and\
  \citenamefont {Wang}}]{PhysRevLett.124.013601}%
  \BibitemOpen
  \bibfield  {author} {\bibinfo {author} {\bibfnamefont {Z.}~\bibnamefont
  {Wang}}, \bibinfo {author} {\bibfnamefont {H.}~\bibnamefont {Li}}, \bibinfo
  {author} {\bibfnamefont {W.}~\bibnamefont {Feng}}, \bibinfo {author}
  {\bibfnamefont {X.}~\bibnamefont {Song}}, \bibinfo {author} {\bibfnamefont
  {C.}~\bibnamefont {Song}}, \bibinfo {author} {\bibfnamefont {W.}~\bibnamefont
  {Liu}}, \bibinfo {author} {\bibfnamefont {Q.}~\bibnamefont {Guo}}, \bibinfo
  {author} {\bibfnamefont {X.}~\bibnamefont {Zhang}}, \bibinfo {author}
  {\bibfnamefont {H.}~\bibnamefont {Dong}}, \bibinfo {author} {\bibfnamefont
  {D.}~\bibnamefont {Zheng}}, \bibinfo {author} {\bibfnamefont
  {H.}~\bibnamefont {Wang}},\ and\ \bibinfo {author} {\bibfnamefont {D.-W.}\
  \bibnamefont {Wang}},\ }\bibfield  {title} {\bibinfo {title} {Controllable
  switching between superradiant and subradiant states in a 10-qubit
  superconducting circuit},\ }\href
  {https://doi.org/10.1103/PhysRevLett.124.013601} {\bibfield  {journal}
  {\bibinfo  {journal} {Phys. Rev. Lett.}\ }\textbf {\bibinfo {volume} {124}},\
  \bibinfo {pages} {013601} (\bibinfo {year} {2020})}\BibitemShut {NoStop}%
\bibitem [{\citenamefont {Hu}\ \emph {et~al.}(2024)\citenamefont {Hu},
  \citenamefont {Wei}, \citenamefont {Liu}, \citenamefont {Che}, \citenamefont
  {Zhou}, \citenamefont {Xie}, \citenamefont {Qin}, \citenamefont {Hu},
  \citenamefont {Yuan}, \citenamefont {Zhou}, \citenamefont {Liu},
  \citenamefont {Tan}, \citenamefont {Xin},\ and\ \citenamefont
  {Yu}}]{PhysRevLett.133.160801}%
  \BibitemOpen
  \bibfield  {author} {\bibinfo {author} {\bibfnamefont {C.-K.}\ \bibnamefont
  {Hu}}, \bibinfo {author} {\bibfnamefont {C.}~\bibnamefont {Wei}}, \bibinfo
  {author} {\bibfnamefont {C.}~\bibnamefont {Liu}}, \bibinfo {author}
  {\bibfnamefont {L.}~\bibnamefont {Che}}, \bibinfo {author} {\bibfnamefont
  {Y.}~\bibnamefont {Zhou}}, \bibinfo {author} {\bibfnamefont {G.}~\bibnamefont
  {Xie}}, \bibinfo {author} {\bibfnamefont {H.}~\bibnamefont {Qin}}, \bibinfo
  {author} {\bibfnamefont {G.}~\bibnamefont {Hu}}, \bibinfo {author}
  {\bibfnamefont {H.}~\bibnamefont {Yuan}}, \bibinfo {author} {\bibfnamefont
  {R.}~\bibnamefont {Zhou}}, \bibinfo {author} {\bibfnamefont {S.}~\bibnamefont
  {Liu}}, \bibinfo {author} {\bibfnamefont {D.}~\bibnamefont {Tan}}, \bibinfo
  {author} {\bibfnamefont {T.}~\bibnamefont {Xin}},\ and\ \bibinfo {author}
  {\bibfnamefont {D.}~\bibnamefont {Yu}},\ }\bibfield  {title} {\bibinfo
  {title} {Experimental sample-efficient quantum state tomography via parallel
  measurements},\ }\href {https://doi.org/10.1103/PhysRevLett.133.160801}
  {\bibfield  {journal} {\bibinfo  {journal} {Phys. Rev. Lett.}\ }\textbf
  {\bibinfo {volume} {133}},\ \bibinfo {pages} {160801} (\bibinfo {year}
  {2024})}\BibitemShut {NoStop}%
\bibitem [{\citenamefont {B{\"a}rtschi}\ and\ \citenamefont
  {Eidenbenz}(2019)}]{bartschi2019deterministic}%
  \BibitemOpen
  \bibfield  {author} {\bibinfo {author} {\bibfnamefont {A.}~\bibnamefont
  {B{\"a}rtschi}}\ and\ \bibinfo {author} {\bibfnamefont {S.}~\bibnamefont
  {Eidenbenz}},\ }\bibfield  {title} {\bibinfo {title} {Deterministic
  preparation of dicke states},\ }in\ \href@noop {} {\emph {\bibinfo
  {booktitle} {Fundamentals of Computation Theory}}}\ (\bibinfo  {publisher}
  {Springer International Publishing},\ \bibinfo {address} {Cham},\ \bibinfo
  {year} {2019})\ pp.\ \bibinfo {pages} {126--139}\BibitemShut {NoStop}%
\bibitem [{\citenamefont {Chen}\ \emph
  {et~al.}(2023{\natexlab{b}})\citenamefont {Chen}, \citenamefont {Lu},
  \citenamefont {Xia}, \citenamefont {Lu}, \citenamefont {Zhu},\ and\
  \citenamefont {Ma}}]{PhysRevLett.130.223601}%
  \BibitemOpen
  \bibfield  {author} {\bibinfo {author} {\bibfnamefont {L.}~\bibnamefont
  {Chen}}, \bibinfo {author} {\bibfnamefont {L.}~\bibnamefont {Lu}}, \bibinfo
  {author} {\bibfnamefont {L.}~\bibnamefont {Xia}}, \bibinfo {author}
  {\bibfnamefont {Y.}~\bibnamefont {Lu}}, \bibinfo {author} {\bibfnamefont
  {S.}~\bibnamefont {Zhu}},\ and\ \bibinfo {author} {\bibfnamefont {X.-s.}\
  \bibnamefont {Ma}},\ }\bibfield  {title} {\bibinfo {title} {On-chip
  generation and collectively coherent control of the superposition of the
  whole family of dicke states},\ }\href
  {https://doi.org/10.1103/PhysRevLett.130.223601} {\bibfield  {journal}
  {\bibinfo  {journal} {Phys. Rev. Lett.}\ }\textbf {\bibinfo {volume} {130}},\
  \bibinfo {pages} {223601} (\bibinfo {year} {2023}{\natexlab{b}})}\BibitemShut
  {NoStop}%
\bibitem [{\citenamefont {Kim}\ \emph {et~al.}(2023)\citenamefont {Kim},
  \citenamefont {Eddins}, \citenamefont {Anand}, \citenamefont {Wei},
  \citenamefont {van~den Berg}, \citenamefont {Rosenblatt}, \citenamefont
  {Nayfeh}, \citenamefont {Wu}, \citenamefont {Zaletel}, \citenamefont
  {Temme},\ and\ \citenamefont {Kandala}}]{kim2023evidence}%
  \BibitemOpen
  \bibfield  {author} {\bibinfo {author} {\bibfnamefont {Y.}~\bibnamefont
  {Kim}}, \bibinfo {author} {\bibfnamefont {A.}~\bibnamefont {Eddins}},
  \bibinfo {author} {\bibfnamefont {S.}~\bibnamefont {Anand}}, \bibinfo
  {author} {\bibfnamefont {K.~X.}\ \bibnamefont {Wei}}, \bibinfo {author}
  {\bibfnamefont {E.}~\bibnamefont {van~den Berg}}, \bibinfo {author}
  {\bibfnamefont {S.}~\bibnamefont {Rosenblatt}}, \bibinfo {author}
  {\bibfnamefont {H.}~\bibnamefont {Nayfeh}}, \bibinfo {author} {\bibfnamefont
  {Y.}~\bibnamefont {Wu}}, \bibinfo {author} {\bibfnamefont {M.}~\bibnamefont
  {Zaletel}}, \bibinfo {author} {\bibfnamefont {K.}~\bibnamefont {Temme}},\
  and\ \bibinfo {author} {\bibfnamefont {A.}~\bibnamefont {Kandala}},\
  }\bibfield  {title} {\bibinfo {title} {Evidence for the utility of quantum
  computing before fault tolerance},\ }\href
  {https://doi.org/10.1038/s41586-023-06096-3} {\bibfield  {journal} {\bibinfo
  {journal} {Nature}\ }\textbf {\bibinfo {volume} {618}},\ \bibinfo {pages}
  {500} (\bibinfo {year} {2023})}\BibitemShut {NoStop}%
\bibitem [{\citenamefont {Daley}\ \emph {et~al.}(2022)\citenamefont {Daley},
  \citenamefont {Bloch}, \citenamefont {Kokail}, \citenamefont {Flannigan},
  \citenamefont {Pearson}, \citenamefont {Troyer},\ and\ \citenamefont
  {Zoller}}]{daley2022practical}%
  \BibitemOpen
  \bibfield  {author} {\bibinfo {author} {\bibfnamefont {A.~J.}\ \bibnamefont
  {Daley}}, \bibinfo {author} {\bibfnamefont {I.}~\bibnamefont {Bloch}},
  \bibinfo {author} {\bibfnamefont {C.}~\bibnamefont {Kokail}}, \bibinfo
  {author} {\bibfnamefont {S.}~\bibnamefont {Flannigan}}, \bibinfo {author}
  {\bibfnamefont {N.}~\bibnamefont {Pearson}}, \bibinfo {author} {\bibfnamefont
  {M.}~\bibnamefont {Troyer}},\ and\ \bibinfo {author} {\bibfnamefont
  {P.}~\bibnamefont {Zoller}},\ }\bibfield  {title} {\bibinfo {title}
  {Practical quantum advantage in quantum simulation},\ }\href
  {https://doi.org/10.1038/s41586-022-04940-6} {\bibfield  {journal} {\bibinfo
  {journal} {Nature}\ }\textbf {\bibinfo {volume} {607}},\ \bibinfo {pages}
  {667} (\bibinfo {year} {2022})}\BibitemShut {NoStop}%
\bibitem [{\citenamefont {Acharya}\ \emph {et~al.}(2024)\citenamefont
  {Acharya}, \citenamefont {Aghababaie-Beni}, \citenamefont {Aleiner},
  \citenamefont {Andersen}, \citenamefont {Ansmann}, \citenamefont {Arute},
  \citenamefont {Arya}, \citenamefont {Asfaw}, \citenamefont {Astrakhantsev},
  \citenamefont {Atalaya} \emph {et~al.}}]{acharya2024quantum}%
  \BibitemOpen
  \bibfield  {author} {\bibinfo {author} {\bibfnamefont {R.}~\bibnamefont
  {Acharya}}, \bibinfo {author} {\bibfnamefont {L.}~\bibnamefont
  {Aghababaie-Beni}}, \bibinfo {author} {\bibfnamefont {I.}~\bibnamefont
  {Aleiner}}, \bibinfo {author} {\bibfnamefont {T.~I.}\ \bibnamefont
  {Andersen}}, \bibinfo {author} {\bibfnamefont {M.}~\bibnamefont {Ansmann}},
  \bibinfo {author} {\bibfnamefont {F.}~\bibnamefont {Arute}}, \bibinfo
  {author} {\bibfnamefont {K.}~\bibnamefont {Arya}}, \bibinfo {author}
  {\bibfnamefont {A.}~\bibnamefont {Asfaw}}, \bibinfo {author} {\bibfnamefont
  {N.}~\bibnamefont {Astrakhantsev}}, \bibinfo {author} {\bibfnamefont
  {J.}~\bibnamefont {Atalaya}}, \emph {et~al.},\ }\bibfield  {title} {\bibinfo
  {title} {Quantum error correction below the surface code threshold},\
  }\href@noop {} {\bibfield  {journal} {\bibinfo  {journal} {arXiv preprint
  arXiv:2408.13687}\ } (\bibinfo {year} {2024})}\BibitemShut {NoStop}%
\bibitem [{\citenamefont {Cao}\ \emph {et~al.}(2023)\citenamefont {Cao},
  \citenamefont {Wu}, \citenamefont {Chen}, \citenamefont {Gong}, \citenamefont
  {Wu}, \citenamefont {Ye}, \citenamefont {Zha}, \citenamefont {Qian},
  \citenamefont {Ying}, \citenamefont {Guo}, \citenamefont {Zhu}, \citenamefont
  {Huang}, \citenamefont {Zhao}, \citenamefont {Li}, \citenamefont {Wang},
  \citenamefont {Yu}, \citenamefont {Fan}, \citenamefont {Wu}, \citenamefont
  {Su}, \citenamefont {Deng}, \citenamefont {Rong}, \citenamefont {Li},
  \citenamefont {Zhang}, \citenamefont {Chung}, \citenamefont {Liang},
  \citenamefont {Lin}, \citenamefont {Xu}, \citenamefont {Sun}, \citenamefont
  {Guo}, \citenamefont {Li}, \citenamefont {Huo}, \citenamefont {Peng},
  \citenamefont {Lu}, \citenamefont {Yuan}, \citenamefont {Zhu},\ and\
  \citenamefont {Pan}}]{cao2023generation}%
  \BibitemOpen
  \bibfield  {author} {\bibinfo {author} {\bibfnamefont {S.}~\bibnamefont
  {Cao}}, \bibinfo {author} {\bibfnamefont {B.}~\bibnamefont {Wu}}, \bibinfo
  {author} {\bibfnamefont {F.}~\bibnamefont {Chen}}, \bibinfo {author}
  {\bibfnamefont {M.}~\bibnamefont {Gong}}, \bibinfo {author} {\bibfnamefont
  {Y.}~\bibnamefont {Wu}}, \bibinfo {author} {\bibfnamefont {Y.}~\bibnamefont
  {Ye}}, \bibinfo {author} {\bibfnamefont {C.}~\bibnamefont {Zha}}, \bibinfo
  {author} {\bibfnamefont {H.}~\bibnamefont {Qian}}, \bibinfo {author}
  {\bibfnamefont {C.}~\bibnamefont {Ying}}, \bibinfo {author} {\bibfnamefont
  {S.}~\bibnamefont {Guo}}, \bibinfo {author} {\bibfnamefont {Q.}~\bibnamefont
  {Zhu}}, \bibinfo {author} {\bibfnamefont {H.-L.}\ \bibnamefont {Huang}},
  \bibinfo {author} {\bibfnamefont {Y.}~\bibnamefont {Zhao}}, \bibinfo {author}
  {\bibfnamefont {S.}~\bibnamefont {Li}}, \bibinfo {author} {\bibfnamefont
  {S.}~\bibnamefont {Wang}}, \bibinfo {author} {\bibfnamefont {J.}~\bibnamefont
  {Yu}}, \bibinfo {author} {\bibfnamefont {D.}~\bibnamefont {Fan}}, \bibinfo
  {author} {\bibfnamefont {D.}~\bibnamefont {Wu}}, \bibinfo {author}
  {\bibfnamefont {H.}~\bibnamefont {Su}}, \bibinfo {author} {\bibfnamefont
  {H.}~\bibnamefont {Deng}}, \bibinfo {author} {\bibfnamefont {H.}~\bibnamefont
  {Rong}}, \bibinfo {author} {\bibfnamefont {Y.}~\bibnamefont {Li}}, \bibinfo
  {author} {\bibfnamefont {K.}~\bibnamefont {Zhang}}, \bibinfo {author}
  {\bibfnamefont {T.-H.}\ \bibnamefont {Chung}}, \bibinfo {author}
  {\bibfnamefont {F.}~\bibnamefont {Liang}}, \bibinfo {author} {\bibfnamefont
  {J.}~\bibnamefont {Lin}}, \bibinfo {author} {\bibfnamefont {Y.}~\bibnamefont
  {Xu}}, \bibinfo {author} {\bibfnamefont {L.}~\bibnamefont {Sun}}, \bibinfo
  {author} {\bibfnamefont {C.}~\bibnamefont {Guo}}, \bibinfo {author}
  {\bibfnamefont {N.}~\bibnamefont {Li}}, \bibinfo {author} {\bibfnamefont
  {Y.-H.}\ \bibnamefont {Huo}}, \bibinfo {author} {\bibfnamefont {C.-Z.}\
  \bibnamefont {Peng}}, \bibinfo {author} {\bibfnamefont {C.-Y.}\ \bibnamefont
  {Lu}}, \bibinfo {author} {\bibfnamefont {X.}~\bibnamefont {Yuan}}, \bibinfo
  {author} {\bibfnamefont {X.}~\bibnamefont {Zhu}},\ and\ \bibinfo {author}
  {\bibfnamefont {J.-W.}\ \bibnamefont {Pan}},\ }\bibfield  {title} {\bibinfo
  {title} {Generation of genuine entanglement up to 51 superconducting
  qubits},\ }\href {https://doi.org/10.1038/s41586-023-06195-1} {\bibfield
  {journal} {\bibinfo  {journal} {Nature}\ }\textbf {\bibinfo {volume} {619}},\
  \bibinfo {pages} {738} (\bibinfo {year} {2023})}\BibitemShut {NoStop}%
\bibitem [{\citenamefont {Zhang}\ \emph {et~al.}(2022)\citenamefont {Zhang},
  \citenamefont {Jiang}, \citenamefont {Deng}, \citenamefont {Wang},
  \citenamefont {Chen}, \citenamefont {Zhang}, \citenamefont {Ren},
  \citenamefont {Dong}, \citenamefont {Xu}, \citenamefont {Gao}, \citenamefont
  {Jin}, \citenamefont {Zhu}, \citenamefont {Guo}, \citenamefont {Li},
  \citenamefont {Song}, \citenamefont {Gorshkov}, \citenamefont {Iadecola},
  \citenamefont {Liu}, \citenamefont {Gong}, \citenamefont {Wang},
  \citenamefont {Deng},\ and\ \citenamefont {Wang}}]{zhang2022digital}%
  \BibitemOpen
  \bibfield  {author} {\bibinfo {author} {\bibfnamefont {X.}~\bibnamefont
  {Zhang}}, \bibinfo {author} {\bibfnamefont {W.}~\bibnamefont {Jiang}},
  \bibinfo {author} {\bibfnamefont {J.}~\bibnamefont {Deng}}, \bibinfo {author}
  {\bibfnamefont {K.}~\bibnamefont {Wang}}, \bibinfo {author} {\bibfnamefont
  {J.}~\bibnamefont {Chen}}, \bibinfo {author} {\bibfnamefont {P.}~\bibnamefont
  {Zhang}}, \bibinfo {author} {\bibfnamefont {W.}~\bibnamefont {Ren}}, \bibinfo
  {author} {\bibfnamefont {H.}~\bibnamefont {Dong}}, \bibinfo {author}
  {\bibfnamefont {S.}~\bibnamefont {Xu}}, \bibinfo {author} {\bibfnamefont
  {Y.}~\bibnamefont {Gao}}, \bibinfo {author} {\bibfnamefont {F.}~\bibnamefont
  {Jin}}, \bibinfo {author} {\bibfnamefont {X.}~\bibnamefont {Zhu}}, \bibinfo
  {author} {\bibfnamefont {Q.}~\bibnamefont {Guo}}, \bibinfo {author}
  {\bibfnamefont {H.}~\bibnamefont {Li}}, \bibinfo {author} {\bibfnamefont
  {C.}~\bibnamefont {Song}}, \bibinfo {author} {\bibfnamefont {A.~V.}\
  \bibnamefont {Gorshkov}}, \bibinfo {author} {\bibfnamefont {T.}~\bibnamefont
  {Iadecola}}, \bibinfo {author} {\bibfnamefont {F.}~\bibnamefont {Liu}},
  \bibinfo {author} {\bibfnamefont {Z.-X.}\ \bibnamefont {Gong}}, \bibinfo
  {author} {\bibfnamefont {Z.}~\bibnamefont {Wang}}, \bibinfo {author}
  {\bibfnamefont {D.-L.}\ \bibnamefont {Deng}},\ and\ \bibinfo {author}
  {\bibfnamefont {H.}~\bibnamefont {Wang}},\ }\bibfield  {title} {\bibinfo
  {title} {Digital quantum simulation of floquet symmetry-protected topological
  phases},\ }\href {https://doi.org/10.1038/s41586-022-04854-3} {\bibfield
  {journal} {\bibinfo  {journal} {Nature}\ }\textbf {\bibinfo {volume} {607}},\
  \bibinfo {pages} {468} (\bibinfo {year} {2022})}\BibitemShut {NoStop}%
\bibitem [{\citenamefont {Yang}\ \emph {et~al.}(2024)\citenamefont {Yang},
  \citenamefont {Ding}, \citenamefont {Ye}, \citenamefont {Huang},
  \citenamefont {Chen},\ and\ \citenamefont {Xie}}]{yang2024_reQISC}%
  \BibitemOpen
  \bibfield  {author} {\bibinfo {author} {\bibfnamefont {Z.}~\bibnamefont
  {Yang}}, \bibinfo {author} {\bibfnamefont {D.}~\bibnamefont {Ding}}, \bibinfo
  {author} {\bibfnamefont {Q.}~\bibnamefont {Ye}}, \bibinfo {author}
  {\bibfnamefont {C.}~\bibnamefont {Huang}}, \bibinfo {author} {\bibfnamefont
  {J.}~\bibnamefont {Chen}},\ and\ \bibinfo {author} {\bibfnamefont
  {Y.}~\bibnamefont {Xie}},\ }\bibfield  {title} {\bibinfo {title}
  {Reconfigurable quantum instruction set computers: Towards ultimate
  performance with realization feasibility},\ }\href@noop {} {\bibfield
  {journal} {\bibinfo  {journal} {Manuscript under review}\ } (\bibinfo {year}
  {2024})}\BibitemShut {NoStop}%
\bibitem [{\citenamefont {Zhou}\ \emph {et~al.}(2024)\citenamefont {Zhou},
  \citenamefont {Zhang}, \citenamefont {Kong},\ and\ \citenamefont
  {Chen}}]{zhou2024_defects}%
  \BibitemOpen
  \bibfield  {author} {\bibinfo {author} {\bibfnamefont {R.}~\bibnamefont
  {Zhou}}, \bibinfo {author} {\bibfnamefont {F.}~\bibnamefont {Zhang}},
  \bibinfo {author} {\bibfnamefont {L.}~\bibnamefont {Kong}},\ and\ \bibinfo
  {author} {\bibfnamefont {J.}~\bibnamefont {Chen}},\ }\bibfield  {title}
  {\bibinfo {title} {Halma, a routing-based technique for defects mitigation in
  quantum error correction},\ }\href@noop {} {\bibfield  {journal} {\bibinfo
  {journal} {arXiv preprint arXiv:2412.21000}\ } (\bibinfo {year}
  {2024})}\BibitemShut {NoStop}%
\bibitem [{\citenamefont {Morvan}\ \emph {et~al.}(2024)\citenamefont {Morvan},
  \citenamefont {Villalonga}, \citenamefont {Mi}, \citenamefont {Mandr{\`a}},
  \citenamefont {Bengtsson}, \citenamefont {Klimov}, \citenamefont {Chen},
  \citenamefont {Hong}, \citenamefont {Erickson}, \citenamefont {Drozdov},
  \citenamefont {Chau}, \citenamefont {Laun}, \citenamefont {Movassagh},
  \citenamefont {Asfaw}, \citenamefont {Brand{\~a}o}, \citenamefont {Peralta},
  \citenamefont {Abanin}, \citenamefont {Acharya}, \citenamefont {Allen},
  \citenamefont {Andersen}, \citenamefont {Anderson}, \citenamefont {Ansmann},
  \citenamefont {Arute}, \citenamefont {Arya}, \citenamefont {Atalaya},
  \citenamefont {Bardin}, \citenamefont {Bilmes}, \citenamefont {Bortoli},
  \citenamefont {Bourassa}, \citenamefont {Bovaird}, \citenamefont {Brill},
  \citenamefont {Broughton}, \citenamefont {Buckley}, \citenamefont {Buell},
  \citenamefont {Burger}, \citenamefont {Burkett}, \citenamefont {Bushnell},
  \citenamefont {Campero}, \citenamefont {Chang}, \citenamefont {Chiaro},
  \citenamefont {Chik}, \citenamefont {Chou}, \citenamefont {Cogan},
  \citenamefont {Collins}, \citenamefont {Conner}, \citenamefont {Courtney},
  \citenamefont {Crook}, \citenamefont {Curtin}, \citenamefont {Debroy},
  \citenamefont {Barba}, \citenamefont {Demura}, \citenamefont {Paolo},
  \citenamefont {Dunsworth}, \citenamefont {Faoro}, \citenamefont {Farhi},
  \citenamefont {Fatemi}, \citenamefont {Ferreira}, \citenamefont {Burgos},
  \citenamefont {Forati}, \citenamefont {Fowler}, \citenamefont {Foxen},
  \citenamefont {Garcia}, \citenamefont {Genois}, \citenamefont {Giang},
  \citenamefont {Gidney}, \citenamefont {Gilboa}, \citenamefont {Giustina},
  \citenamefont {Gosula}, \citenamefont {Dau}, \citenamefont {Gross},
  \citenamefont {Habegger}, \citenamefont {Hamilton}, \citenamefont {Hansen},
  \citenamefont {Harrigan}, \citenamefont {Harrington}, \citenamefont {Heu},
  \citenamefont {Hoffmann}, \citenamefont {Huang}, \citenamefont {Huff},
  \citenamefont {Huggins}, \citenamefont {Ioffe}, \citenamefont {Isakov},
  \citenamefont {Iveland}, \citenamefont {Jeffrey}, \citenamefont {Jiang},
  \citenamefont {Jones}, \citenamefont {Juhas}, \citenamefont {Kafri},
  \citenamefont {Khattar}, \citenamefont {Khezri}, \citenamefont
  {Kieferov{\'a}}, \citenamefont {Kim}, \citenamefont {Kitaev}, \citenamefont
  {Klots}, \citenamefont {Korotkov}, \citenamefont {Kostritsa}, \citenamefont
  {Kreikebaum}, \citenamefont {Landhuis}, \citenamefont {Laptev}, \citenamefont
  {Lau}, \citenamefont {Laws}, \citenamefont {Lee}, \citenamefont {Lee},
  \citenamefont {Lensky}, \citenamefont {Lester}, \citenamefont {Lill},
  \citenamefont {Liu}, \citenamefont {Livingston}, \citenamefont {Locharla},
  \citenamefont {Malone}, \citenamefont {Martin}, \citenamefont {Martin},
  \citenamefont {McClean}, \citenamefont {McEwen}, \citenamefont {Miao},
  \citenamefont {Mieszala}, \citenamefont {Montazeri}, \citenamefont
  {Mruczkiewicz}, \citenamefont {Naaman}, \citenamefont {Neeley}, \citenamefont
  {Neill}, \citenamefont {Nersisyan}, \citenamefont {Newman}, \citenamefont
  {Ng}, \citenamefont {Nguyen}, \citenamefont {Nguyen}, \citenamefont {Niu},
  \citenamefont {O'Brien}, \citenamefont {Omonije}, \citenamefont {Opremcak},
  \citenamefont {Petukhov}, \citenamefont {Potter}, \citenamefont {Pryadko},
  \citenamefont {Quintana}, \citenamefont {Rhodes}, \citenamefont {Rocque},
  \citenamefont {Rosenberg}, \citenamefont {Rubin}, \citenamefont {Saei},
  \citenamefont {Sank}, \citenamefont {Sankaragomathi}, \citenamefont
  {Satzinger}, \citenamefont {Schurkus}, \citenamefont {Schuster},
  \citenamefont {Shearn}, \citenamefont {Shorter}, \citenamefont {Shutty},
  \citenamefont {Shvarts}, \citenamefont {Sivak}, \citenamefont {Skruzny},
  \citenamefont {Smith}, \citenamefont {Somma}, \citenamefont {Sterling},
  \citenamefont {Strain}, \citenamefont {Szalay}, \citenamefont {Thor},
  \citenamefont {Torres}, \citenamefont {Vidal}, \citenamefont {Heidweiller},
  \citenamefont {White}, \citenamefont {Woo}, \citenamefont {Xing},
  \citenamefont {Yao}, \citenamefont {Yeh}, \citenamefont {Yoo}, \citenamefont
  {Young}, \citenamefont {Zalcman}, \citenamefont {Zhang}, \citenamefont {Zhu},
  \citenamefont {Zobrist}, \citenamefont {Rieffel}, \citenamefont {Biswas},
  \citenamefont {Babbush}, \citenamefont {Bacon}, \citenamefont {Hilton},
  \citenamefont {Lucero}, \citenamefont {Neven}, \citenamefont {Megrant},
  \citenamefont {Kelly}, \citenamefont {Roushan}, \citenamefont {Aleiner},
  \citenamefont {Smelyanskiy}, \citenamefont {Kechedzhi}, \citenamefont
  {Chen},\ and\ \citenamefont {Boixo}}]{morvan_phase_2024}%
  \BibitemOpen
  \bibfield  {author} {\bibinfo {author} {\bibfnamefont {A.}~\bibnamefont
  {Morvan}}, \bibinfo {author} {\bibfnamefont {B.}~\bibnamefont {Villalonga}},
  \bibinfo {author} {\bibfnamefont {X.}~\bibnamefont {Mi}}, \bibinfo {author}
  {\bibfnamefont {S.}~\bibnamefont {Mandr{\`a}}}, \bibinfo {author}
  {\bibfnamefont {A.}~\bibnamefont {Bengtsson}}, \bibinfo {author}
  {\bibfnamefont {P.~V.}\ \bibnamefont {Klimov}}, \bibinfo {author}
  {\bibfnamefont {Z.}~\bibnamefont {Chen}}, \bibinfo {author} {\bibfnamefont
  {S.}~\bibnamefont {Hong}}, \bibinfo {author} {\bibfnamefont {C.}~\bibnamefont
  {Erickson}}, \bibinfo {author} {\bibfnamefont {I.~K.}\ \bibnamefont
  {Drozdov}}, \bibinfo {author} {\bibfnamefont {J.}~\bibnamefont {Chau}},
  \bibinfo {author} {\bibfnamefont {G.}~\bibnamefont {Laun}}, \bibinfo {author}
  {\bibfnamefont {R.}~\bibnamefont {Movassagh}}, \bibinfo {author}
  {\bibfnamefont {A.}~\bibnamefont {Asfaw}}, \bibinfo {author} {\bibfnamefont
  {L.~T. A.~N.}\ \bibnamefont {Brand{\~a}o}}, \bibinfo {author} {\bibfnamefont
  {R.}~\bibnamefont {Peralta}}, \bibinfo {author} {\bibfnamefont
  {D.}~\bibnamefont {Abanin}}, \bibinfo {author} {\bibfnamefont
  {R.}~\bibnamefont {Acharya}}, \bibinfo {author} {\bibfnamefont
  {R.}~\bibnamefont {Allen}}, \bibinfo {author} {\bibfnamefont {T.~I.}\
  \bibnamefont {Andersen}}, \bibinfo {author} {\bibfnamefont {K.}~\bibnamefont
  {Anderson}}, \bibinfo {author} {\bibfnamefont {M.}~\bibnamefont {Ansmann}},
  \bibinfo {author} {\bibfnamefont {F.}~\bibnamefont {Arute}}, \bibinfo
  {author} {\bibfnamefont {K.}~\bibnamefont {Arya}}, \bibinfo {author}
  {\bibfnamefont {J.}~\bibnamefont {Atalaya}}, \bibinfo {author} {\bibfnamefont
  {J.~C.}\ \bibnamefont {Bardin}}, \bibinfo {author} {\bibfnamefont
  {A.}~\bibnamefont {Bilmes}}, \bibinfo {author} {\bibfnamefont
  {G.}~\bibnamefont {Bortoli}}, \bibinfo {author} {\bibfnamefont
  {A.}~\bibnamefont {Bourassa}}, \bibinfo {author} {\bibfnamefont
  {J.}~\bibnamefont {Bovaird}}, \bibinfo {author} {\bibfnamefont
  {L.}~\bibnamefont {Brill}}, \bibinfo {author} {\bibfnamefont
  {M.}~\bibnamefont {Broughton}}, \bibinfo {author} {\bibfnamefont {B.~B.}\
  \bibnamefont {Buckley}}, \bibinfo {author} {\bibfnamefont {D.~A.}\
  \bibnamefont {Buell}}, \bibinfo {author} {\bibfnamefont {T.}~\bibnamefont
  {Burger}}, \bibinfo {author} {\bibfnamefont {B.}~\bibnamefont {Burkett}},
  \bibinfo {author} {\bibfnamefont {N.}~\bibnamefont {Bushnell}}, \bibinfo
  {author} {\bibfnamefont {J.}~\bibnamefont {Campero}}, \bibinfo {author}
  {\bibfnamefont {H.-S.}\ \bibnamefont {Chang}}, \bibinfo {author}
  {\bibfnamefont {B.}~\bibnamefont {Chiaro}}, \bibinfo {author} {\bibfnamefont
  {D.}~\bibnamefont {Chik}}, \bibinfo {author} {\bibfnamefont {C.}~\bibnamefont
  {Chou}}, \bibinfo {author} {\bibfnamefont {J.}~\bibnamefont {Cogan}},
  \bibinfo {author} {\bibfnamefont {R.}~\bibnamefont {Collins}}, \bibinfo
  {author} {\bibfnamefont {P.}~\bibnamefont {Conner}}, \bibinfo {author}
  {\bibfnamefont {W.}~\bibnamefont {Courtney}}, \bibinfo {author}
  {\bibfnamefont {A.~L.}\ \bibnamefont {Crook}}, \bibinfo {author}
  {\bibfnamefont {B.}~\bibnamefont {Curtin}}, \bibinfo {author} {\bibfnamefont
  {D.~M.}\ \bibnamefont {Debroy}}, \bibinfo {author} {\bibfnamefont {A.~D.~T.}\
  \bibnamefont {Barba}}, \bibinfo {author} {\bibfnamefont {S.}~\bibnamefont
  {Demura}}, \bibinfo {author} {\bibfnamefont {A.~D.}\ \bibnamefont {Paolo}},
  \bibinfo {author} {\bibfnamefont {A.}~\bibnamefont {Dunsworth}}, \bibinfo
  {author} {\bibfnamefont {L.}~\bibnamefont {Faoro}}, \bibinfo {author}
  {\bibfnamefont {E.}~\bibnamefont {Farhi}}, \bibinfo {author} {\bibfnamefont
  {R.}~\bibnamefont {Fatemi}}, \bibinfo {author} {\bibfnamefont {V.~S.}\
  \bibnamefont {Ferreira}}, \bibinfo {author} {\bibfnamefont {L.~F.}\
  \bibnamefont {Burgos}}, \bibinfo {author} {\bibfnamefont {E.}~\bibnamefont
  {Forati}}, \bibinfo {author} {\bibfnamefont {A.~G.}\ \bibnamefont {Fowler}},
  \bibinfo {author} {\bibfnamefont {B.}~\bibnamefont {Foxen}}, \bibinfo
  {author} {\bibfnamefont {G.}~\bibnamefont {Garcia}}, \bibinfo {author}
  {\bibfnamefont {{\'E}.}~\bibnamefont {Genois}}, \bibinfo {author}
  {\bibfnamefont {W.}~\bibnamefont {Giang}}, \bibinfo {author} {\bibfnamefont
  {C.}~\bibnamefont {Gidney}}, \bibinfo {author} {\bibfnamefont
  {D.}~\bibnamefont {Gilboa}}, \bibinfo {author} {\bibfnamefont
  {M.}~\bibnamefont {Giustina}}, \bibinfo {author} {\bibfnamefont
  {R.}~\bibnamefont {Gosula}}, \bibinfo {author} {\bibfnamefont {A.~G.}\
  \bibnamefont {Dau}}, \bibinfo {author} {\bibfnamefont {J.~A.}\ \bibnamefont
  {Gross}}, \bibinfo {author} {\bibfnamefont {S.}~\bibnamefont {Habegger}},
  \bibinfo {author} {\bibfnamefont {M.~C.}\ \bibnamefont {Hamilton}}, \bibinfo
  {author} {\bibfnamefont {M.}~\bibnamefont {Hansen}}, \bibinfo {author}
  {\bibfnamefont {M.~P.}\ \bibnamefont {Harrigan}}, \bibinfo {author}
  {\bibfnamefont {S.~D.}\ \bibnamefont {Harrington}}, \bibinfo {author}
  {\bibfnamefont {P.}~\bibnamefont {Heu}}, \bibinfo {author} {\bibfnamefont
  {M.~R.}\ \bibnamefont {Hoffmann}}, \bibinfo {author} {\bibfnamefont
  {T.}~\bibnamefont {Huang}}, \bibinfo {author} {\bibfnamefont
  {A.}~\bibnamefont {Huff}}, \bibinfo {author} {\bibfnamefont {W.~J.}\
  \bibnamefont {Huggins}}, \bibinfo {author} {\bibfnamefont {L.~B.}\
  \bibnamefont {Ioffe}}, \bibinfo {author} {\bibfnamefont {S.~V.}\ \bibnamefont
  {Isakov}}, \bibinfo {author} {\bibfnamefont {J.}~\bibnamefont {Iveland}},
  \bibinfo {author} {\bibfnamefont {E.}~\bibnamefont {Jeffrey}}, \bibinfo
  {author} {\bibfnamefont {Z.}~\bibnamefont {Jiang}}, \bibinfo {author}
  {\bibfnamefont {C.}~\bibnamefont {Jones}}, \bibinfo {author} {\bibfnamefont
  {P.}~\bibnamefont {Juhas}}, \bibinfo {author} {\bibfnamefont
  {D.}~\bibnamefont {Kafri}}, \bibinfo {author} {\bibfnamefont
  {T.}~\bibnamefont {Khattar}}, \bibinfo {author} {\bibfnamefont
  {M.}~\bibnamefont {Khezri}}, \bibinfo {author} {\bibfnamefont
  {M.}~\bibnamefont {Kieferov{\'a}}}, \bibinfo {author} {\bibfnamefont
  {S.}~\bibnamefont {Kim}}, \bibinfo {author} {\bibfnamefont {A.}~\bibnamefont
  {Kitaev}}, \bibinfo {author} {\bibfnamefont {A.~R.}\ \bibnamefont {Klots}},
  \bibinfo {author} {\bibfnamefont {A.~N.}\ \bibnamefont {Korotkov}}, \bibinfo
  {author} {\bibfnamefont {F.}~\bibnamefont {Kostritsa}}, \bibinfo {author}
  {\bibfnamefont {J.~M.}\ \bibnamefont {Kreikebaum}}, \bibinfo {author}
  {\bibfnamefont {D.}~\bibnamefont {Landhuis}}, \bibinfo {author}
  {\bibfnamefont {P.}~\bibnamefont {Laptev}}, \bibinfo {author} {\bibfnamefont
  {K.-M.}\ \bibnamefont {Lau}}, \bibinfo {author} {\bibfnamefont
  {L.}~\bibnamefont {Laws}}, \bibinfo {author} {\bibfnamefont {J.}~\bibnamefont
  {Lee}}, \bibinfo {author} {\bibfnamefont {K.~W.}\ \bibnamefont {Lee}},
  \bibinfo {author} {\bibfnamefont {Y.~D.}\ \bibnamefont {Lensky}}, \bibinfo
  {author} {\bibfnamefont {B.~J.}\ \bibnamefont {Lester}}, \bibinfo {author}
  {\bibfnamefont {A.~T.}\ \bibnamefont {Lill}}, \bibinfo {author}
  {\bibfnamefont {W.}~\bibnamefont {Liu}}, \bibinfo {author} {\bibfnamefont
  {W.~P.}\ \bibnamefont {Livingston}}, \bibinfo {author} {\bibfnamefont
  {A.}~\bibnamefont {Locharla}}, \bibinfo {author} {\bibfnamefont {F.~D.}\
  \bibnamefont {Malone}}, \bibinfo {author} {\bibfnamefont {O.}~\bibnamefont
  {Martin}}, \bibinfo {author} {\bibfnamefont {S.}~\bibnamefont {Martin}},
  \bibinfo {author} {\bibfnamefont {J.~R.}\ \bibnamefont {McClean}}, \bibinfo
  {author} {\bibfnamefont {M.}~\bibnamefont {McEwen}}, \bibinfo {author}
  {\bibfnamefont {K.~C.}\ \bibnamefont {Miao}}, \bibinfo {author}
  {\bibfnamefont {A.}~\bibnamefont {Mieszala}}, \bibinfo {author}
  {\bibfnamefont {S.}~\bibnamefont {Montazeri}}, \bibinfo {author}
  {\bibfnamefont {W.}~\bibnamefont {Mruczkiewicz}}, \bibinfo {author}
  {\bibfnamefont {O.}~\bibnamefont {Naaman}}, \bibinfo {author} {\bibfnamefont
  {M.}~\bibnamefont {Neeley}}, \bibinfo {author} {\bibfnamefont
  {C.}~\bibnamefont {Neill}}, \bibinfo {author} {\bibfnamefont
  {A.}~\bibnamefont {Nersisyan}}, \bibinfo {author} {\bibfnamefont
  {M.}~\bibnamefont {Newman}}, \bibinfo {author} {\bibfnamefont {J.~H.}\
  \bibnamefont {Ng}}, \bibinfo {author} {\bibfnamefont {A.}~\bibnamefont
  {Nguyen}}, \bibinfo {author} {\bibfnamefont {M.}~\bibnamefont {Nguyen}},
  \bibinfo {author} {\bibfnamefont {M.~Y.}\ \bibnamefont {Niu}}, \bibinfo
  {author} {\bibfnamefont {T.~E.}\ \bibnamefont {O'Brien}}, \bibinfo {author}
  {\bibfnamefont {S.}~\bibnamefont {Omonije}}, \bibinfo {author} {\bibfnamefont
  {A.}~\bibnamefont {Opremcak}}, \bibinfo {author} {\bibfnamefont
  {A.}~\bibnamefont {Petukhov}}, \bibinfo {author} {\bibfnamefont
  {R.}~\bibnamefont {Potter}}, \bibinfo {author} {\bibfnamefont {L.~P.}\
  \bibnamefont {Pryadko}}, \bibinfo {author} {\bibfnamefont {C.}~\bibnamefont
  {Quintana}}, \bibinfo {author} {\bibfnamefont {D.~M.}\ \bibnamefont
  {Rhodes}}, \bibinfo {author} {\bibfnamefont {C.}~\bibnamefont {Rocque}},
  \bibinfo {author} {\bibfnamefont {E.}~\bibnamefont {Rosenberg}}, \bibinfo
  {author} {\bibfnamefont {N.~C.}\ \bibnamefont {Rubin}}, \bibinfo {author}
  {\bibfnamefont {N.}~\bibnamefont {Saei}}, \bibinfo {author} {\bibfnamefont
  {D.}~\bibnamefont {Sank}}, \bibinfo {author} {\bibfnamefont {K.}~\bibnamefont
  {Sankaragomathi}}, \bibinfo {author} {\bibfnamefont {K.~J.}\ \bibnamefont
  {Satzinger}}, \bibinfo {author} {\bibfnamefont {H.~F.}\ \bibnamefont
  {Schurkus}}, \bibinfo {author} {\bibfnamefont {C.}~\bibnamefont {Schuster}},
  \bibinfo {author} {\bibfnamefont {M.~J.}\ \bibnamefont {Shearn}}, \bibinfo
  {author} {\bibfnamefont {A.}~\bibnamefont {Shorter}}, \bibinfo {author}
  {\bibfnamefont {N.}~\bibnamefont {Shutty}}, \bibinfo {author} {\bibfnamefont
  {V.}~\bibnamefont {Shvarts}}, \bibinfo {author} {\bibfnamefont
  {V.}~\bibnamefont {Sivak}}, \bibinfo {author} {\bibfnamefont
  {J.}~\bibnamefont {Skruzny}}, \bibinfo {author} {\bibfnamefont {W.~C.}\
  \bibnamefont {Smith}}, \bibinfo {author} {\bibfnamefont {R.~D.}\ \bibnamefont
  {Somma}}, \bibinfo {author} {\bibfnamefont {G.}~\bibnamefont {Sterling}},
  \bibinfo {author} {\bibfnamefont {D.}~\bibnamefont {Strain}}, \bibinfo
  {author} {\bibfnamefont {M.}~\bibnamefont {Szalay}}, \bibinfo {author}
  {\bibfnamefont {D.}~\bibnamefont {Thor}}, \bibinfo {author} {\bibfnamefont
  {A.}~\bibnamefont {Torres}}, \bibinfo {author} {\bibfnamefont
  {G.}~\bibnamefont {Vidal}}, \bibinfo {author} {\bibfnamefont {C.~V.}\
  \bibnamefont {Heidweiller}}, \bibinfo {author} {\bibfnamefont
  {T.}~\bibnamefont {White}}, \bibinfo {author} {\bibfnamefont {B.~W.~K.}\
  \bibnamefont {Woo}}, \bibinfo {author} {\bibfnamefont {C.}~\bibnamefont
  {Xing}}, \bibinfo {author} {\bibfnamefont {Z.~J.}\ \bibnamefont {Yao}},
  \bibinfo {author} {\bibfnamefont {P.}~\bibnamefont {Yeh}}, \bibinfo {author}
  {\bibfnamefont {J.}~\bibnamefont {Yoo}}, \bibinfo {author} {\bibfnamefont
  {G.}~\bibnamefont {Young}}, \bibinfo {author} {\bibfnamefont
  {A.}~\bibnamefont {Zalcman}}, \bibinfo {author} {\bibfnamefont
  {Y.}~\bibnamefont {Zhang}}, \bibinfo {author} {\bibfnamefont
  {N.}~\bibnamefont {Zhu}}, \bibinfo {author} {\bibfnamefont {N.}~\bibnamefont
  {Zobrist}}, \bibinfo {author} {\bibfnamefont {E.~G.}\ \bibnamefont
  {Rieffel}}, \bibinfo {author} {\bibfnamefont {R.}~\bibnamefont {Biswas}},
  \bibinfo {author} {\bibfnamefont {R.}~\bibnamefont {Babbush}}, \bibinfo
  {author} {\bibfnamefont {D.}~\bibnamefont {Bacon}}, \bibinfo {author}
  {\bibfnamefont {J.}~\bibnamefont {Hilton}}, \bibinfo {author} {\bibfnamefont
  {E.}~\bibnamefont {Lucero}}, \bibinfo {author} {\bibfnamefont
  {H.}~\bibnamefont {Neven}}, \bibinfo {author} {\bibfnamefont
  {A.}~\bibnamefont {Megrant}}, \bibinfo {author} {\bibfnamefont
  {J.}~\bibnamefont {Kelly}}, \bibinfo {author} {\bibfnamefont
  {P.}~\bibnamefont {Roushan}}, \bibinfo {author} {\bibfnamefont
  {I.}~\bibnamefont {Aleiner}}, \bibinfo {author} {\bibfnamefont
  {V.}~\bibnamefont {Smelyanskiy}}, \bibinfo {author} {\bibfnamefont
  {K.}~\bibnamefont {Kechedzhi}}, \bibinfo {author} {\bibfnamefont
  {Y.}~\bibnamefont {Chen}},\ and\ \bibinfo {author} {\bibfnamefont
  {S.}~\bibnamefont {Boixo}},\ }\bibfield  {title} {\bibinfo {title} {Phase
  transitions in random circuit sampling},\ }\href
  {https://doi.org/10.1038/s41586-024-07998-6} {\bibfield  {journal} {\bibinfo
  {journal} {Nature}\ }\textbf {\bibinfo {volume} {634}},\ \bibinfo {pages}
  {328} (\bibinfo {year} {2024})}\BibitemShut {NoStop}%
\bibitem [{\citenamefont {Klimov}\ \emph {et~al.}(2024)\citenamefont {Klimov},
  \citenamefont {Bengtsson}, \citenamefont {Quintana}, \citenamefont
  {Bourassa}, \citenamefont {Hong}, \citenamefont {Dunsworth}, \citenamefont
  {Satzinger}, \citenamefont {Livingston}, \citenamefont {Sivak}, \citenamefont
  {Niu}, \citenamefont {Andersen}, \citenamefont {Zhang}, \citenamefont {Chik},
  \citenamefont {Chen}, \citenamefont {Neill}, \citenamefont {Erickson},
  \citenamefont {Grajales~Dau}, \citenamefont {Megrant}, \citenamefont
  {Roushan}, \citenamefont {Korotkov}, \citenamefont {Kelly}, \citenamefont
  {Smelyanskiy}, \citenamefont {Chen},\ and\ \citenamefont
  {Neven}}]{klimov_optimizing_2024}%
  \BibitemOpen
  \bibfield  {author} {\bibinfo {author} {\bibfnamefont {P.~V.}\ \bibnamefont
  {Klimov}}, \bibinfo {author} {\bibfnamefont {A.}~\bibnamefont {Bengtsson}},
  \bibinfo {author} {\bibfnamefont {C.}~\bibnamefont {Quintana}}, \bibinfo
  {author} {\bibfnamefont {A.}~\bibnamefont {Bourassa}}, \bibinfo {author}
  {\bibfnamefont {S.}~\bibnamefont {Hong}}, \bibinfo {author} {\bibfnamefont
  {A.}~\bibnamefont {Dunsworth}}, \bibinfo {author} {\bibfnamefont {K.~J.}\
  \bibnamefont {Satzinger}}, \bibinfo {author} {\bibfnamefont {W.~P.}\
  \bibnamefont {Livingston}}, \bibinfo {author} {\bibfnamefont
  {V.}~\bibnamefont {Sivak}}, \bibinfo {author} {\bibfnamefont {M.~Y.}\
  \bibnamefont {Niu}}, \bibinfo {author} {\bibfnamefont {T.~I.}\ \bibnamefont
  {Andersen}}, \bibinfo {author} {\bibfnamefont {Y.}~\bibnamefont {Zhang}},
  \bibinfo {author} {\bibfnamefont {D.}~\bibnamefont {Chik}}, \bibinfo {author}
  {\bibfnamefont {Z.}~\bibnamefont {Chen}}, \bibinfo {author} {\bibfnamefont
  {C.}~\bibnamefont {Neill}}, \bibinfo {author} {\bibfnamefont
  {C.}~\bibnamefont {Erickson}}, \bibinfo {author} {\bibfnamefont
  {A.}~\bibnamefont {Grajales~Dau}}, \bibinfo {author} {\bibfnamefont
  {A.}~\bibnamefont {Megrant}}, \bibinfo {author} {\bibfnamefont
  {P.}~\bibnamefont {Roushan}}, \bibinfo {author} {\bibfnamefont {A.~N.}\
  \bibnamefont {Korotkov}}, \bibinfo {author} {\bibfnamefont {J.}~\bibnamefont
  {Kelly}}, \bibinfo {author} {\bibfnamefont {V.}~\bibnamefont {Smelyanskiy}},
  \bibinfo {author} {\bibfnamefont {Y.}~\bibnamefont {Chen}},\ and\ \bibinfo
  {author} {\bibfnamefont {H.}~\bibnamefont {Neven}},\ }\bibfield  {title}
  {\bibinfo {title} {Optimizing quantum gates towards the scale of logical
  qubits},\ }\href {https://doi.org/10.1038/s41467-024-46623-y} {\bibfield
  {journal} {\bibinfo  {journal} {Nature Communications}\ }\textbf {\bibinfo
  {volume} {15}},\ \bibinfo {pages} {2442} (\bibinfo {year}
  {2024})}\BibitemShut {NoStop}%
\bibitem [{\citenamefont {Eickbusch}\ \emph {et~al.}(2024)\citenamefont
  {Eickbusch}, \citenamefont {McEwen}, \citenamefont {Sivak}, \citenamefont
  {Bourassa}, \citenamefont {Atalaya}, \citenamefont {Claes}, \citenamefont
  {Kafri}, \citenamefont {Gidney}, \citenamefont {Warren}, \citenamefont
  {Gross}, \citenamefont {Opremcak}, \citenamefont {Miao}, \citenamefont
  {Roberts}, \citenamefont {Satzinger}, \citenamefont {Bengtsson},
  \citenamefont {Neeley}, \citenamefont {Livingston}, \citenamefont {Greene},
  \citenamefont {Rajeev}, \citenamefont {Acharya}, \citenamefont {Beni},
  \citenamefont {Aigeldinger}, \citenamefont {Alcaraz}, \citenamefont
  {Andersen}, \citenamefont {Ansmann}, \citenamefont {Frank}, \citenamefont
  {Arute}, \citenamefont {Arya}, \citenamefont {Asfaw}, \citenamefont
  {Babbush}, \citenamefont {Ballard}, \citenamefont {Bardin}, \citenamefont
  {Bilmes}, \citenamefont {Jenna}, \citenamefont {Bovaird}, \citenamefont
  {Bowers}, \citenamefont {Brill}, \citenamefont {Broughton}, \citenamefont
  {Browne}, \citenamefont {Buchea}, \citenamefont {Buckley}, \citenamefont
  {Tim}, \citenamefont {Burger}, \citenamefont {Burkett}, \citenamefont
  {Bushnell}, \citenamefont {Cabrera}, \citenamefont {Campero}, \citenamefont
  {Chang}, \citenamefont {Chiaro}, \citenamefont {Chih}, \citenamefont
  {Cleland}, \citenamefont {Cogan}, \citenamefont {Collins}, \citenamefont
  {Conner}, \citenamefont {Courtney}, \citenamefont {Alexander}, \citenamefont
  {Crook}, \citenamefont {Curtin}, \citenamefont {Das}, \citenamefont {Barba},
  \citenamefont {Demura}, \citenamefont {Lorenzo}, \citenamefont {Paolo},
  \citenamefont {Donohoe}, \citenamefont {Drozdov}, \citenamefont {Dunsworth},
  \citenamefont {Elbag}, \citenamefont {Elzouka}, \citenamefont {Erickson},
  \citenamefont {Ferreira}, \citenamefont {Burgos}, \citenamefont {Forati},
  \citenamefont {Fowler}, \citenamefont {Foxen}, \citenamefont {Ganjam},
  \citenamefont {Gonzalo}, \citenamefont {Garcia}, \citenamefont {Gasca},
  \citenamefont {Élie Genois}, \citenamefont {Giang}, \citenamefont {Gilboa},
  \citenamefont {Gosula}, \citenamefont {Dau}, \citenamefont {Dietrich},
  \citenamefont {Graumann}, \citenamefont {Ha}, \citenamefont {Habegger},
  \citenamefont {Hansen}, \citenamefont {Harrigan}, \citenamefont {Harrington},
  \citenamefont {Heslin}, \citenamefont {Heu}, \citenamefont {Higgott},
  \citenamefont {Hiltermann}, \citenamefont {Hilton}, \citenamefont {Huang},
  \citenamefont {Huff}, \citenamefont {Huggins}, \citenamefont {Jeffrey},
  \citenamefont {Jiang}, \citenamefont {Jin}, \citenamefont {Jones},
  \citenamefont {Joshi}, \citenamefont {Juhas}, \citenamefont {Kabel},
  \citenamefont {Kang}, \citenamefont {Amir}, \citenamefont {Karamlou},
  \citenamefont {Kechedzhi}, \citenamefont {Khaire}, \citenamefont {Khattar},
  \citenamefont {Khezri}, \citenamefont {Kim}, \citenamefont {Kobrin},
  \citenamefont {Korotkov}, \citenamefont {Kostritsa}, \citenamefont
  {Kreikebaum}, \citenamefont {Kurilovich}, \citenamefont {Landhuis},
  \citenamefont {Tiano}, \citenamefont {Lange-Dei}, \citenamefont {Langley},
  \citenamefont {Lau}, \citenamefont {Ledford}, \citenamefont {Lee},
  \citenamefont {Lester}, \citenamefont {Guevel}, \citenamefont {Wing},
  \citenamefont {Li}, \citenamefont {Lill}, \citenamefont {Locharla},
  \citenamefont {Lucero}, \citenamefont {Lundahl}, \citenamefont {Lunt},
  \citenamefont {Madhuk}, \citenamefont {Maloney}, \citenamefont {Mandrà},
  \citenamefont {Martin}, \citenamefont {Martin}, \citenamefont {Maxfield},
  \citenamefont {McClean}, \citenamefont {Meeks}, \citenamefont {Anthony},
  \citenamefont {Megrant}, \citenamefont {Molavi}, \citenamefont {Molina},
  \citenamefont {Montazeri}, \citenamefont {Movassagh}, \citenamefont {Newman},
  \citenamefont {Nguyen}, \citenamefont {Nguyen}, \citenamefont {Ni},
  \citenamefont {Oas}, \citenamefont {Orosco}, \citenamefont {Ottosson},
  \citenamefont {Pizzuto}, \citenamefont {Potter}, \citenamefont {Pritchard},
  \citenamefont {Quintana}, \citenamefont {Ramachandran}, \citenamefont
  {Reagor}, \citenamefont {Rhodes}, \citenamefont {Rosenberg}, \citenamefont
  {Rossi}, \citenamefont {Sankaragomathi}, \citenamefont {Schurkus},
  \citenamefont {Shearn}, \citenamefont {Shorter}, \citenamefont {Shutty},
  \citenamefont {Shvarts}, \citenamefont {Small}, \citenamefont {Smith},
  \citenamefont {Springer}, \citenamefont {Sterling}, \citenamefont {Suchard},
  \citenamefont {Szasz}, \citenamefont {Sztein}, \citenamefont {Thor},
  \citenamefont {Tomita}, \citenamefont {Torres}, \citenamefont {Torunbalci},
  \citenamefont {Vaishnav}, \citenamefont {Vargas}, \citenamefont {Sergey},
  \citenamefont {Vdovichev}, \citenamefont {Vidal}, \citenamefont
  {Heidweiller}, \citenamefont {Waltman}, \citenamefont {Waltz}, \citenamefont
  {Wang}, \citenamefont {Ware}, \citenamefont {Weidel}, \citenamefont {White},
  \citenamefont {Wong}, \citenamefont {Woo}, \citenamefont {Woodson},
  \citenamefont {Xing}, \citenamefont {Yao}, \citenamefont {Yeh}, \citenamefont
  {Ying}, \citenamefont {Yoo}, \citenamefont {Yosri}, \citenamefont {Young},
  \citenamefont {Zalcman}, \citenamefont {Yaxing}, \citenamefont {Zhang},
  \citenamefont {Zhu}, \citenamefont {Boixo}, \citenamefont {Kelly},
  \citenamefont {Smelyanskiy}, \citenamefont {Neven}, \citenamefont {Bacon},
  \citenamefont {Chen}, \citenamefont {Klimov}, \citenamefont {Roushan},
  \citenamefont {Neill}, \citenamefont {Chen},\ and\ \citenamefont
  {Morvan}}]{eickbusch2024demonstratingdynamicsurfacecodes}%
  \BibitemOpen
  \bibfield  {author} {\bibinfo {author} {\bibfnamefont {A.}~\bibnamefont
  {Eickbusch}}, \bibinfo {author} {\bibfnamefont {M.}~\bibnamefont {McEwen}},
  \bibinfo {author} {\bibfnamefont {V.}~\bibnamefont {Sivak}}, \bibinfo
  {author} {\bibfnamefont {A.}~\bibnamefont {Bourassa}}, \bibinfo {author}
  {\bibfnamefont {J.}~\bibnamefont {Atalaya}}, \bibinfo {author} {\bibfnamefont
  {J.}~\bibnamefont {Claes}}, \bibinfo {author} {\bibfnamefont
  {D.}~\bibnamefont {Kafri}}, \bibinfo {author} {\bibfnamefont
  {C.}~\bibnamefont {Gidney}}, \bibinfo {author} {\bibfnamefont {C.~W.}\
  \bibnamefont {Warren}}, \bibinfo {author} {\bibfnamefont {J.}~\bibnamefont
  {Gross}}, \bibinfo {author} {\bibfnamefont {A.}~\bibnamefont {Opremcak}},
  \bibinfo {author} {\bibfnamefont {N.~Z. K.~C.}\ \bibnamefont {Miao}},
  \bibinfo {author} {\bibfnamefont {G.}~\bibnamefont {Roberts}}, \bibinfo
  {author} {\bibfnamefont {K.~J.}\ \bibnamefont {Satzinger}}, \bibinfo {author}
  {\bibfnamefont {A.}~\bibnamefont {Bengtsson}}, \bibinfo {author}
  {\bibfnamefont {M.}~\bibnamefont {Neeley}}, \bibinfo {author} {\bibfnamefont
  {W.~P.}\ \bibnamefont {Livingston}}, \bibinfo {author} {\bibfnamefont
  {A.}~\bibnamefont {Greene}}, \bibinfo {author} {\bibnamefont {Rajeev}},
  \bibinfo {author} {\bibnamefont {Acharya}}, \bibinfo {author} {\bibfnamefont
  {L.~A.}\ \bibnamefont {Beni}}, \bibinfo {author} {\bibfnamefont
  {G.}~\bibnamefont {Aigeldinger}}, \bibinfo {author} {\bibfnamefont
  {R.}~\bibnamefont {Alcaraz}}, \bibinfo {author} {\bibfnamefont {T.~I.}\
  \bibnamefont {Andersen}}, \bibinfo {author} {\bibfnamefont {M.}~\bibnamefont
  {Ansmann}}, \bibinfo {author} {\bibnamefont {Frank}}, \bibinfo {author}
  {\bibnamefont {Arute}}, \bibinfo {author} {\bibfnamefont {K.}~\bibnamefont
  {Arya}}, \bibinfo {author} {\bibfnamefont {A.}~\bibnamefont {Asfaw}},
  \bibinfo {author} {\bibfnamefont {R.}~\bibnamefont {Babbush}}, \bibinfo
  {author} {\bibfnamefont {B.}~\bibnamefont {Ballard}}, \bibinfo {author}
  {\bibfnamefont {J.~C.}\ \bibnamefont {Bardin}}, \bibinfo {author}
  {\bibfnamefont {A.}~\bibnamefont {Bilmes}}, \bibinfo {author} {\bibnamefont
  {Jenna}}, \bibinfo {author} {\bibnamefont {Bovaird}}, \bibinfo {author}
  {\bibfnamefont {D.}~\bibnamefont {Bowers}}, \bibinfo {author} {\bibfnamefont
  {L.}~\bibnamefont {Brill}}, \bibinfo {author} {\bibfnamefont
  {M.}~\bibnamefont {Broughton}}, \bibinfo {author} {\bibfnamefont {D.~A.}\
  \bibnamefont {Browne}}, \bibinfo {author} {\bibfnamefont {B.}~\bibnamefont
  {Buchea}}, \bibinfo {author} {\bibfnamefont {B.~B.}\ \bibnamefont {Buckley}},
  \bibinfo {author} {\bibnamefont {Tim}}, \bibinfo {author} {\bibnamefont
  {Burger}}, \bibinfo {author} {\bibfnamefont {B.}~\bibnamefont {Burkett}},
  \bibinfo {author} {\bibfnamefont {N.}~\bibnamefont {Bushnell}}, \bibinfo
  {author} {\bibfnamefont {A.}~\bibnamefont {Cabrera}}, \bibinfo {author}
  {\bibfnamefont {J.}~\bibnamefont {Campero}}, \bibinfo {author} {\bibfnamefont
  {H.-S.}\ \bibnamefont {Chang}}, \bibinfo {author} {\bibfnamefont
  {B.}~\bibnamefont {Chiaro}}, \bibinfo {author} {\bibfnamefont {L.-Y.}\
  \bibnamefont {Chih}}, \bibinfo {author} {\bibfnamefont {A.~Y.}\ \bibnamefont
  {Cleland}}, \bibinfo {author} {\bibfnamefont {J.}~\bibnamefont {Cogan}},
  \bibinfo {author} {\bibfnamefont {R.}~\bibnamefont {Collins}}, \bibinfo
  {author} {\bibfnamefont {P.}~\bibnamefont {Conner}}, \bibinfo {author}
  {\bibfnamefont {W.}~\bibnamefont {Courtney}}, \bibinfo {author} {\bibnamefont
  {Alexander}}, \bibinfo {author} {\bibfnamefont {L.}~\bibnamefont {Crook}},
  \bibinfo {author} {\bibfnamefont {B.}~\bibnamefont {Curtin}}, \bibinfo
  {author} {\bibfnamefont {S.}~\bibnamefont {Das}}, \bibinfo {author}
  {\bibfnamefont {A.~D.~T.}\ \bibnamefont {Barba}}, \bibinfo {author}
  {\bibfnamefont {S.}~\bibnamefont {Demura}}, \bibinfo {author} {\bibfnamefont
  {L.~D.}\ \bibnamefont {Lorenzo}}, \bibinfo {author} {\bibfnamefont {A.~D.}\
  \bibnamefont {Paolo}}, \bibinfo {author} {\bibfnamefont {P.}~\bibnamefont
  {Donohoe}}, \bibinfo {author} {\bibfnamefont {I.~K.}\ \bibnamefont
  {Drozdov}}, \bibinfo {author} {\bibfnamefont {A.}~\bibnamefont {Dunsworth}},
  \bibinfo {author} {\bibfnamefont {A.~M.}\ \bibnamefont {Elbag}}, \bibinfo
  {author} {\bibfnamefont {M.}~\bibnamefont {Elzouka}}, \bibinfo {author}
  {\bibfnamefont {C.}~\bibnamefont {Erickson}}, \bibinfo {author}
  {\bibfnamefont {V.~S.}\ \bibnamefont {Ferreira}}, \bibinfo {author}
  {\bibfnamefont {L.~F.}\ \bibnamefont {Burgos}}, \bibinfo {author}
  {\bibfnamefont {E.}~\bibnamefont {Forati}}, \bibinfo {author} {\bibfnamefont
  {A.~G.}\ \bibnamefont {Fowler}}, \bibinfo {author} {\bibfnamefont
  {B.}~\bibnamefont {Foxen}}, \bibinfo {author} {\bibfnamefont
  {S.}~\bibnamefont {Ganjam}}, \bibinfo {author} {\bibnamefont {Gonzalo}},
  \bibinfo {author} {\bibnamefont {Garcia}}, \bibinfo {author} {\bibfnamefont
  {R.}~\bibnamefont {Gasca}}, \bibinfo {author} {\bibnamefont {Élie Genois}},
  \bibinfo {author} {\bibfnamefont {W.}~\bibnamefont {Giang}}, \bibinfo
  {author} {\bibfnamefont {D.}~\bibnamefont {Gilboa}}, \bibinfo {author}
  {\bibfnamefont {R.}~\bibnamefont {Gosula}}, \bibinfo {author} {\bibfnamefont
  {A.~G.}\ \bibnamefont {Dau}}, \bibinfo {author} {\bibnamefont {Dietrich}},
  \bibinfo {author} {\bibnamefont {Graumann}}, \bibinfo {author} {\bibfnamefont
  {T.}~\bibnamefont {Ha}}, \bibinfo {author} {\bibfnamefont {S.}~\bibnamefont
  {Habegger}}, \bibinfo {author} {\bibfnamefont {M.}~\bibnamefont {Hansen}},
  \bibinfo {author} {\bibfnamefont {M.~P.}\ \bibnamefont {Harrigan}}, \bibinfo
  {author} {\bibfnamefont {S.~D.}\ \bibnamefont {Harrington}}, \bibinfo
  {author} {\bibfnamefont {S.}~\bibnamefont {Heslin}}, \bibinfo {author}
  {\bibfnamefont {P.}~\bibnamefont {Heu}}, \bibinfo {author} {\bibfnamefont
  {O.}~\bibnamefont {Higgott}}, \bibinfo {author} {\bibfnamefont
  {R.}~\bibnamefont {Hiltermann}}, \bibinfo {author} {\bibfnamefont
  {J.}~\bibnamefont {Hilton}}, \bibinfo {author} {\bibfnamefont {H.-Y.}\
  \bibnamefont {Huang}}, \bibinfo {author} {\bibfnamefont {A.}~\bibnamefont
  {Huff}}, \bibinfo {author} {\bibfnamefont {W.~J.}\ \bibnamefont {Huggins}},
  \bibinfo {author} {\bibfnamefont {E.}~\bibnamefont {Jeffrey}}, \bibinfo
  {author} {\bibfnamefont {Z.}~\bibnamefont {Jiang}}, \bibinfo {author}
  {\bibfnamefont {X.}~\bibnamefont {Jin}}, \bibinfo {author} {\bibfnamefont
  {C.}~\bibnamefont {Jones}}, \bibinfo {author} {\bibfnamefont
  {C.}~\bibnamefont {Joshi}}, \bibinfo {author} {\bibfnamefont
  {P.}~\bibnamefont {Juhas}}, \bibinfo {author} {\bibfnamefont
  {A.}~\bibnamefont {Kabel}}, \bibinfo {author} {\bibfnamefont
  {H.}~\bibnamefont {Kang}}, \bibinfo {author} {\bibnamefont {Amir}}, \bibinfo
  {author} {\bibfnamefont {H.}~\bibnamefont {Karamlou}}, \bibinfo {author}
  {\bibfnamefont {K.}~\bibnamefont {Kechedzhi}}, \bibinfo {author}
  {\bibfnamefont {T.}~\bibnamefont {Khaire}}, \bibinfo {author} {\bibfnamefont
  {T.}~\bibnamefont {Khattar}}, \bibinfo {author} {\bibfnamefont
  {M.}~\bibnamefont {Khezri}}, \bibinfo {author} {\bibfnamefont
  {S.}~\bibnamefont {Kim}}, \bibinfo {author} {\bibfnamefont {B.}~\bibnamefont
  {Kobrin}}, \bibinfo {author} {\bibfnamefont {A.~N.}\ \bibnamefont
  {Korotkov}}, \bibinfo {author} {\bibfnamefont {F.}~\bibnamefont {Kostritsa}},
  \bibinfo {author} {\bibfnamefont {J.~M.}\ \bibnamefont {Kreikebaum}},
  \bibinfo {author} {\bibfnamefont {V.~D.}\ \bibnamefont {Kurilovich}},
  \bibinfo {author} {\bibfnamefont {D.}~\bibnamefont {Landhuis}}, \bibinfo
  {author} {\bibnamefont {Tiano}}, \bibinfo {author} {\bibnamefont
  {Lange-Dei}}, \bibinfo {author} {\bibfnamefont {B.~W.}\ \bibnamefont
  {Langley}}, \bibinfo {author} {\bibfnamefont {K.-M.}\ \bibnamefont {Lau}},
  \bibinfo {author} {\bibfnamefont {J.}~\bibnamefont {Ledford}}, \bibinfo
  {author} {\bibfnamefont {K.}~\bibnamefont {Lee}}, \bibinfo {author}
  {\bibfnamefont {B.~J.}\ \bibnamefont {Lester}}, \bibinfo {author}
  {\bibfnamefont {L.~L.}\ \bibnamefont {Guevel}}, \bibinfo {author}
  {\bibnamefont {Wing}}, \bibinfo {author} {\bibfnamefont {Y.}~\bibnamefont
  {Li}}, \bibinfo {author} {\bibfnamefont {A.~T.}\ \bibnamefont {Lill}},
  \bibinfo {author} {\bibfnamefont {A.}~\bibnamefont {Locharla}}, \bibinfo
  {author} {\bibfnamefont {E.}~\bibnamefont {Lucero}}, \bibinfo {author}
  {\bibfnamefont {D.}~\bibnamefont {Lundahl}}, \bibinfo {author} {\bibfnamefont
  {A.}~\bibnamefont {Lunt}}, \bibinfo {author} {\bibfnamefont {S.}~\bibnamefont
  {Madhuk}}, \bibinfo {author} {\bibfnamefont {A.}~\bibnamefont {Maloney}},
  \bibinfo {author} {\bibfnamefont {S.}~\bibnamefont {Mandrà}}, \bibinfo
  {author} {\bibfnamefont {L.~S.}\ \bibnamefont {Martin}}, \bibinfo {author}
  {\bibfnamefont {O.}~\bibnamefont {Martin}}, \bibinfo {author} {\bibfnamefont
  {C.}~\bibnamefont {Maxfield}}, \bibinfo {author} {\bibfnamefont {J.~R.}\
  \bibnamefont {McClean}}, \bibinfo {author} {\bibfnamefont {S.}~\bibnamefont
  {Meeks}}, \bibinfo {author} {\bibnamefont {Anthony}}, \bibinfo {author}
  {\bibnamefont {Megrant}}, \bibinfo {author} {\bibfnamefont {R.}~\bibnamefont
  {Molavi}}, \bibinfo {author} {\bibfnamefont {S.}~\bibnamefont {Molina}},
  \bibinfo {author} {\bibfnamefont {S.}~\bibnamefont {Montazeri}}, \bibinfo
  {author} {\bibfnamefont {R.}~\bibnamefont {Movassagh}}, \bibinfo {author}
  {\bibfnamefont {M.}~\bibnamefont {Newman}}, \bibinfo {author} {\bibfnamefont
  {A.}~\bibnamefont {Nguyen}}, \bibinfo {author} {\bibfnamefont
  {M.}~\bibnamefont {Nguyen}}, \bibinfo {author} {\bibfnamefont {C.-H.}\
  \bibnamefont {Ni}}, \bibinfo {author} {\bibfnamefont {L.}~\bibnamefont
  {Oas}}, \bibinfo {author} {\bibfnamefont {R.}~\bibnamefont {Orosco}},
  \bibinfo {author} {\bibfnamefont {K.}~\bibnamefont {Ottosson}}, \bibinfo
  {author} {\bibfnamefont {A.}~\bibnamefont {Pizzuto}}, \bibinfo {author}
  {\bibfnamefont {R.}~\bibnamefont {Potter}}, \bibinfo {author} {\bibfnamefont
  {O.}~\bibnamefont {Pritchard}}, \bibinfo {author} {\bibfnamefont
  {C.}~\bibnamefont {Quintana}}, \bibinfo {author} {\bibfnamefont
  {G.}~\bibnamefont {Ramachandran}}, \bibinfo {author} {\bibfnamefont {M.~J.}\
  \bibnamefont {Reagor}}, \bibinfo {author} {\bibfnamefont {D.~M.}\
  \bibnamefont {Rhodes}}, \bibinfo {author} {\bibfnamefont {E.}~\bibnamefont
  {Rosenberg}}, \bibinfo {author} {\bibfnamefont {E.}~\bibnamefont {Rossi}},
  \bibinfo {author} {\bibfnamefont {K.}~\bibnamefont {Sankaragomathi}},
  \bibinfo {author} {\bibfnamefont {H.~F.}\ \bibnamefont {Schurkus}}, \bibinfo
  {author} {\bibfnamefont {M.~J.}\ \bibnamefont {Shearn}}, \bibinfo {author}
  {\bibfnamefont {A.}~\bibnamefont {Shorter}}, \bibinfo {author} {\bibfnamefont
  {N.}~\bibnamefont {Shutty}}, \bibinfo {author} {\bibfnamefont
  {V.}~\bibnamefont {Shvarts}}, \bibinfo {author} {\bibfnamefont
  {S.}~\bibnamefont {Small}}, \bibinfo {author} {\bibfnamefont {W.~C.}\
  \bibnamefont {Smith}}, \bibinfo {author} {\bibfnamefont {S.}~\bibnamefont
  {Springer}}, \bibinfo {author} {\bibfnamefont {G.}~\bibnamefont {Sterling}},
  \bibinfo {author} {\bibfnamefont {J.}~\bibnamefont {Suchard}}, \bibinfo
  {author} {\bibfnamefont {A.}~\bibnamefont {Szasz}}, \bibinfo {author}
  {\bibfnamefont {A.}~\bibnamefont {Sztein}}, \bibinfo {author} {\bibfnamefont
  {D.}~\bibnamefont {Thor}}, \bibinfo {author} {\bibfnamefont {E.}~\bibnamefont
  {Tomita}}, \bibinfo {author} {\bibfnamefont {A.}~\bibnamefont {Torres}},
  \bibinfo {author} {\bibfnamefont {M.~M.}\ \bibnamefont {Torunbalci}},
  \bibinfo {author} {\bibfnamefont {A.}~\bibnamefont {Vaishnav}}, \bibinfo
  {author} {\bibfnamefont {J.}~\bibnamefont {Vargas}}, \bibinfo {author}
  {\bibnamefont {Sergey}}, \bibinfo {author} {\bibnamefont {Vdovichev}},
  \bibinfo {author} {\bibfnamefont {G.}~\bibnamefont {Vidal}}, \bibinfo
  {author} {\bibfnamefont {C.~V.}\ \bibnamefont {Heidweiller}}, \bibinfo
  {author} {\bibfnamefont {S.}~\bibnamefont {Waltman}}, \bibinfo {author}
  {\bibfnamefont {J.}~\bibnamefont {Waltz}}, \bibinfo {author} {\bibfnamefont
  {S.~X.}\ \bibnamefont {Wang}}, \bibinfo {author} {\bibfnamefont
  {B.}~\bibnamefont {Ware}}, \bibinfo {author} {\bibfnamefont {T.}~\bibnamefont
  {Weidel}}, \bibinfo {author} {\bibfnamefont {T.}~\bibnamefont {White}},
  \bibinfo {author} {\bibfnamefont {K.}~\bibnamefont {Wong}}, \bibinfo {author}
  {\bibfnamefont {B.~W.~K.}\ \bibnamefont {Woo}}, \bibinfo {author}
  {\bibfnamefont {M.}~\bibnamefont {Woodson}}, \bibinfo {author} {\bibfnamefont
  {C.}~\bibnamefont {Xing}}, \bibinfo {author} {\bibfnamefont {Z.~J.}\
  \bibnamefont {Yao}}, \bibinfo {author} {\bibfnamefont {P.}~\bibnamefont
  {Yeh}}, \bibinfo {author} {\bibfnamefont {B.}~\bibnamefont {Ying}}, \bibinfo
  {author} {\bibfnamefont {J.}~\bibnamefont {Yoo}}, \bibinfo {author}
  {\bibfnamefont {N.}~\bibnamefont {Yosri}}, \bibinfo {author} {\bibfnamefont
  {G.}~\bibnamefont {Young}}, \bibinfo {author} {\bibfnamefont
  {A.}~\bibnamefont {Zalcman}}, \bibinfo {author} {\bibnamefont {Yaxing}},
  \bibinfo {author} {\bibnamefont {Zhang}}, \bibinfo {author} {\bibfnamefont
  {N.}~\bibnamefont {Zhu}}, \bibinfo {author} {\bibfnamefont {S.}~\bibnamefont
  {Boixo}}, \bibinfo {author} {\bibfnamefont {J.}~\bibnamefont {Kelly}},
  \bibinfo {author} {\bibfnamefont {V.}~\bibnamefont {Smelyanskiy}}, \bibinfo
  {author} {\bibfnamefont {H.}~\bibnamefont {Neven}}, \bibinfo {author}
  {\bibfnamefont {D.}~\bibnamefont {Bacon}}, \bibinfo {author} {\bibfnamefont
  {Z.}~\bibnamefont {Chen}}, \bibinfo {author} {\bibfnamefont {P.~V.}\
  \bibnamefont {Klimov}}, \bibinfo {author} {\bibfnamefont {P.}~\bibnamefont
  {Roushan}}, \bibinfo {author} {\bibfnamefont {C.}~\bibnamefont {Neill}},
  \bibinfo {author} {\bibfnamefont {Y.}~\bibnamefont {Chen}},\ and\ \bibinfo
  {author} {\bibfnamefont {A.}~\bibnamefont {Morvan}},\ }\bibfield  {title}
  {\bibinfo {title} {Demonstrating dynamic surface codes},\ }\href@noop {}
  {\bibfield  {journal} {\bibinfo  {journal} {arXiv preprint arXiv:2412.14360}\
  } (\bibinfo {year} {2024})}\BibitemShut {NoStop}%
\bibitem [{\citenamefont {Nguyen}\ \emph {et~al.}(2019)\citenamefont {Nguyen},
  \citenamefont {Lin}, \citenamefont {Somoroff}, \citenamefont {Mencia},
  \citenamefont {Grabon},\ and\ \citenamefont {Manucharyan}}]{nguyen2019high}%
  \BibitemOpen
  \bibfield  {author} {\bibinfo {author} {\bibfnamefont {L.~B.}\ \bibnamefont
  {Nguyen}}, \bibinfo {author} {\bibfnamefont {Y.-H.}\ \bibnamefont {Lin}},
  \bibinfo {author} {\bibfnamefont {A.}~\bibnamefont {Somoroff}}, \bibinfo
  {author} {\bibfnamefont {R.}~\bibnamefont {Mencia}}, \bibinfo {author}
  {\bibfnamefont {N.}~\bibnamefont {Grabon}},\ and\ \bibinfo {author}
  {\bibfnamefont {V.~E.}\ \bibnamefont {Manucharyan}},\ }\bibfield  {title}
  {\bibinfo {title} {High-coherence fluxonium qubit},\ }\href
  {https://doi.org/10.1103/PhysRevX.9.041041} {\bibfield  {journal} {\bibinfo
  {journal} {Phys. Rev. X}\ }\textbf {\bibinfo {volume} {9}},\ \bibinfo {pages}
  {041041} (\bibinfo {year} {2019})}\BibitemShut {NoStop}%
\bibitem [{\citenamefont {Bao}\ \emph {et~al.}(2022)\citenamefont {Bao},
  \citenamefont {Deng}, \citenamefont {Ding}, \citenamefont {Gao},
  \citenamefont {Gao}, \citenamefont {Huang}, \citenamefont {Jiang},
  \citenamefont {Ku}, \citenamefont {Li}, \citenamefont {Ma}, \citenamefont
  {Ni}, \citenamefont {Qin}, \citenamefont {Song}, \citenamefont {Sun},
  \citenamefont {Tang}, \citenamefont {Wang}, \citenamefont {Wu}, \citenamefont
  {Xia}, \citenamefont {Yu}, \citenamefont {Zhang}, \citenamefont {Zhang},
  \citenamefont {Zhang}, \citenamefont {Zhou}, \citenamefont {Zhu},
  \citenamefont {Shi}, \citenamefont {Chen}, \citenamefont {Zhao},\ and\
  \citenamefont {Deng}}]{bao_fluxonium_2022}%
  \BibitemOpen
  \bibfield  {author} {\bibinfo {author} {\bibfnamefont {F.}~\bibnamefont
  {Bao}}, \bibinfo {author} {\bibfnamefont {H.}~\bibnamefont {Deng}}, \bibinfo
  {author} {\bibfnamefont {D.}~\bibnamefont {Ding}}, \bibinfo {author}
  {\bibfnamefont {R.}~\bibnamefont {Gao}}, \bibinfo {author} {\bibfnamefont
  {X.}~\bibnamefont {Gao}}, \bibinfo {author} {\bibfnamefont {C.}~\bibnamefont
  {Huang}}, \bibinfo {author} {\bibfnamefont {X.}~\bibnamefont {Jiang}},
  \bibinfo {author} {\bibfnamefont {H.-S.}\ \bibnamefont {Ku}}, \bibinfo
  {author} {\bibfnamefont {Z.}~\bibnamefont {Li}}, \bibinfo {author}
  {\bibfnamefont {X.}~\bibnamefont {Ma}}, \bibinfo {author} {\bibfnamefont
  {X.}~\bibnamefont {Ni}}, \bibinfo {author} {\bibfnamefont {J.}~\bibnamefont
  {Qin}}, \bibinfo {author} {\bibfnamefont {Z.}~\bibnamefont {Song}}, \bibinfo
  {author} {\bibfnamefont {H.}~\bibnamefont {Sun}}, \bibinfo {author}
  {\bibfnamefont {C.}~\bibnamefont {Tang}}, \bibinfo {author} {\bibfnamefont
  {T.}~\bibnamefont {Wang}}, \bibinfo {author} {\bibfnamefont {F.}~\bibnamefont
  {Wu}}, \bibinfo {author} {\bibfnamefont {T.}~\bibnamefont {Xia}}, \bibinfo
  {author} {\bibfnamefont {W.}~\bibnamefont {Yu}}, \bibinfo {author}
  {\bibfnamefont {F.}~\bibnamefont {Zhang}}, \bibinfo {author} {\bibfnamefont
  {G.}~\bibnamefont {Zhang}}, \bibinfo {author} {\bibfnamefont
  {X.}~\bibnamefont {Zhang}}, \bibinfo {author} {\bibfnamefont
  {J.}~\bibnamefont {Zhou}}, \bibinfo {author} {\bibfnamefont {X.}~\bibnamefont
  {Zhu}}, \bibinfo {author} {\bibfnamefont {Y.}~\bibnamefont {Shi}}, \bibinfo
  {author} {\bibfnamefont {J.}~\bibnamefont {Chen}}, \bibinfo {author}
  {\bibfnamefont {H.-H.}\ \bibnamefont {Zhao}},\ and\ \bibinfo {author}
  {\bibfnamefont {C.}~\bibnamefont {Deng}},\ }\bibfield  {title} {\bibinfo
  {title} {Fluxonium: An alternative qubit platform for high-fidelity
  operations},\ }\href {https://doi.org/10.1103/PhysRevLett.129.010502}
  {\bibfield  {journal} {\bibinfo  {journal} {Phys. Rev. Lett.}\ }\textbf
  {\bibinfo {volume} {129}},\ \bibinfo {pages} {010502} (\bibinfo {year}
  {2022})}\BibitemShut {NoStop}%
\bibitem [{\citenamefont {Ding}\ \emph {et~al.}(2023)\citenamefont {Ding},
  \citenamefont {Hays}, \citenamefont {Sung}, \citenamefont {Kannan},
  \citenamefont {An}, \citenamefont {Di~Paolo}, \citenamefont {Karamlou},
  \citenamefont {Hazard}, \citenamefont {Azar}, \citenamefont {Kim},
  \citenamefont {Niedzielski}, \citenamefont {Melville}, \citenamefont
  {Schwartz}, \citenamefont {Yoder}, \citenamefont {Orlando}, \citenamefont
  {Gustavsson}, \citenamefont {Grover}, \citenamefont {Serniak},\ and\
  \citenamefont {Oliver}}]{ding2023high}%
  \BibitemOpen
  \bibfield  {author} {\bibinfo {author} {\bibfnamefont {L.}~\bibnamefont
  {Ding}}, \bibinfo {author} {\bibfnamefont {M.}~\bibnamefont {Hays}}, \bibinfo
  {author} {\bibfnamefont {Y.}~\bibnamefont {Sung}}, \bibinfo {author}
  {\bibfnamefont {B.}~\bibnamefont {Kannan}}, \bibinfo {author} {\bibfnamefont
  {J.}~\bibnamefont {An}}, \bibinfo {author} {\bibfnamefont {A.}~\bibnamefont
  {Di~Paolo}}, \bibinfo {author} {\bibfnamefont {A.~H.}\ \bibnamefont
  {Karamlou}}, \bibinfo {author} {\bibfnamefont {T.~M.}\ \bibnamefont
  {Hazard}}, \bibinfo {author} {\bibfnamefont {K.}~\bibnamefont {Azar}},
  \bibinfo {author} {\bibfnamefont {D.~K.}\ \bibnamefont {Kim}}, \bibinfo
  {author} {\bibfnamefont {B.~M.}\ \bibnamefont {Niedzielski}}, \bibinfo
  {author} {\bibfnamefont {A.}~\bibnamefont {Melville}}, \bibinfo {author}
  {\bibfnamefont {M.~E.}\ \bibnamefont {Schwartz}}, \bibinfo {author}
  {\bibfnamefont {J.~L.}\ \bibnamefont {Yoder}}, \bibinfo {author}
  {\bibfnamefont {T.~P.}\ \bibnamefont {Orlando}}, \bibinfo {author}
  {\bibfnamefont {S.}~\bibnamefont {Gustavsson}}, \bibinfo {author}
  {\bibfnamefont {J.~A.}\ \bibnamefont {Grover}}, \bibinfo {author}
  {\bibfnamefont {K.}~\bibnamefont {Serniak}},\ and\ \bibinfo {author}
  {\bibfnamefont {W.~D.}\ \bibnamefont {Oliver}},\ }\bibfield  {title}
  {\bibinfo {title} {High-fidelity, frequency-flexible two-qubit fluxonium
  gates with a transmon coupler},\ }\href
  {https://doi.org/10.1103/PhysRevX.13.031035} {\bibfield  {journal} {\bibinfo
  {journal} {Phys. Rev. X}\ }\textbf {\bibinfo {volume} {13}},\ \bibinfo
  {pages} {031035} (\bibinfo {year} {2023})}\BibitemShut {NoStop}%
\bibitem [{\citenamefont {Dijkema}\ \emph {et~al.}(2024)\citenamefont
  {Dijkema}, \citenamefont {Xue}, \citenamefont {Harvey-Collard}, \citenamefont
  {Rimbach-Russ}, \citenamefont {de~Snoo}, \citenamefont {Zheng}, \citenamefont
  {Sammak}, \citenamefont {Scappucci},\ and\ \citenamefont
  {Vandersypen}}]{dijkema2024cavity}%
  \BibitemOpen
  \bibfield  {author} {\bibinfo {author} {\bibfnamefont {J.}~\bibnamefont
  {Dijkema}}, \bibinfo {author} {\bibfnamefont {X.}~\bibnamefont {Xue}},
  \bibinfo {author} {\bibfnamefont {P.}~\bibnamefont {Harvey-Collard}},
  \bibinfo {author} {\bibfnamefont {M.}~\bibnamefont {Rimbach-Russ}}, \bibinfo
  {author} {\bibfnamefont {S.~L.}\ \bibnamefont {de~Snoo}}, \bibinfo {author}
  {\bibfnamefont {G.}~\bibnamefont {Zheng}}, \bibinfo {author} {\bibfnamefont
  {A.}~\bibnamefont {Sammak}}, \bibinfo {author} {\bibfnamefont
  {G.}~\bibnamefont {Scappucci}},\ and\ \bibinfo {author} {\bibfnamefont
  {L.~M.~K.}\ \bibnamefont {Vandersypen}},\ }\bibfield  {title} {\bibinfo
  {title} {Cavity-mediated iswap oscillations between distant spins},\ }\href
  {https://doi.org/10.1038/s41567-024-02694-8} {\bibfield  {journal} {\bibinfo
  {journal} {Nature Physics}\ } (\bibinfo {year} {2024})}\BibitemShut {NoStop}%
\bibitem [{\citenamefont {Bluvstein}\ \emph {et~al.}(2024)\citenamefont
  {Bluvstein}, \citenamefont {Evered}, \citenamefont {Geim}, \citenamefont
  {Li}, \citenamefont {Zhou}, \citenamefont {Manovitz}, \citenamefont {Ebadi},
  \citenamefont {Cain}, \citenamefont {Kalinowski}, \citenamefont {Hangleiter},
  \citenamefont {Bonilla~Ataides}, \citenamefont {Maskara}, \citenamefont
  {Cong}, \citenamefont {Gao}, \citenamefont {Sales~Rodriguez}, \citenamefont
  {Karolyshyn}, \citenamefont {Semeghini}, \citenamefont {Gullans},
  \citenamefont {Greiner}, \citenamefont {Vuleti{\'{c}}},\ and\ \citenamefont
  {Lukin}}]{bluvstein_logical_2024}%
  \BibitemOpen
  \bibfield  {author} {\bibinfo {author} {\bibfnamefont {D.}~\bibnamefont
  {Bluvstein}}, \bibinfo {author} {\bibfnamefont {S.~J.}\ \bibnamefont
  {Evered}}, \bibinfo {author} {\bibfnamefont {A.~A.}\ \bibnamefont {Geim}},
  \bibinfo {author} {\bibfnamefont {S.~H.}\ \bibnamefont {Li}}, \bibinfo
  {author} {\bibfnamefont {H.}~\bibnamefont {Zhou}}, \bibinfo {author}
  {\bibfnamefont {T.}~\bibnamefont {Manovitz}}, \bibinfo {author}
  {\bibfnamefont {S.}~\bibnamefont {Ebadi}}, \bibinfo {author} {\bibfnamefont
  {M.}~\bibnamefont {Cain}}, \bibinfo {author} {\bibfnamefont {M.}~\bibnamefont
  {Kalinowski}}, \bibinfo {author} {\bibfnamefont {D.}~\bibnamefont
  {Hangleiter}}, \bibinfo {author} {\bibfnamefont {J.~P.}\ \bibnamefont
  {Bonilla~Ataides}}, \bibinfo {author} {\bibfnamefont {N.}~\bibnamefont
  {Maskara}}, \bibinfo {author} {\bibfnamefont {I.}~\bibnamefont {Cong}},
  \bibinfo {author} {\bibfnamefont {X.}~\bibnamefont {Gao}}, \bibinfo {author}
  {\bibfnamefont {P.}~\bibnamefont {Sales~Rodriguez}}, \bibinfo {author}
  {\bibfnamefont {T.}~\bibnamefont {Karolyshyn}}, \bibinfo {author}
  {\bibfnamefont {G.}~\bibnamefont {Semeghini}}, \bibinfo {author}
  {\bibfnamefont {M.~J.}\ \bibnamefont {Gullans}}, \bibinfo {author}
  {\bibfnamefont {M.}~\bibnamefont {Greiner}}, \bibinfo {author} {\bibfnamefont
  {V.}~\bibnamefont {Vuleti{\'{c}}}},\ and\ \bibinfo {author} {\bibfnamefont
  {M.~D.}\ \bibnamefont {Lukin}},\ }\bibfield  {title} {\bibinfo {title}
  {Logical quantum processor based on reconfigurable atom arrays},\ }\href
  {https://doi.org/10.1038/s41586-023-06927-3} {\bibfield  {journal} {\bibinfo
  {journal} {Nature}\ }\textbf {\bibinfo {volume} {626}},\ \bibinfo {pages}
  {58} (\bibinfo {year} {2024})}\BibitemShut {NoStop}%
\bibitem [{\citenamefont {Picard}\ \emph {et~al.}(2024)\citenamefont {Picard},
  \citenamefont {Park}, \citenamefont {Patenotte}, \citenamefont
  {Gebretsadkan}, \citenamefont {Wellnitz}, \citenamefont {Rey},\ and\
  \citenamefont {Ni}}]{picard2024entanglement}%
  \BibitemOpen
  \bibfield  {author} {\bibinfo {author} {\bibfnamefont {L.~R.~B.}\
  \bibnamefont {Picard}}, \bibinfo {author} {\bibfnamefont {A.~J.}\
  \bibnamefont {Park}}, \bibinfo {author} {\bibfnamefont {G.~E.}\ \bibnamefont
  {Patenotte}}, \bibinfo {author} {\bibfnamefont {S.}~\bibnamefont
  {Gebretsadkan}}, \bibinfo {author} {\bibfnamefont {D.}~\bibnamefont
  {Wellnitz}}, \bibinfo {author} {\bibfnamefont {A.~M.}\ \bibnamefont {Rey}},\
  and\ \bibinfo {author} {\bibfnamefont {K.-K.}\ \bibnamefont {Ni}},\
  }\bibfield  {title} {\bibinfo {title} {Entanglement and iswap gate between
  molecular qubits},\ }\href {https://doi.org/10.1038/s41586-024-08177-3}
  {\bibfield  {journal} {\bibinfo  {journal} {Nature}\ } (\bibinfo {year}
  {2024})}\BibitemShut {NoStop}%
\bibitem [{\citenamefont {Monz}\ \emph {et~al.}(2009)\citenamefont {Monz},
  \citenamefont {Kim}, \citenamefont {H\"ansel}, \citenamefont {Riebe},
  \citenamefont {Villar}, \citenamefont {Schindler}, \citenamefont {Chwalla},
  \citenamefont {Hennrich},\ and\ \citenamefont
  {Blatt}}]{PhysRevLett.102.040501}%
  \BibitemOpen
  \bibfield  {author} {\bibinfo {author} {\bibfnamefont {T.}~\bibnamefont
  {Monz}}, \bibinfo {author} {\bibfnamefont {K.}~\bibnamefont {Kim}}, \bibinfo
  {author} {\bibfnamefont {W.}~\bibnamefont {H\"ansel}}, \bibinfo {author}
  {\bibfnamefont {M.}~\bibnamefont {Riebe}}, \bibinfo {author} {\bibfnamefont
  {A.~S.}\ \bibnamefont {Villar}}, \bibinfo {author} {\bibfnamefont
  {P.}~\bibnamefont {Schindler}}, \bibinfo {author} {\bibfnamefont
  {M.}~\bibnamefont {Chwalla}}, \bibinfo {author} {\bibfnamefont
  {M.}~\bibnamefont {Hennrich}},\ and\ \bibinfo {author} {\bibfnamefont
  {R.}~\bibnamefont {Blatt}},\ }\bibfield  {title} {\bibinfo {title}
  {Realization of the quantum toffoli gate with trapped ions},\ }\href
  {https://doi.org/10.1103/PhysRevLett.102.040501} {\bibfield  {journal}
  {\bibinfo  {journal} {Phys. Rev. Lett.}\ }\textbf {\bibinfo {volume} {102}},\
  \bibinfo {pages} {040501} (\bibinfo {year} {2009})}\BibitemShut {NoStop}%
\end{thebibliography}%


\begin{thebibliography}{14}%
\makeatletter
\providecommand \@ifxundefined [1]{%
 \@ifx{#1\undefined}
}%
\providecommand \@ifnum [1]{%
 \ifnum #1\expandafter \@firstoftwo
 \else \expandafter \@secondoftwo
 \fi
}%
\providecommand \@ifx [1]{%
 \ifx #1\expandafter \@firstoftwo
 \else \expandafter \@secondoftwo
 \fi
}%
\providecommand \natexlab [1]{#1}%
\providecommand \enquote  [1]{``#1''}%
\providecommand \bibnamefont  [1]{#1}%
\providecommand \bibfnamefont [1]{#1}%
\providecommand \citenamefont [1]{#1}%
\providecommand \href@noop [0]{\@secondoftwo}%
\providecommand \href [0]{\begingroup \@sanitize@url \@href}%
\providecommand \@href[1]{\@@startlink{#1}\@@href}%
\providecommand \@@href[1]{\endgroup#1\@@endlink}%
\providecommand \@sanitize@url [0]{\catcode `\\12\catcode `\$12\catcode
  `\&12\catcode `\#12\catcode `\^12\catcode `\_12\catcode `\%12\relax}%
\providecommand \@@startlink[1]{}%
\providecommand \@@endlink[0]{}%
\providecommand \url  [0]{\begingroup\@sanitize@url \@url }%
\providecommand \@url [1]{\endgroup\@href {#1}{\urlprefix }}%
\providecommand \urlprefix  [0]{URL }%
\providecommand \Eprint [0]{\href }%
\providecommand \doibase [0]{https://doi.org/}%
\providecommand \selectlanguage [0]{\@gobble}%
\providecommand \bibinfo  [0]{\@secondoftwo}%
\providecommand \bibfield  [0]{\@secondoftwo}%
\providecommand \translation [1]{[#1]}%
\providecommand \BibitemOpen [0]{}%
\providecommand \bibitemStop [0]{}%
\providecommand \bibitemNoStop [0]{.\EOS\space}%
\providecommand \EOS [0]{\spacefactor3000\relax}%
\providecommand \BibitemShut  [1]{\csname bibitem#1\endcsname}%
\let\auto@bib@innerbib\@empty
\bibitem [{\citenamefont {Tucci}(2005)}]{tucci2005introduction}%
  \BibitemOpen
  \bibfield  {author} {\bibinfo {author} {\bibfnamefont {R.~R.}\ \bibnamefont
  {Tucci}},\ }\bibfield  {title} {\bibinfo {title} {An introduction to cartan's
  kak decomposition for qc programmers},\ }\href@noop {} {\bibfield  {journal}
  {\bibinfo  {journal} {arXiv preprint quant-ph/0507171}\ } (\bibinfo {year}
  {2005})}\BibitemShut {NoStop}%
\bibitem [{\citenamefont {Chen}\ \emph {et~al.}(2024)\citenamefont {Chen},
  \citenamefont {Ding}, \citenamefont {Gong}, \citenamefont {Huang},\ and\
  \citenamefont {Ye}}]{chen2024one}%
  \BibitemOpen
  \bibfield  {author} {\bibinfo {author} {\bibfnamefont {J.}~\bibnamefont
  {Chen}}, \bibinfo {author} {\bibfnamefont {D.}~\bibnamefont {Ding}}, \bibinfo
  {author} {\bibfnamefont {W.}~\bibnamefont {Gong}}, \bibinfo {author}
  {\bibfnamefont {C.}~\bibnamefont {Huang}},\ and\ \bibinfo {author}
  {\bibfnamefont {Q.}~\bibnamefont {Ye}},\ }\bibfield  {title} {\bibinfo
  {title} {One gate scheme to rule them all: Introducing a complex yet reduced
  instruction set for quantum computing},\ }in\ \href
  {https://doi.org/10.1145/3620665.3640386} {\emph {\bibinfo {booktitle}
  {Proceedings of the 29th ACM International Conference on Architectural
  Support for Programming Languages and Operating Systems, Volume 2}}},\
  \bibinfo {series and number} {ASPLOS '24}\ (\bibinfo  {publisher}
  {Association for Computing Machinery},\ \bibinfo {address} {New York, NY,
  USA},\ \bibinfo {year} {2024})\ p.\ \bibinfo {pages} {779–796}\BibitemShut
  {NoStop}%
\bibitem [{\citenamefont {Zhang}\ \emph {et~al.}(2004)\citenamefont {Zhang},
  \citenamefont {Vala}, \citenamefont {Sastry},\ and\ \citenamefont
  {Whaley}}]{PhysRevLett.93.020502}%
  \BibitemOpen
  \bibfield  {author} {\bibinfo {author} {\bibfnamefont {J.}~\bibnamefont
  {Zhang}}, \bibinfo {author} {\bibfnamefont {J.}~\bibnamefont {Vala}},
  \bibinfo {author} {\bibfnamefont {S.}~\bibnamefont {Sastry}},\ and\ \bibinfo
  {author} {\bibfnamefont {K.~B.}\ \bibnamefont {Whaley}},\ }\bibfield  {title}
  {\bibinfo {title} {Minimum construction of two-qubit quantum operations},\
  }\href {https://doi.org/10.1103/PhysRevLett.93.020502} {\bibfield  {journal}
  {\bibinfo  {journal} {Phys. Rev. Lett.}\ }\textbf {\bibinfo {volume} {93}},\
  \bibinfo {pages} {020502} (\bibinfo {year} {2004})}\BibitemShut {NoStop}%
\bibitem [{\citenamefont {Sete}\ \emph {et~al.}(2021)\citenamefont {Sete},
  \citenamefont {Chen}, \citenamefont {Manenti}, \citenamefont {Kulshreshtha},\
  and\ \citenamefont {Poletto}}]{sete2021floating}%
  \BibitemOpen
  \bibfield  {author} {\bibinfo {author} {\bibfnamefont {E.~A.}\ \bibnamefont
  {Sete}}, \bibinfo {author} {\bibfnamefont {A.~Q.}\ \bibnamefont {Chen}},
  \bibinfo {author} {\bibfnamefont {R.}~\bibnamefont {Manenti}}, \bibinfo
  {author} {\bibfnamefont {S.}~\bibnamefont {Kulshreshtha}},\ and\ \bibinfo
  {author} {\bibfnamefont {S.}~\bibnamefont {Poletto}},\ }\bibfield  {title}
  {\bibinfo {title} {Floating tunable coupler for scalable quantum computing
  architectures},\ }\href {https://doi.org/10.1103/PhysRevApplied.15.064063}
  {\bibfield  {journal} {\bibinfo  {journal} {Phys. Rev. Appl.}\ }\textbf
  {\bibinfo {volume} {15}},\ \bibinfo {pages} {064063} (\bibinfo {year}
  {2021})}\BibitemShut {NoStop}%
\bibitem [{\citenamefont {Krantz}\ \emph {et~al.}(2019)\citenamefont {Krantz},
  \citenamefont {Kjaergaard}, \citenamefont {Yan}, \citenamefont {Orlando},
  \citenamefont {Gustavsson},\ and\ \citenamefont
  {Oliver}}]{krantz2019quantum}%
  \BibitemOpen
  \bibfield  {author} {\bibinfo {author} {\bibfnamefont {P.}~\bibnamefont
  {Krantz}}, \bibinfo {author} {\bibfnamefont {M.}~\bibnamefont {Kjaergaard}},
  \bibinfo {author} {\bibfnamefont {F.}~\bibnamefont {Yan}}, \bibinfo {author}
  {\bibfnamefont {T.~P.}\ \bibnamefont {Orlando}}, \bibinfo {author}
  {\bibfnamefont {S.}~\bibnamefont {Gustavsson}},\ and\ \bibinfo {author}
  {\bibfnamefont {W.~D.}\ \bibnamefont {Oliver}},\ }\bibfield  {title}
  {\bibinfo {title} {A quantum engineer's guide to superconducting qubits},\
  }\href {https://doi.org/10.1063/1.5089550} {\bibfield  {journal} {\bibinfo
  {journal} {Applied Physics Reviews}\ }\textbf {\bibinfo {volume} {6}},\
  \bibinfo {pages} {021318} (\bibinfo {year} {2019})}\BibitemShut {NoStop}%
\bibitem [{\citenamefont {Yan}\ \emph {et~al.}(2018)\citenamefont {Yan},
  \citenamefont {Krantz}, \citenamefont {Sung}, \citenamefont {Kjaergaard},
  \citenamefont {Campbell}, \citenamefont {Orlando}, \citenamefont
  {Gustavsson},\ and\ \citenamefont {Oliver}}]{yan2018tunable}%
  \BibitemOpen
  \bibfield  {author} {\bibinfo {author} {\bibfnamefont {F.}~\bibnamefont
  {Yan}}, \bibinfo {author} {\bibfnamefont {P.}~\bibnamefont {Krantz}},
  \bibinfo {author} {\bibfnamefont {Y.}~\bibnamefont {Sung}}, \bibinfo {author}
  {\bibfnamefont {M.}~\bibnamefont {Kjaergaard}}, \bibinfo {author}
  {\bibfnamefont {D.~L.}\ \bibnamefont {Campbell}}, \bibinfo {author}
  {\bibfnamefont {T.~P.}\ \bibnamefont {Orlando}}, \bibinfo {author}
  {\bibfnamefont {S.}~\bibnamefont {Gustavsson}},\ and\ \bibinfo {author}
  {\bibfnamefont {W.~D.}\ \bibnamefont {Oliver}},\ }\bibfield  {title}
  {\bibinfo {title} {Tunable coupling scheme for implementing high-fidelity
  two-qubit gates},\ }\href {https://doi.org/10.1103/PhysRevApplied.10.054062}
  {\bibfield  {journal} {\bibinfo  {journal} {Phys. Rev. Appl.}\ }\textbf
  {\bibinfo {volume} {10}},\ \bibinfo {pages} {054062} (\bibinfo {year}
  {2018})}\BibitemShut {NoStop}%
\bibitem [{\citenamefont {Wang}\ \emph {et~al.}(2022)\citenamefont {Wang},
  \citenamefont {Li}, \citenamefont {Xu}, \citenamefont {Li}, \citenamefont
  {Wang}, \citenamefont {Yang}, \citenamefont {Mi}, \citenamefont {Liang},
  \citenamefont {Su}, \citenamefont {Yang}, \citenamefont {Wang}, \citenamefont
  {Wang}, \citenamefont {Li}, \citenamefont {Chen}, \citenamefont {Li},
  \citenamefont {Linghu}, \citenamefont {Han}, \citenamefont {Zhang},
  \citenamefont {Feng}, \citenamefont {Song}, \citenamefont {Ma}, \citenamefont
  {Zhang}, \citenamefont {Wang}, \citenamefont {Zhao}, \citenamefont {Liu},
  \citenamefont {Xue}, \citenamefont {Jin},\ and\ \citenamefont
  {Yu}}]{wang2022towards}%
  \BibitemOpen
  \bibfield  {author} {\bibinfo {author} {\bibfnamefont {C.}~\bibnamefont
  {Wang}}, \bibinfo {author} {\bibfnamefont {X.}~\bibnamefont {Li}}, \bibinfo
  {author} {\bibfnamefont {H.}~\bibnamefont {Xu}}, \bibinfo {author}
  {\bibfnamefont {Z.}~\bibnamefont {Li}}, \bibinfo {author} {\bibfnamefont
  {J.}~\bibnamefont {Wang}}, \bibinfo {author} {\bibfnamefont {Z.}~\bibnamefont
  {Yang}}, \bibinfo {author} {\bibfnamefont {Z.}~\bibnamefont {Mi}}, \bibinfo
  {author} {\bibfnamefont {X.}~\bibnamefont {Liang}}, \bibinfo {author}
  {\bibfnamefont {T.}~\bibnamefont {Su}}, \bibinfo {author} {\bibfnamefont
  {C.}~\bibnamefont {Yang}}, \bibinfo {author} {\bibfnamefont {G.}~\bibnamefont
  {Wang}}, \bibinfo {author} {\bibfnamefont {W.}~\bibnamefont {Wang}}, \bibinfo
  {author} {\bibfnamefont {Y.}~\bibnamefont {Li}}, \bibinfo {author}
  {\bibfnamefont {M.}~\bibnamefont {Chen}}, \bibinfo {author} {\bibfnamefont
  {C.}~\bibnamefont {Li}}, \bibinfo {author} {\bibfnamefont {K.}~\bibnamefont
  {Linghu}}, \bibinfo {author} {\bibfnamefont {J.}~\bibnamefont {Han}},
  \bibinfo {author} {\bibfnamefont {Y.}~\bibnamefont {Zhang}}, \bibinfo
  {author} {\bibfnamefont {Y.}~\bibnamefont {Feng}}, \bibinfo {author}
  {\bibfnamefont {Y.}~\bibnamefont {Song}}, \bibinfo {author} {\bibfnamefont
  {T.}~\bibnamefont {Ma}}, \bibinfo {author} {\bibfnamefont {J.}~\bibnamefont
  {Zhang}}, \bibinfo {author} {\bibfnamefont {R.}~\bibnamefont {Wang}},
  \bibinfo {author} {\bibfnamefont {P.}~\bibnamefont {Zhao}}, \bibinfo {author}
  {\bibfnamefont {W.}~\bibnamefont {Liu}}, \bibinfo {author} {\bibfnamefont
  {G.}~\bibnamefont {Xue}}, \bibinfo {author} {\bibfnamefont {Y.}~\bibnamefont
  {Jin}},\ and\ \bibinfo {author} {\bibfnamefont {H.}~\bibnamefont {Yu}},\
  }\bibfield  {title} {\bibinfo {title} {Towards practical quantum computers:
  transmon qubit with a lifetime approaching 0.5 milliseconds},\ }\href
  {https://doi.org/10.1038/s41534-021-00510-2} {\bibfield  {journal} {\bibinfo
  {journal} {npj Quantum Information}\ }\textbf {\bibinfo {volume} {8}},\
  \bibinfo {pages} {3} (\bibinfo {year} {2022})}\BibitemShut {NoStop}%
\bibitem [{\citenamefont {Li}\ \emph {et~al.}(2024)\citenamefont {Li},
  \citenamefont {Xu}, \citenamefont {Wang}, \citenamefont {Tang}, \citenamefont
  {Zhang}, \citenamefont {Yang}, \citenamefont {Su}, \citenamefont {Wang},
  \citenamefont {Mi}, \citenamefont {Sun}, \citenamefont {Liang}, \citenamefont
  {Chen}, \citenamefont {Li}, \citenamefont {Zhang}, \citenamefont {Linghu},
  \citenamefont {Han}, \citenamefont {Liu}, \citenamefont {Feng}, \citenamefont
  {Liu}, \citenamefont {Xue}, \citenamefont {Zhang}, \citenamefont {Jin},
  \citenamefont {Zhu}, \citenamefont {Yu}, \citenamefont {Zhao},\ and\
  \citenamefont {Xue}}]{PhysRevResearch.6.L042038}%
  \BibitemOpen
  \bibfield  {author} {\bibinfo {author} {\bibfnamefont {X.}~\bibnamefont
  {Li}}, \bibinfo {author} {\bibfnamefont {H.}~\bibnamefont {Xu}}, \bibinfo
  {author} {\bibfnamefont {J.}~\bibnamefont {Wang}}, \bibinfo {author}
  {\bibfnamefont {L.-Z.}\ \bibnamefont {Tang}}, \bibinfo {author}
  {\bibfnamefont {D.-W.}\ \bibnamefont {Zhang}}, \bibinfo {author}
  {\bibfnamefont {C.}~\bibnamefont {Yang}}, \bibinfo {author} {\bibfnamefont
  {T.}~\bibnamefont {Su}}, \bibinfo {author} {\bibfnamefont {C.}~\bibnamefont
  {Wang}}, \bibinfo {author} {\bibfnamefont {Z.}~\bibnamefont {Mi}}, \bibinfo
  {author} {\bibfnamefont {W.}~\bibnamefont {Sun}}, \bibinfo {author}
  {\bibfnamefont {X.}~\bibnamefont {Liang}}, \bibinfo {author} {\bibfnamefont
  {M.}~\bibnamefont {Chen}}, \bibinfo {author} {\bibfnamefont {C.}~\bibnamefont
  {Li}}, \bibinfo {author} {\bibfnamefont {Y.}~\bibnamefont {Zhang}}, \bibinfo
  {author} {\bibfnamefont {K.}~\bibnamefont {Linghu}}, \bibinfo {author}
  {\bibfnamefont {J.}~\bibnamefont {Han}}, \bibinfo {author} {\bibfnamefont
  {W.}~\bibnamefont {Liu}}, \bibinfo {author} {\bibfnamefont {Y.}~\bibnamefont
  {Feng}}, \bibinfo {author} {\bibfnamefont {P.}~\bibnamefont {Liu}}, \bibinfo
  {author} {\bibfnamefont {G.}~\bibnamefont {Xue}}, \bibinfo {author}
  {\bibfnamefont {J.}~\bibnamefont {Zhang}}, \bibinfo {author} {\bibfnamefont
  {Y.}~\bibnamefont {Jin}}, \bibinfo {author} {\bibfnamefont {S.-L.}\
  \bibnamefont {Zhu}}, \bibinfo {author} {\bibfnamefont {H.}~\bibnamefont
  {Yu}}, \bibinfo {author} {\bibfnamefont {S.~P.}\ \bibnamefont {Zhao}},\ and\
  \bibinfo {author} {\bibfnamefont {Q.-K.}\ \bibnamefont {Xue}},\ }\bibfield
  {title} {\bibinfo {title} {Mapping the topology-localization phase diagram
  with quasiperiodic disorder using a programmable superconducting simulator},\
  }\href {https://doi.org/10.1103/PhysRevResearch.6.L042038} {\bibfield
  {journal} {\bibinfo  {journal} {Phys. Rev. Res.}\ }\textbf {\bibinfo {volume}
  {6}},\ \bibinfo {pages} {L042038} (\bibinfo {year} {2024})}\BibitemShut
  {NoStop}%
\bibitem [{\citenamefont {Sung}\ \emph {et~al.}(2021)\citenamefont {Sung},
  \citenamefont {Ding}, \citenamefont {Braum\"uller}, \citenamefont
  {Veps\"al\"ainen}, \citenamefont {Kannan}, \citenamefont {Kjaergaard},
  \citenamefont {Greene}, \citenamefont {Samach}, \citenamefont {McNally},
  \citenamefont {Kim}, \citenamefont {Melville}, \citenamefont {Niedzielski},
  \citenamefont {Schwartz}, \citenamefont {Yoder}, \citenamefont {Orlando},
  \citenamefont {Gustavsson},\ and\ \citenamefont
  {Oliver}}]{sung2021realization}%
  \BibitemOpen
  \bibfield  {author} {\bibinfo {author} {\bibfnamefont {Y.}~\bibnamefont
  {Sung}}, \bibinfo {author} {\bibfnamefont {L.}~\bibnamefont {Ding}}, \bibinfo
  {author} {\bibfnamefont {J.}~\bibnamefont {Braum\"uller}}, \bibinfo {author}
  {\bibfnamefont {A.}~\bibnamefont {Veps\"al\"ainen}}, \bibinfo {author}
  {\bibfnamefont {B.}~\bibnamefont {Kannan}}, \bibinfo {author} {\bibfnamefont
  {M.}~\bibnamefont {Kjaergaard}}, \bibinfo {author} {\bibfnamefont
  {A.}~\bibnamefont {Greene}}, \bibinfo {author} {\bibfnamefont {G.~O.}\
  \bibnamefont {Samach}}, \bibinfo {author} {\bibfnamefont {C.}~\bibnamefont
  {McNally}}, \bibinfo {author} {\bibfnamefont {D.}~\bibnamefont {Kim}},
  \bibinfo {author} {\bibfnamefont {A.}~\bibnamefont {Melville}}, \bibinfo
  {author} {\bibfnamefont {B.~M.}\ \bibnamefont {Niedzielski}}, \bibinfo
  {author} {\bibfnamefont {M.~E.}\ \bibnamefont {Schwartz}}, \bibinfo {author}
  {\bibfnamefont {J.~L.}\ \bibnamefont {Yoder}}, \bibinfo {author}
  {\bibfnamefont {T.~P.}\ \bibnamefont {Orlando}}, \bibinfo {author}
  {\bibfnamefont {S.}~\bibnamefont {Gustavsson}},\ and\ \bibinfo {author}
  {\bibfnamefont {W.~D.}\ \bibnamefont {Oliver}},\ }\bibfield  {title}
  {\bibinfo {title} {Realization of high-fidelity cz and $zz$-free iswap gates
  with a tunable coupler},\ }\href {https://doi.org/10.1103/PhysRevX.11.021058}
  {\bibfield  {journal} {\bibinfo  {journal} {Phys. Rev. X}\ }\textbf {\bibinfo
  {volume} {11}},\ \bibinfo {pages} {021058} (\bibinfo {year}
  {2021})}\BibitemShut {NoStop}%
\bibitem [{\citenamefont {Song}\ \emph {et~al.}(2017)\citenamefont {Song},
  \citenamefont {Xu}, \citenamefont {Liu}, \citenamefont {Yang}, \citenamefont
  {Zheng}, \citenamefont {Deng}, \citenamefont {Xie}, \citenamefont {Huang},
  \citenamefont {Guo}, \citenamefont {Zhang}, \citenamefont {Zhang},
  \citenamefont {Xu}, \citenamefont {Zheng}, \citenamefont {Zhu}, \citenamefont
  {Wang}, \citenamefont {Chen}, \citenamefont {Lu}, \citenamefont {Han},\ and\
  \citenamefont {Pan}}]{song201710}%
  \BibitemOpen
  \bibfield  {author} {\bibinfo {author} {\bibfnamefont {C.}~\bibnamefont
  {Song}}, \bibinfo {author} {\bibfnamefont {K.}~\bibnamefont {Xu}}, \bibinfo
  {author} {\bibfnamefont {W.}~\bibnamefont {Liu}}, \bibinfo {author}
  {\bibfnamefont {C.-p.}\ \bibnamefont {Yang}}, \bibinfo {author}
  {\bibfnamefont {S.-B.}\ \bibnamefont {Zheng}}, \bibinfo {author}
  {\bibfnamefont {H.}~\bibnamefont {Deng}}, \bibinfo {author} {\bibfnamefont
  {Q.}~\bibnamefont {Xie}}, \bibinfo {author} {\bibfnamefont {K.}~\bibnamefont
  {Huang}}, \bibinfo {author} {\bibfnamefont {Q.}~\bibnamefont {Guo}}, \bibinfo
  {author} {\bibfnamefont {L.}~\bibnamefont {Zhang}}, \bibinfo {author}
  {\bibfnamefont {P.}~\bibnamefont {Zhang}}, \bibinfo {author} {\bibfnamefont
  {D.}~\bibnamefont {Xu}}, \bibinfo {author} {\bibfnamefont {D.}~\bibnamefont
  {Zheng}}, \bibinfo {author} {\bibfnamefont {X.}~\bibnamefont {Zhu}}, \bibinfo
  {author} {\bibfnamefont {H.}~\bibnamefont {Wang}}, \bibinfo {author}
  {\bibfnamefont {Y.-A.}\ \bibnamefont {Chen}}, \bibinfo {author}
  {\bibfnamefont {C.-Y.}\ \bibnamefont {Lu}}, \bibinfo {author} {\bibfnamefont
  {S.}~\bibnamefont {Han}},\ and\ \bibinfo {author} {\bibfnamefont {J.-W.}\
  \bibnamefont {Pan}},\ }\bibfield  {title} {\bibinfo {title} {10-qubit
  entanglement and parallel logic operations with a superconducting circuit},\
  }\href {https://doi.org/10.1103/PhysRevLett.119.180511} {\bibfield  {journal}
  {\bibinfo  {journal} {Phys. Rev. Lett.}\ }\textbf {\bibinfo {volume} {119}},\
  \bibinfo {pages} {180511} (\bibinfo {year} {2017})}\BibitemShut {NoStop}%
\bibitem [{\citenamefont {Jozsa}(1994)}]{jozsa1994fidelity}%
  \BibitemOpen
  \bibfield  {author} {\bibinfo {author} {\bibfnamefont {R.}~\bibnamefont
  {Jozsa}},\ }\bibfield  {title} {\bibinfo {title} {Fidelity for mixed quantum
  states},\ }\href {https://doi.org/10.1080/09500349414552171} {\bibfield
  {journal} {\bibinfo  {journal} {Journal of Modern Optics}\ }\textbf {\bibinfo
  {volume} {41}},\ \bibinfo {pages} {2315} (\bibinfo {year}
  {1994})}\BibitemShut {NoStop}%
\bibitem [{\citenamefont {Vandersypen}\ and\ \citenamefont
  {Chuang}(2005)}]{vandersypen2004nmr}%
  \BibitemOpen
  \bibfield  {author} {\bibinfo {author} {\bibfnamefont {L.~M.~K.}\
  \bibnamefont {Vandersypen}}\ and\ \bibinfo {author} {\bibfnamefont {I.~L.}\
  \bibnamefont {Chuang}},\ }\bibfield  {title} {\bibinfo {title} {Nmr
  techniques for quantum control and computation},\ }\href
  {https://doi.org/10.1103/RevModPhys.76.1037} {\bibfield  {journal} {\bibinfo
  {journal} {Rev. Mod. Phys.}\ }\textbf {\bibinfo {volume} {76}},\ \bibinfo
  {pages} {1037} (\bibinfo {year} {2005})}\BibitemShut {NoStop}%
\bibitem [{\citenamefont {Sastry}(2013)}]{sastry2013nonlinear}%
  \BibitemOpen
  \bibfield  {author} {\bibinfo {author} {\bibfnamefont {S.}~\bibnamefont
  {Sastry}},\ }\href@noop {} {\emph {\bibinfo {title} {Nonlinear systems:
  analysis, stability, and control}}},\ Vol.~\bibinfo {volume} {10}\ (\bibinfo
  {publisher} {Springer Science \& Business Media},\ \bibinfo {year}
  {2013})\BibitemShut {NoStop}%
\bibitem [{\citenamefont {Hammerer}\ \emph {et~al.}(2002)\citenamefont
  {Hammerer}, \citenamefont {Vidal},\ and\ \citenamefont
  {Cirac}}]{hammerer2002characterization}%
  \BibitemOpen
  \bibfield  {author} {\bibinfo {author} {\bibfnamefont {K.}~\bibnamefont
  {Hammerer}}, \bibinfo {author} {\bibfnamefont {G.}~\bibnamefont {Vidal}},\
  and\ \bibinfo {author} {\bibfnamefont {J.~I.}\ \bibnamefont {Cirac}},\
  }\bibfield  {title} {\bibinfo {title} {Characterization of nonlocal gates},\
  }\href {https://doi.org/10.1103/PhysRevA.66.062321} {\bibfield  {journal}
  {\bibinfo  {journal} {Phys. Rev. A}\ }\textbf {\bibinfo {volume} {66}},\
  \bibinfo {pages} {062321} (\bibinfo {year} {2002})}\BibitemShut {NoStop}%
\end{thebibliography}%

\end{document}



\title{Supplementary Materials for:\\Efficient Implementation of Arbitrary Two-Qubit Gates via Unified Control}

\newcommand{\BAQIS}{\affiliation{1}{Beijing Academy of Quantum Information Sciences, Beijing 100193, China}}
\newcommand{\IOP}{\affiliation{2}{Institute of Physics, Chinese Academy of Science, Beijing 100190, China}}
\newcommand{\UCAS}{\affiliation{3}{University of Chinese Academy of Sciences, Beijing 101408, China}}
\newcommand{\THUCS}{\affiliation{4}{Department of Computer Science and Technology, Tsinghua University, Beijing 100084, China}}

\author{Zhen Chen}
\thanks{These authors have contributed equally to this work.}
\affiliation{\BAQIS}

\author{Weiyang Liu}
\thanks{These authors have contributed equally to this work.}
\affiliation{\BAQIS}

\author{Yanjun Ma}
\affiliation{\BAQIS}
\author{Weijie Sun}
\affiliation{\BAQIS}
\author{Ruixia Wang}
\affiliation{\BAQIS}
\author{He Wang}
\affiliation{\BAQIS}
\author{Huikai Xu}
\affiliation{\BAQIS}
\author{Guangming Xue}
\affiliation{\BAQIS}
\author{Haisheng Yan}
\affiliation{\BAQIS}
\author{Zhen Yang}
\affiliation{\BAQIS}

\author{Jiayu Ding}
\affiliation{\BAQIS}
\author{Yang Gao}
\affiliation{\BAQIS}
\affiliation{\IOP}
\affiliation{\UCAS}
\author{Feiyu Li}
\affiliation{\BAQIS}
\affiliation{\IOP}
\affiliation{\UCAS}
\author{Yujia Zhang}
\affiliation{\BAQIS}
\affiliation{\IOP}
\affiliation{\UCAS}
\author{Zikang Zhang}
\affiliation{\THUCS}
\author{Yirong Jin}
\affiliation{\BAQIS}
\author{Haifeng Yu}
\affiliation{\BAQIS}

\author{Jianxin Chen}
\email{jianxinchen@acm.org}
\affiliation{\THUCS}

\author{Fei Yan}
\email{yanfei@baqis.ac.cn}
\affiliation{\BAQIS}


\maketitle


\section{\label{sec:Cartan}Cartan decomposition and the Weyl chamber}
Since an arbitrary two-qubit operation $\textit{U}$ $\in$ U(4) can be expressed as the product of an operation in $\SU{4}$ and a global phase shift which has no physical meaning in the context here, we focus on the study of the SU(4) group throughout this work. 
The SU(4) group can be broken down into local and non-local components, with local components belonging to SU(2) $\otimes$ SU(2) and representing single-qubit operations, and non-local components representing entangling operations that are essential to two-qubit systems.

A useful tool for analyzing the two-qubit unitary is a special case of the Cartan decomposition, also known as the KAK decomposition~\cite{tucci2005introduction}, which says that, for an arbitrary \textit{U} $\in$ SU(4), there exists a set of three parameters $a,b,c \in \mathbb{R}$, a set of single qubit gates $K_1, K_2, K_3, K_4$ $\in$ SU(2) and a global phase $e^{i\theta}$, such that,
\begin{equation}
 U = e^{i\theta} \cdot (K_1\otimes K_2)\cdot \mathrm{exp}[i(a \cdot XX+b \cdot YY+c \cdot ZZ)]\cdot(K_3\otimes K_4),
\label{KAK0}
\end{equation}
where $XX = \sigma_x \otimes \sigma_x$, $YY = \sigma_y \otimes \sigma_y$, $ZZ = \sigma_z \otimes \sigma_z$, and $\sigma_{x}, \sigma_{y}, \sigma_{z}$ are Pauli operators, 
\begin{eqnarray}
 \sigma_{x} = \left( \begin{array}{cc}
 0&\ 1\\
 1&\ 0
 \end{array}
 \right),
  \sigma_{y} = \left( \begin{array}{cc}
 0&-i\\
 i&0
 \end{array}
 \right), 
 \sigma_{z} = \left( \begin{array}{cc}
 1&0\\
 0&-1
 \end{array}
 \right).
\label{pauli}
\end{eqnarray}

\begin{figure}[hbt]
\includegraphics[scale=1]{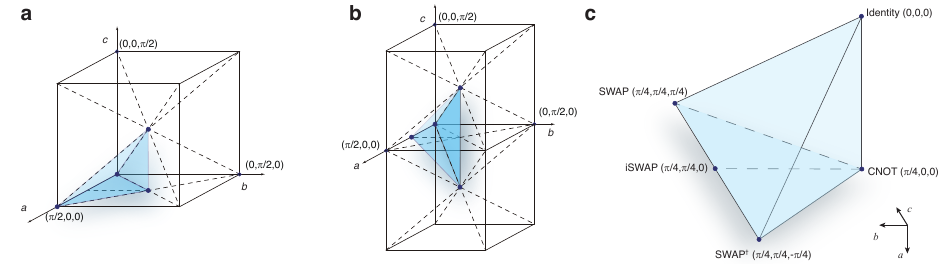}
\caption{ \textbf{The Weyl chamber}. \textbf{a}, The Weyl chamber is a tetrahedron spanned by vertices located at $(0,0,0)$, $(\pi/4,\pi/4,\pi/4)$, $(\pi/2,0,0)$ and $(\pi/4,\pi/4,0)$, which occupies one twenty-fourth of a cube.  \textbf{b}, The Weyl chamber used in this work spanned by vertices located at $(0,0,0)$, $(\pi/4,\pi/4,\pi/4)$, $(\pi/4,0,0)$ and $(\pi/4,\pi/4,-\pi/4)$. \textbf{c}, A rotated display of the same Weyl chamber as in \textbf{b}.}
\label{figWeyl}
\end{figure}

To visualize the geometric structure of the real vector $(a,b,c)$ representing the non-local component in Eq.~\ref{KAK0}, we consider a cube with side length of $\pi/2$, as indicated in Fig.~\ref{figWeyl}a. As per equivalent rules such as $(a,b,c)\sim (a\pm \frac{\pi}{2}, b, c), (-a, -b, c), (b, a, c)$ or their symmetric variants,
the cube in Fig.~\ref{figWeyl}a can be reduced to a tetrahedron, which is called the Weyl chamber, as shown by the blue tetrahedron in Fig.~\ref{figWeyl}a. 

There are many ways to define the Weyl chamber, in this work, we employ the following region as the Weyl chamber, as shown in Fig.~\ref{figWeyl}b and c,
\begin{equation}
\label{kak}
\pi/4 \geq a \geq b \geq |c| \ \mathrm{and}\  c \geq 0 \ \mathrm{if} \ a = \pi/4.
\end{equation}

For arbitrary two-qubit gates ${U}_{1}$, ${U}_{2}$, they are called locally equivalent if they satisfy ${U}_{1} = (K_1\otimes K_2)U_{2}(K_3\otimes K_4)$, for some $K_1, K_2, K_3, K_4$ $\in$ SU(2). Accordingly, every point in the Weyl chamber represents a unique local equivalence class of two-qubit gates (except for the Identity point and the $\mathrm{SWAP^{\dagger}}$ - $\iSWAP$ - $\CNOT$ plane, because the latter is equivalent to the $\SWAP$ - $\iSWAP$ - $\CNOT$ plane). For example, the $\CZ$ gate and the $\CNOT$ gate are locally equivalent and both are located at the point ($\pi/4$,0,0) in Fig.~\ref{figWeyl}b and c. 

\section{\label{sec:AshN}The AshN scheme}

\begin{figure}[hbt]
\includegraphics[scale=1]{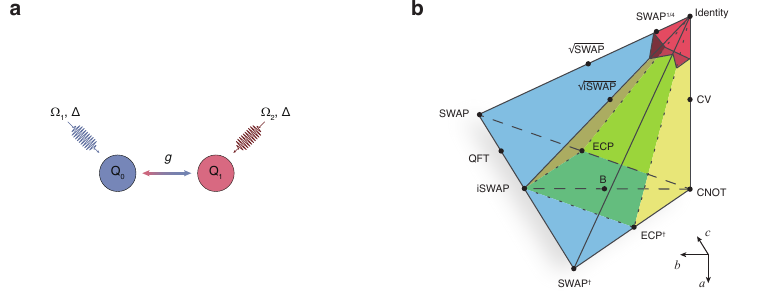}
\caption{ \textbf{The AshN scheme}. \textbf{a}, The schematic of the AshN implementation.  \textbf{b}, An arbitrary two-qubit gate can be implemented via one of the AshN-ND (yellow), AshN-EA (blue), and AshN-ND-EXT (red) algorithms.}
\label{figashn}
\end{figure}

The AshN scheme offers a useful protocol for generating arbitrary two-qubit gates natively~\cite{chen2024one}. It promises up to a $3\times$ reduction in circuit depth for certain quantum algorithms, while maintaining accuracy comparable to traditional gates. The AshN scheme can be conveniently implemented in the tunable transmon-qubit architecture. As shown in Fig.~\ref{figashn}a, consider a pair of qubits with coupling strength $g$ and each driven by a pulse with the same frequency $\omega_d$. The Hamiltonian of the system can be expressed as (set $\hbar \equiv 1$):
\begin{eqnarray}
H  = &&\ \frac{\omega_{1}}{2}Z I +\frac{\omega_{2}}{2}IZ+gYY\nonumber\\
&&+A_1 \mathrm{sin}(\omega_{d}t+\phi_{1})YI + A_2 \mathrm{sin}(\omega_{d}t+\phi_{2})IY, 
\label{ashn}
\end{eqnarray}
where $I$ is the Pauli identity and $\omega_{1}$ ($\omega_{2}$) is the frequency of the first (second) qubit. Without loss of generality, we treat the pulses to be sinusoidal with a square envelope with amplitudes $A_{1} = \Omega_{1}$ and $A_{2} = \Omega_{2}$ and duration $\tau$. 

In the rotating frame with the driving and under the rotating wave approximation, Eq.~\ref{ashn} can be transformed to:
\begin{eqnarray}
H'  = && \ \frac{\Delta}{2}(ZI+IZ)+\frac{g}{2}(XX+YY) \nonumber \\ 
&&+\frac{\Omega_{1}}{2}(\mathrm{cos}\phi_{1}X I+\mathrm{sin}\phi_{1}YI) +\frac{\Omega_{2}}{2}(\mathrm{cos}\phi_{2}IX +\mathrm{sin}\phi_{2}IY).
\label{rotate}
\end{eqnarray}
Here we bring the two qubits into resonance ($\omega_{1} = \omega_{2} = \omega$), and $\Delta = \omega-\omega_{d}$ is the frequency detuning between the qubit and the driving.  Setting $\phi_1 = \phi_2 = 0$, then the Hamiltonian can be expressed in terms of $g, \Omega_{1}, \Omega_{2}, \Delta$,
\begin{eqnarray}
H'  = && \ \frac{\Delta}{2}(ZI+IZ)+\frac{g}{2}(XX+YY) \nonumber \\ 
&&+\frac{\Omega_{1}}{2}XI
+ \frac{\Omega_{2}}{2} IX.
\label{rotate1}
\end{eqnarray}
Therefore the two-qubit gate can be written as,
\begin{eqnarray}
U  = \mathrm{exp}(-i\cdot H' \cdot \tau).
\label{unitary}
\end{eqnarray}

It is proven that by choosing proper parameters $ g, \Omega_{1}, \Omega_{2}, \Delta, \tau$, Eq.~\ref{unitary} can be employed to realize two-qubit gates that correspond to any Weyl chamber coordinates $(a,b,c)$ with certain single-qubit gate corrections. The AshN scheme can be further divided into three algorithms: AshN-ND (no tuning, i.e. $\Delta = 0$, yellow region in Fig.~\ref{figashn}b); AshN-EA (equal amplitude, i.e. $\Omega_1 =  \Omega_2$, blue region in Fig.~\ref{figashn}b); AshN-ND-EXT (no tuning with extend gate time, red region in Fig.~\ref{figashn}b).

Given the Hamiltonian Eq.~\ref{rotate1} that describes our quantum device, we include the details of the AshN scheme from~\cite{chen2024one} here for reference, with necessary notational revisions to ensure compatibility with this paper. More specifically, the AshN scheme is given explicitly in Alg.~\ref{alg:ashn} in the last section of the Supplementary Materials, which gives the gate time $\tau$, amplitudes $\Omega_1, \Omega_2$, and detuning $\Delta$ given the desired Weyl chamber coordinates $(x,y,z)$. Its subschemes, AshN-ND, AshN-ND-EXT and AshN-EA are described in Alg.~\ref{alg:ashna0}, Alg.~\ref{alg:ashna1} and Alg.~\ref{alg:ashnb} (and its symmetric counterpart Alg.~\ref{alg:ashnc}), respectively. For more details, particularly regarding the theoretical aspects of the AshN protocol, please refer to~\cite{chen2024one}. Please note that the control parameters $\Omega_1'$, $\Omega_2'$, and $\delta'$ used in~\cite{chen2024one} correspond to $\frac{\Omega_1 + \Omega_2}{4}$, $\frac{\Omega_1 - \Omega_2}{4}$, and $\frac{\Delta}{2}$, respectively, where $\Omega_1$, $\Omega_2$, and $\Delta$ are the parameters used in this paper.

\section{\label{sec:Bgate} The B gate}
The B gate is particularly notable for its high symmetry, positioned at the center of the region of gates capable of generating the maximum amount of entanglement, as well as its greater efficiency than conventional two-qubit gates such as CNOT and iSWAP. It is particularly efficient for implementing any arbitrary two-qubit quantum operation with just two applications of the B gate, complemented by certain single-qubit gates, as illustrated in Fig. S3a. Within the Weyl chamber coordinates, the B gate can be expressed as:
\begin{align}
U_{\mathrm{B}} =& \  \mathrm{exp}[i(\frac{\pi}{4}XX +  \frac{\pi}{8}YY)]  \nonumber\\
=& \ \left(
\begin{matrix}
0.924 & 0 & 0 & 0.383i \\
0 & 0.383 & 0.924i & 0 \\
0 & 0.924i & 0.383 & 0 \\
0.383i & 0 & 0 & 0.924 \\   
\end{matrix}
\right) .
\label{Ub}
\end{align}

\begin{figure}[hbt]
\includegraphics[scale=2]{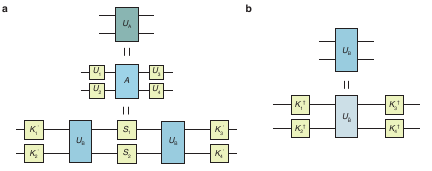}
\caption{\textbf{The B gate}. \textbf{a}, Circuit to generate arbitrary two-qubit gate $U_{\mathrm{A}}$ using two applications of the B gate. \textbf{b}, The AshN generated B gate $U_{\tilde{\mathrm{B}}}$ is locally equivalent to the B gate $U_{\mathrm{B}}$. With certain single-qubit corrections, $U_{\tilde{\mathrm{B}}}$ can be transformed to $U_{\mathrm{B}}$.}
\label{figsb}
\end{figure}

The AshN-ND protocol can generate a gate that is locally equivalent to the B gate, as indicated in Fig.~\ref{figsb}b which is: 
\begin{eqnarray}
H_{\tilde{\mathrm{B}}} =\  &&\frac{g}{2}(XX +YY) + 1.119gXI,\nonumber\\
U_{\tilde{\mathrm{B}}} =\  &&\mathrm{exp}(-iH_{\tilde{\mathrm{B}}}\frac{\pi}{2g})  \nonumber\\
= \ && \left(
\begin{matrix}
0.025 & -0.605 & -0.605i & 0.516i \\
-0.605 & -0.516 & -0.025i & -0.605i \\
-0.605i & -0.025i & -0.516 & -0.605 \\
0.516i & -0.605i & -0.605 & 0.025 \\
\end{matrix}
\right).
\label{BAshN}
\end{eqnarray}
Via the KAK decomposition, $U_{\tilde{\mathrm{B}}}$ can transfer to $U_{\mathrm{\mathrm{B}}}$,
\begin{eqnarray}
U_{\tilde{\mathrm{B}}} = \ && e^{i\theta} (K_{1}\otimes K_{2})U_{\mathrm{B}}(K_{3}\otimes K_{4}) \nonumber \\
= \ && \left(\left( \begin{matrix}
0.829i&0.559\\
-0.559&-0.829i
\end{matrix}
\right) \otimes
\left(\begin{matrix}
0&-i\\
-i&0
\end{matrix}
\right )\right)
U_{\mathrm{B}}
\left(\left( \begin{matrix}
-0.829i&0.559\\
-0.559&0.829i
\end{matrix}
\right ) \otimes
\left( \begin{matrix}
0&-i\\
-i&0
\end{matrix}
\right ) \right), \\ \nonumber
\label{B}
\end{eqnarray} 
We can then obtain $U_{\mathrm{B}}$ using:
\begin{eqnarray}
U_{\mathrm{B}} = \ && e^{-i\theta} (K_{1}^\dagger\otimes K_{2}^\dagger)U_{\tilde{\mathrm{B}}}(K_{3}^\dagger\otimes K_{4}^\dagger).  
\label{B2}
\end{eqnarray}

To implement an arbitrary two-qubit gate $U_{\mathrm{A}}$ by the B gate, we can utilize two applications of the B gate and certain single-qubit gates. As shown in Fig.~\ref{figsb}a, the single-qubit gates $S_1, S_2$ are~\cite{PhysRevLett.93.020502}:
\begin{eqnarray}
&&S_{1} =
\begin{cases} 
\mathrm{exp}(i a \sigma_{y}), & \text{if } c \geq 0, \\
\mathrm{exp}[i (\pi/2 - a) \sigma_{y}], & \text{otherwise.}
\end{cases} \nonumber\\
&&S_{2} = \mathrm{exp}(i\beta_2 \sigma_{z}) \cdot \mathrm{exp}(i\beta_1 \sigma_{y}) \cdot \mathrm{exp}(i\beta_2 \sigma_{z}),\nonumber\\
&&\mathrm{cos}(\beta_1) = 1 - 4 \mathrm{sin^2}(b)\mathrm{cos^2}(c), \nonumber\\
&&\mathrm{sin}(\beta_2) = \sqrt{\frac{\mathrm{cos}(2b)\mathrm{cos}(2c)}{1-2\mathrm{sin^2}(b)\mathrm{cos^2}(c)}},
\label{bsu4}
\end{eqnarray}    
where $(a,b,c)$ represents the Weyl chamber coordinates. The single-qubit gates $K_{i},K_{i}^{'}, i = 1, 2, 3, 4$ can be obtained via the KAK decomposition, respectively. 

\section{\label{sec: Device}Device design}
\begin{figure}[hbt]
\includegraphics[scale=2]{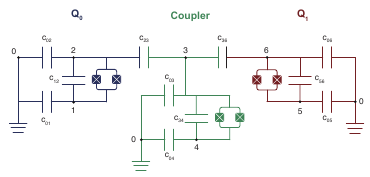}
\caption{ \textbf{A lumped-element circuit representation of the two transmon tuned by a tunable coupler}. }
\label{figscircuit}
\end{figure}

The qubit-coupler-qubit(QCQ) system base on capacitive coupling used in the main text is shown in Fig.~\ref{figscircuit}, both the qubits and coupler are floating. We numerically calculate the QCQ system capacitance from electromagnetic simulations: $C_{01} = 137$ fF, $C_{02} = 119$ fF,  $C_{12} = 23$ fF, $C_{03} = 133$ fF, $C_{04} = 165$ fF, $C_{34} = 28$ fF, $C_{05} = 137$ fF, $C_{06} = 119$ fF, $C_{56} = 23$ fF, $C_{23} = 26$ fF, $C_{36} = 26$ fF. 
The QCQ system's Lagrangian reads:
\begin{eqnarray}
\mathcal{L} &=& T - U, \nonumber \\ 
T &=& \frac{1}{2}\dot{\mathbf{\Phi}}^T_M\mathbf{M}\dot{\mathbf{\Phi}}_M, \nonumber\\ 
U &=& \sum_{k\in\{q_0,q_1,c\}}E_{Jk}\left(1-\cos\left(\frac{2\pi\Phi_{\Delta_k}}{\Phi_0}\right)\right),
\end{eqnarray}
where $T$ and $U$ are the kinetic and potential energy respectively; $\dot{\mathbf{\Phi}}_M=(\dot{\Phi}_{1}, \dot{\Phi}_{2}, ...,\dot{\Phi}_{6})^T$ where $\Phi_\mu$ $(\mu=1,2,...,6)$ is the node flux of the $\mu$-th metal plate and $\dot{\Phi}_\mu$ corresponds to its node potential.

$\mathbf{M}$ is the Maxwell capacitance matrix where the diagonal element $\mathbf{M}_{\mu\mu}$ is the sum of self-capacitance and all the mutual capacitances for the $\mu$-th metal plate (e.g. $\mathbf{M}_{11}=C_{01}+C_{13}$, and the off-diagonal element $\mathbf{M}_{\mu\nu}$ $(\mu,\nu=1,2,...,\mu\neq\nu)$ is the negative value of the mutual capacitance between the $\mu$-th and $\nu$-th metal plates (e.g. $\mathbf{M}_{12}=-C_{12}$; $q_1,q_2,c$ refer to two qubits and the coupler respectively; $\Phi_{\Delta_k}$ is the flux through the element's Josephson junction (or SQUID), which we call the junction flux (e.g. $\Phi_{\Delta_c}=\Phi_3-\Phi_4$). Rather than the node flux $\Phi_\mu$, the junction flux $\Phi_{\Delta_k}$ is the one who contributes to the energy oscillation. Therefore, we will use junction fluxes as the primary variables in the Hamiltonian. Introducing an auxiliary flux $\Phi_{\Sigma_k}=\Phi_{k1}+\Phi_{k2}$, the node fluxes can be rewritten as,
\begin{eqnarray}
    \begin{pmatrix}
         \Phi_{k1} \\
         \Phi_{k2}
    \end{pmatrix}
    = \begin{pmatrix}
         {1}/{2} & {1}/{2} \\
         {1}/{2} & -{1}/{2}
    \end{pmatrix}
    \begin{pmatrix}
         \Phi_{\Sigma_k} \\
         \Phi_{\Delta_k}
    \end{pmatrix}.
\end{eqnarray}
Thus the kinetic energy rereads 
\begin{eqnarray}
    T = \frac{1}{2}\dot{\mathbf{\Phi}}^T\mathbf{C}\dot{\mathbf{\Phi}},
\end{eqnarray}
where $\mathbf{C} = \mathbf{S}^T\mathbf{M}\mathbf{S}$ and
\begin{eqnarray}
    \dot{\mathbf{\Phi}}=
    \begin{pmatrix}
    \dot{\Phi}_{\Sigma_{q_1}} \\
    \dot{\Phi}_{\Delta_{q_1}} \\
    \dot{\Phi}_{\Sigma_{c}} \\
    \dot{\Phi}_{\Delta_{c}} \\
    \dot{\Phi}_{\Sigma_{q_2}} \\
    \dot{\Phi}_{\Delta_{q_2}} 
    \end{pmatrix},  
    \mathbf{S} = \begin{pmatrix}
    1/2 & 1/2 & 0 & 0 & 0 & 0 \\
    1/2 & -1/2 & 0 & 0 & 0 & 0 \\
    0 & 0 & {1}/{2} & {1}/{2} & 0 & 0 \\
    0 & 0 & {1}/{2} & -{1}/{2} & 0 & 0 \\
    0 & 0 & 0 & 0 & 1/2 & 1/2 \\
    0 & 0 & 0 & 0 & 1/2 & -1/2
    \end{pmatrix}.
\end{eqnarray}

Introducing the charge variables,
\begin{eqnarray}
    Q_{\Delta_k} = \frac{\partial\mathcal{L}}{\partial\dot{\Phi}_{\Delta_k}}, \quad Q_{\Sigma_k} = \frac{\partial\mathcal{L}}{\partial\dot{\Phi}_{\Sigma_k}},
\end{eqnarray}
the system Hamiltonian is obtained as: 
\begin{eqnarray}
    H=&&T+U, \nonumber \\
    T =&& \frac{1}{2}\mathbf{Q}^T\mathbf{C}^{-1}\mathbf{Q},\nonumber \\
    U=&& \sum_{k\in\{q_1,q_2,c\}}E_{Jk}\left(1-\cos\left(\frac{2\pi\Phi_{\Delta_k}}{\Phi_0}\right)\right),
\end{eqnarray}
where $\mathbf{Q}=(Q_{\Sigma_{q_1}},Q_{\Delta_{q_1}}, Q_{\Sigma_c}, Q_{\Delta_c}, Q_{\Sigma_{q_2}},Q_{\Delta_{q_2}})^T$. Again we introduce new variables, the Cooper pair number operators as $\hat{n}=Q/2e$, the kinetic energy further reads
\begin{eqnarray}
    T = 4e^2 \cdot \frac{1}{2}\mathbf{N}^T\mathbf{C}^{-1}\mathbf{N},
\end{eqnarray}
where $\mathbf{N} = (\hat{n}_{\Sigma_{q_1}}, \hat{n}_{\Delta_{q_1}}, \hat{n}_{\Sigma_c}, \hat{n}_{\Delta_c}, \hat{n}_{\Sigma_{q_2}}, \hat{n}_{\Delta_{q_2}})^T$. As we mentioned before, only $\Phi_{\Delta_k}$ is engaged in the energy oscillation, thus $\hat{n}_{\Delta_k}$ terms stand for the element modes while  $\hat{n}_{\Sigma_k}$ terms can be omitted~\cite{sete2021floating}. By expanding $T$ and taking $\hat{n}_k=\hat{n}_{\Delta_k}, k\in\{q_1,q_2,c\}$, we obtain 
\begin{eqnarray}
    T = \sum_{k\in\{q_1,q_2,c\}}4E_{Ck}\hat{n}_k^2+4E_{q_1q_2}\hat{n}_{q_1}\hat{n}_{q_2}+4E_{q_1c}\hat{n}_{q_1}\hat{n}_{c}+4E_{q_2c}\hat{n}_{q_2}\hat{n}_{c},
\end{eqnarray}
here 
\begin{eqnarray}
E_{Ck} &=& e^2\frac{\mathbf{C}^{-1}_{\Delta_k,\Delta_k}}{2}, \label{ECk} \nonumber \\ 
E_{ij} &=& e^2\mathbf{C}^{-1}_{\Delta_i,\Delta_j}, \label{Eij} \quad i,j,k\in\{q_1,q_2,c\}, i\neq j.
\end{eqnarray}
Defining the reduced flux or the phase operator as $\hat{\phi}_k=\frac{2\pi\Phi_{\Delta_k}}{\Phi_0}$, the quantized system Hamiltonian reads,
\begin{eqnarray}
    H  = && 4 E_{C q_1} \hat{n}_1^2+4 E_{C q_2} \hat{n}_2^2+4 E_{C c} \hat{n}_c^2 \\ \nonumber
&& +4 E_{q_1q_2} \hat{n}_1 \hat{n}_2+4 E_{q_1 c} \hat{n}_1 \hat{n}_c+4 E_{q_2 c} \hat{n}_2 \hat{n}_c \\ \nonumber
&& -\sum_{k \in\{q_1,q_2, c\}} E_{J k} \cos \left(\hat{\phi}_k\right).
\end{eqnarray}

Expanding $\cos \left(\hat{\phi}_k\right)$ to the second order approximately, and taking~\cite{krantz2019quantum} 
\begin{eqnarray}
    \hat{n}=i n_{\text{zpf}}(a-a^\dag),\;\hat{\phi}=\phi_{\text{zpf}}(a+a^\dag),
\end{eqnarray}
where $a^\dag$($a$) is the creation (annihilation) operator of the harmonic oscillator basis and $n_{\text{zpf}}=\left[E_J/(32E_C)\right]^{1/4}, \phi_{\text{zpf}}=\left[2E_C/E_J\right]^{1/4}$ are the ``zero-point fluctuations'' of the charge and phase variables, respectively.
The system Hamiltonian can be expressed as the second quantized form
\begin{eqnarray}
    H=&&\sum_{k\in\{q_1,q_2,n\}}(\omega_k a_k^\dag a_k + \frac{\alpha_k}{2}a_k^\dag a_k^\dag a_ka_k) \\ \nonumber
    &&+ \sum_{i,j\in\{q_1,q_2,n\},i<j}g_{ij}(a_i+a_i^\dag)(a_j+a_j^\dag),
\end{eqnarray}
where $\omega_k=\sqrt{8E_{Ck}E_{Jk}}-E_{Ck}, \alpha_k=-E_{Ck}, g_{ij}=E_{ij}/\sqrt{2}\cdot\left[(E_{Ji}E_{Jj})/(E_{Ci}E_{Cj})\right]^{1/4}$. The effective coupling between two qubits via coupler is~\cite{yan2018tunable}:
\begin{eqnarray}    
g=g_{q_1q_2}+\frac12g_{q_1c}g_{q_2c}\left(\frac{1}{\omega_{q_1}-\omega_c}+\frac{1}{\omega_{q_2}-\omega_c}\right).
\label{effg}
\end{eqnarray}

In our design, the typical parameter values are: $\omega_{q_{1}}/2\pi = \omega_{q_{2}}/2\pi=4.1 $ GHz, $\alpha_{q_{1}}/2\pi=\alpha_{q_{2}}/2\pi=-185 $ MHz, $g_{q_{1c}}/2\pi=g_{q_{2c}}/2\pi = 90 $ MHz, $g_{q_{1}q_{2}}/2\pi = 5$ MHz, and $\omega_{c}/2\pi \approx 5.2$ GHz when the effective qubit-qubit coupling is zero.

\section{\label{sec:fab}Device Fabrication}
In our superconducting quantum processor, we employ a flip-chip architecture. The coherent components, including qubits, couplers, and readout resonators, are patterned on the top chip, while the transmission lines for control and readout are placed on the bottom carrier chip. The top and carrier chips are fabricated and tested separately before flip-chip bonding, minimizing additional processing on each chip and improving overall yield.

For the top wafer, we deposit 200-nm thick tantalum (Ta) films on a pre-annealed sapphire substrate, while the carrier wafer is made from a high-resistivity silicon substrate. The base circuits are defined using direct laser photolithography, and the patterns are transferred to the Ta films by reactive ion etching after development.

Following the removal of residual photoresist, we fabricate the Al/AlOx/Al Josephson junctions using the double-angle shadow evaporation technique in the Dolan-bridge style. This is done via electron beam lithography with double layers of PMMA A4/LOR 10B photoresist~\cite{wang2022towards}. The top wafer is then diced into the desired size and soaked in an NMP bath for at least 16 hours to ensure thorough lift-off.

On the carrier wafer, we use the reflowing process as described in Ref.~\cite{PhysRevResearch.6.L042038} to fabricate tunnel airbridges along the transmission lines, which helps to reduce crosstalk. Next, indium bumps, approximately 10 $\mu$m in height, are grown by thermal evaporation. 
We perform flip-chip bonding using a bonder with a bonding force of approximately 80~N.

\section{\label{sec: setup}Measurement setup}

\begin{figure}[hbt]
\includegraphics[scale=2]{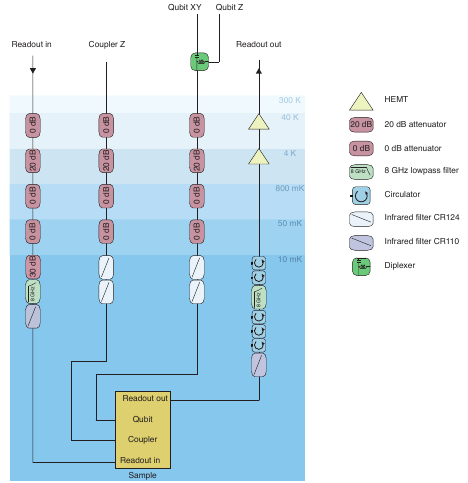}
\caption{ \textbf{Schematic of the measurement setup.} }
\label{figssetup}
\end{figure}

As shown in the schematic in Fig.~\ref{figssetup}, the experiments are carried out in a BlueFors LD-1000 dilution refrigerator with magnetic shielding at a base temperature of 10 mK. The sample package is enclosed by an additional \textmu-metal shield.

The readout signal is amplified by two high-electron mobility transistor (HEMT) amplifiers, one at the 4K stage and another at the 40K stage.
Filtering and attenuation of the control lines are identical for all qubits and couplers. Double infrared filters are installed at the base-temperature place to prevent radiation of higher-temperature stages from reaching the device.

At room temperature, signals for qubit XY control and readout are processed by direct digital synthesis (DDS) systems without extra frequency mixing processes. The Z signals are generated by arbitrary waveform generators. For qubit control, the XY and Z signals are combined using diplexers before entering the fridge. 

\section{\label{sec: charact} Device characterization}

\begin{table}[!htb]
    \centering
    \begin{tabular}{ccccccccccccc}
    \hline
    \hline
        Qubit & $Q_{0}$ & $Q_{1}$ & $Q_{2}$ & $Q_{3}$ & $Q_{4}$ & $Q_{5}$ & $Q_{6}$ & $Q_{7}$ & $Q_{8}$ & $Q_{9}$ & $Q_{11}$ & $Q_{12}$\\ \hline
         $f_{\mathrm{max}}$ (GHz)& 4.112 & 4.175 & 4.145 & 4.243 & 4.115 & 4.094 & 4.102& 4.146 & 4.171 & 4.247 & 4.029 &4.092 \\ 
        $f_{\mathrm{idle}}$ (GHz)& 3.896 & 4.018 & 4.038 & 4.052 & 4.115 & 4.000&4.102& 4.044 &4.100 & 4.039 & 4.029 & 4.092\\ 
        $\alpha$ (MHz)& -179.1 & -192.0 & -200.4 & -194.5 & -195.4 &-187.9 & -192.1 & -194.7 & -186.4 & -200.2 & -194.1 & -192.3\\ 
        
        $f_{\mathrm{read}}$ (GHz)& 7.112 & 6.866 & 7.035 & 6.767 & 7.148 &6.891 & 7.063 & 6.893 & 7.135 & 6.736 & 7.068 & 6.816\\ 
        
        $T_{1}$  ($\mu$s) & 55.5 & 76.4 & 39.7 & 48.7 & 82.4 & 60.4 & 87.8 & 82.1 & 66.3 & 65.9 &108.2&69.9 \\ 
        
        $T_{\mathrm{2Ramsey}}$ ($\mu$s)  & 2.4 & 4.7& 4.1 & 3.4 & 4.8 & 3.0 & 3.5 & 2.8 & 4.0 & 2.6 & 18.1 &11 \\ 

        $\epsilon_{\mathrm{read}}^*$ $\ket{0}$ (\%)& 5.8 & 4.6 & 3.9 & 4.3 & 6.0 &5.4 & 5.9 & 5.9 & 4.3 & 6.6 & 4.0 & 4.5\\ 

        $\epsilon_{\mathrm{read}}^*$ $\ket{1}$  (\%)& 6.0 & 8.9 & 8.3 & 6.3 & 9.4 &6.8 & 7.2 & 7.9 & 10.2 & 6.6 & 9.2 & 8.6\\ 

         $\epsilon_{\mathrm{1Q}}^*$  (\%)& 0.26 & 0.32 & 0.20 & 0.19 & 0.18 &0.22 & 0.21 & 0.27 & 0.20 & 0.19 & 0.08 & 0.15\\ 
         $\epsilon_{\mathrm{iSWAP}}$ (\%)&   \multicolumn{10}{c}{\quad 0.40 \quad  0.13 \quad 0.59 \quad 0.58 \quad 1.06 \quad 0.71\quad 0.56  \quad 1.17 \quad 0.88 \quad}&   \multicolumn{2}{c}{0.38}  \\ 
 
         \hline
         \hline
    \end{tabular}
    \caption{\label{tab1} \textbf{System parameters.} Coherence times shown here are measured at the idling point. The star symbols over the gate or readout error rates ($\epsilon$) indicate that they are benchmarked simultaneously. }
\end{table}

In Table~\ref{tab1}, we list the characterization results for the qubits used in this work including the ten qubits ($Q_{0}$-$Q_{9}$) used in the entangled state preparation experiment and the two qubits ($Q_{11}$-$Q_{12}$) used to demonstrate the AshN and $\BGate$ gates. 
At the idling bias, the average relaxation time $T_1$ for these qubits is 70.3 $\mu$s, with a standard deviation of 17.8 $\mu$s, and the average Ramsey time $T_{\mathrm{2Ramsey}}$ is 5.4 $\mu$s, with a standard deviation of 4.4 $\mu$s. The readout errors are also characterized: the average error is 5.1\% with a standard deviation of 0.9\% for $\ket{0}$ and 7.9\% with a standard deviation of 1.3\% for $\ket{1}$.

To verify our gate performance, we utilize cross-entropy benchmarking for the single- qubit gates and two-qubit $\iSWAP$ gates. The single-qubit gates are benchmarked simultaneously, with the average gate error being 0.21\% and a standard deviation of 0.06\%, while the $\iSWAP$ gates are performed individually, with the average gate error being 0.65\% and a standard deviation of 0.30\%.

\section{\label{sec: mwphase} Phase matching between driving pulses}
In the AshN Hamiltonian, the phase $\phi$ between two microwave drives should be strictly equal to $2n\pi, n\in  \mathbb{Z}$. Nevertheless, in the experiment, a non-zero phase may exist due to the imperfection of the experimental system such as cable delay and crosstalk. Furthermore, this phase has a significant impact on the gate generated. In this section, we introduce a simple method for calibrating the phase between two drives using a specific gate realized through the AshN-ND scheme.

\begin{figure}[hbt]
\includegraphics[scale=1.2]{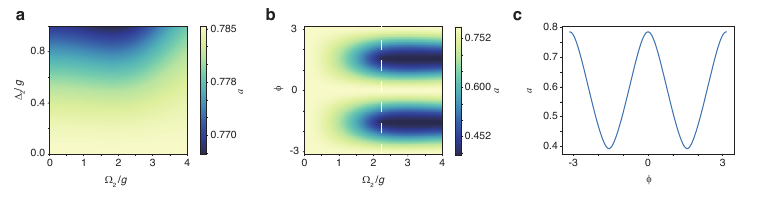}
\caption{\textbf{Strategy for calibrating the microwave phase between two drives}. {\bf{a}}, Numerical calculation of the coefficient $a$ as a function of the coefficient $\Omega_{2}$ and $\Delta_{2}$. {\bf{b}}, The coefficient $a$ as a function of $\Omega_{2}$ and $\theta$, as $\Delta_2 = 0$. The white dashed line corresponds to the B gate amplitude $2.238g$. {\bf{c}}, The line cut in {\bf{b}} at $\Omega_{2} = 2.238g$.}
\label{figsmphase}
\end{figure}

For the AshN gate with one drive, e.g., such as the $\BGate$ gate, $\phi$ does not affect the gate generated. If we add an analogous drive to another qubit, the coefficient $a$ of the $XX$ term in the resulting gate will remain $\pi/4$, as long as the phase $\phi$ is zero or $\pi$. Fig.~\ref{figsmphase}a shows the calculated $a$ 
as a function of the second drive amplitude $\Omega_{2}$ and the drive detuning $\Delta_{2}$ (likely caused by the AC-Stark shift). It can be seen that a second drive with amplitude $\Omega_{2}$ up to $4g$ and detuning $\Delta_{2}$ up to $g$, $a$ will always be around $\pi/4$.

If we fix the detuning $\Delta_{2}$, and alter the phase $\phi$, results will be different. As illustrated in Figs.~\ref{figsmphase}b and c, as $\Omega_{2}$ is closed to the driving amplitude of the $\BGate$ gate ($\Omega_{1} \approx 2.238g$), $a$ will be far from $\pi/4$ unless $\phi = 0 \ \mathrm{or} \ \pi$.
In the experiment, after calibrating the B gate, we add a second drive such that $\Omega_2 \approx 2.238g$. We perform quantum process tomography (QPT) of this gate and then perform the KAK decomposition. Finally We vary the phase $\phi$ and monitor the coefficient $a$ until it reaches its maximum value close to $\pi/4$. 

\section{\label{sec:localz}  Corrections for the local phase shift}
Like $\iSWAP$ gate, most two-qubit gates contain energy-exchanging components. During the gate operation, two qubits exchange their state in certain ways. However, because each qubit has its own rotating frame, performing these gates will cause a nontrivial time-dependent local phase shift. Putting qubits into the same rotating frame is an effective method to avoid this issue, however, careful microwave crosstalk cancellation must be applied~\cite{sung2021realization}. In the following, we will introduce a universal strategy to resolve this local phase shift.              

Considering two uncoupled qubits (i.e., two qubits at their idle frequencies with the coupling turned off), the system can be well described by the lab frame Hamiltonian,
\begin{eqnarray}
H_{\mathrm{lab}}  = \frac{1}{2}\omega_{1} ZI +\frac{1}{2}\omega_{2} IZ.
\label{lab}
\end{eqnarray}
For an arbitrary state $|\psi(t)\rangle$, its evolution under Eq.~\ref{lab} will be:
\begin{eqnarray}
|\psi(t)\rangle  =  \left( \begin{array}{c}
 c_{00}(t)\\
 c_{01}(t)\\
 c_{10}(t)\\
 c_{11}(t)
\end{array}
\right )
= \left( \begin{array}{c}
 c_{00}(0)\\
e^{i\omega_2t}c_{01}(0)\\
e^{i\omega_1t}c_{10}(0)\\
e^{i(\omega_2+\omega_1)t}c_{11}(0),
\end{array}
\right ),
\label{lab0}
\end{eqnarray}
here $c_{m}(t)$ refers to the probability amplitude associated with the state $|m\rangle \in \{|00\rangle,|01\rangle,|10\rangle,|11\rangle\}$ at time $t$. From the perspective of the doubly rotating frame, the state $|\psi(t)\rangle$ becomes:
\begin{eqnarray}
|\tilde{\psi}(t)\rangle  =  \left( \begin{array}{c}
 \tilde{c}_{00}\\
 \tilde{c}_{01}\\
 \tilde{c}_{10}\\
 \tilde{c}_{11}
\end{array}
\right ) 
&&= \left( \begin{array}{c}
 c_{00}(0)\\
e^{-i\omega_2t}e^{i\omega_2t}c_{01}(0)\\
e^{-i\omega_1t}e^{i\omega_1t}c_{10}(0)\\
e^{-i(\omega_2+\omega_1)t}e^{i(\omega_2+\omega_1)t}c_{11}(0)
\end{array}
\right ) \nonumber \\
&&= \left( \begin{array}{c}
 c_{00}(0)\\
c_{01}(0)\\
c_{10}(0)\\
c_{11}(0),
\end{array}
\right ),
\label{logic1}
\end{eqnarray}
$|\tilde{\psi}(t)\rangle$ is the state of the system in the computational basis. If we apply an arbitrary two-qubit gate $U$ at time $t = \tau$, in the lab frame, the system will be:
\begin{eqnarray}
|\psi(\tau)\rangle  &&=  \left( \begin{array}{cccc}
 p_{0}&q_{0}&r_{0}&s_{0}\\
 p_{1}&q_{1}&r_{1}&s_{1}\\
 p_{2}&q_{2}&r_{2}&s_{2}\\
 p_{3}&q_{3}&r_{3}&s_{3}
\end{array}
\right )
\left( \begin{array}{c}
 c_{00}(\tau)\\
 c_{01}(\tau)\\
 c_{10}(\tau)\\
 c_{11}(\tau)
\end{array}
\right ) \nonumber \\
&&= \left( \begin{array}{cccc}
p_{0}c_{00}(0)+q_{0}c_{01}(0)e^{i\omega_2\tau}+r_{0}c_{10}(0)e^{i\omega_1\tau}+s_{0}c_{11}(0)e^{i(\omega_2+\omega_1)\tau}\\
p_{1}c_{00}(0)+q_{1}c_{01}(0)e^{i\omega_2\tau}+r_{1}c_{10}(0)e^{i\omega_1\tau}+s_{1}c_{11}(0)e^{i(\omega_2+\omega_1)\tau}\\
p_{2}c_{00}(0)+q_{2}c_{01}(0)e^{i\omega_2\tau}+r_{2}c_{10}(0)e^{i\omega_1\tau}+s_{2}c_{11}(0)e^{i(\omega_2+\omega_1)\tau}\\
p_{3}c_{00}(0)+q_{3}c_{01}(0)e^{i\omega_2\tau}+r_{3}c_{10}(0)e^{i\omega_1\tau}+s_{3}c_{11}(0)e^{i(\omega_2+\omega_1)\tau}
\end{array}
\right ).
\label{lab1}
\end{eqnarray}
In the doubly rotating frame, Eq.~\ref{lab1} becomes:
\begin{eqnarray}
|\tilde{\psi}(\tau)\rangle 
&&= \left( \begin{array}{cccc}
p_{0}c_{00}(0)+q_{0}c_{01}(0)e^{i\omega_2\tau}+r_{0}c_{10}(0)e^{i\omega_1\tau}+s_{0}c_{11}(0)e^{i(\omega_2+\omega_1)\tau}\\
p_{1}c_{00}(0)e^{-i\omega_2\tau}+q_{1}c_{01}(0)+r_{1}c_{10}(0)e^{-i(\omega_2-\omega_1)\tau}+s_{1}c_{11}(0)e^{i\omega_1\tau}\\
p_{2}c_{00}(0)e^{-i\omega_1\tau}+q_{2}c_{01}(0)e^{-i(\omega_1-\omega_2)\tau}+r_{2}c_{10}(0)+s_{2}c_{11}(0)e^{i\omega_2\tau}\\
p_{3}c_{00}(0)e^{-i(\omega_1+\omega_2)}+q_{3}c_{01}(0)e^{-i\omega_1\tau}+r_{3}c_{10}(0)e^{-i\omega_2\tau}+s_{3}c_{11}(0)
\end{array}
\right ) \nonumber \\ 
&&= \left( \begin{array}{cccc}
p_{0}&q_{0}e^{i\omega_2\tau}&r_{0}e^{i\omega_1\tau}&s_{0}e^{i(\omega_2+\omega_1)\tau}\\
p_{1}e^{-i\omega_2\tau}&q_{1}&r_{1}e^{i(\omega_1-\omega_2)\tau}&s_{1}e^{i\omega_1\tau}\\
p_{2}e^{-i\omega_1\tau}&q_{2}e^{i(\omega_2-\omega_1)\tau}&r_{2}&s_{2}e^{i\omega_2\tau}\\
p_{3}e^{-i(\omega_1+\omega_2)}&q_{3}e^{-i\omega_1\tau}&r_{3}e^{-i\omega_2\tau}&s_{3}
\end{array}
\right )
\left( \begin{array}{c}
 c_{00}(0)\\
c_{01}(0)\\
c_{10}(0)\\
c_{11}(0)
\end{array}
\right ) \nonumber \\
&& = \tilde{U}|\tilde{\psi}(t)\rangle.
\label{logic3}
\end{eqnarray}
Eq.~\ref{logic3} reveals that in the rotating frame, the gate $U$ becomes $\tilde{U}$ with certain time-dependent phase shifts. Fortunately, we can separate the phase shifts as:
\begin{eqnarray}
\tilde{U} 
&& = \left( \begin{array}{cccc}
p_{0}&q_{0}e^{i\omega_2\tau}&r_{0}e^{i\omega_1\tau}&s_{0}e^{i(\omega_2+\omega_1)\tau}\\
p_{1}e^{-i\omega_2\tau}&q_{1}&r_{1}e^{i(\omega_1-\omega_2)\tau}&s_{1}e^{i\omega_1\tau}\\
p_{2}e^{-i\omega_1\tau}&q_{2}e^{i(\omega_2-\omega_1)\tau}&r_{2}&s_{2}e^{i\omega_2\tau}\\
p_{3}e^{-i(\omega_1+\omega_2)}&q_{3}e^{-i\omega_1\tau}&r_{3}e^{-i\omega_2\tau}&s_{3}
\end{array}
\right )\nonumber \\
&& = \left( \begin{array}{cccc}
1&0&0&0\\
0&e^{-i\omega_2\tau}&0&0\\
0&0&e^{-i\omega_1\tau}&0\\
0&0&0&e^{-i(\omega_1+\omega_2)\tau}
\end{array}
\right )
\left( \begin{array}{cccc}
 p_{0}&q_{0}&r_{0}&s_{0}\\
 p_{1}&q_{1}&r_{1}&s_{1}\\
 p_{2}&q_{2}&r_{2}&s_{2}\\
 p_{3}&q_{3}&r_{3}&s_{3}
\end{array}
\right )
\left( \begin{array}{cccc}
1&0&0&0\\
0&e^{i\omega_2\tau}&0&0\\
0&0&e^{i\omega_1\tau}&0\\
0&0&0&e^{i(\omega_1+\omega_2)\tau}
\end{array}
\right ) \nonumber \\
&& = \left( \begin{array}{cc}
1&0\\
0&e^{-i\omega_1\tau}
\end{array}
\right ) \otimes
\left( \begin{array}{cc}
1&0\\
0&e^{-i\omega_2\tau}
\end{array}
\right )
\left( \begin{array}{cccc}
 p_{0}&q_{0}&r_{0}&s_{0}\\
 p_{1}&q_{1}&r_{1}&s_{1}\\
 p_{2}&q_{2}&r_{2}&s_{2}\\
 p_{3}&q_{3}&r_{3}&s_{3}
\end{array}
\right )
\left( \begin{array}{cc}
1&0\\
0&e^{i\omega_1\tau}
\end{array}
\right ) \otimes
\left( \begin{array}{cc}
1&0\\
0&e^{i\omega_2\tau}
\end{array}
\right ) \nonumber \\
&& = (Z_{3} \otimes
Z_{4})
U
(Z_{1} \otimes
Z_{2}).
\label{phase2}
\end{eqnarray}
Hence, for the frequency-tunable qubit system,  we can compensate the phase shift caused by frame transform by four time-dependent local phases $Z_{1}^\dagger, Z_{2}^\dagger, Z_{3}^\dagger, Z_{4}^\dagger$, which satisfy
\begin{eqnarray}
U 
=  (Z_{3}^\dagger \otimes
Z_{4}^\dagger)
\tilde{U} 
(Z_{1}^\dagger \otimes
Z_{2}^\dagger).
\label{phase1}
\end{eqnarray}
\begin{figure}[hbt]
\centering
\includegraphics[scale=1.8]{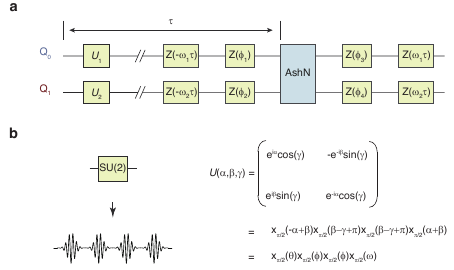}
\caption{ \textbf{Compensating Z gates and PMW-4 pulses}. \textbf{a}, Strategy for compensating the nontrivial local phase shift. \textbf{b}, We use PMW-4 pulses to generate arbitrary single-qubit gates. Any single-qubit gate $U(\alpha, \beta, \gamma)$ ($\alpha, \beta, \gamma$ are Euler angles) in SU(2) can be decomposed into four $\pi/2$ pulses $\mathrm{X}_{\pi/2}(\theta)\mathrm{X}_{\pi/2}(\phi)\mathrm{X}_{\pi/2}(\phi)\mathrm{X}_{\pi/2}(\omega)$, where $\theta = -\alpha+\beta, \phi = \beta-\gamma+\pi, \omega = \alpha+\beta$ refer to the phase of the $\pi/2$ pulses. }
\label{figszphase}
\end{figure}

Fig.~\ref {figszphase}a shows the strategy of local phase shift compensation for the AshN gate. The compensatory gates have two parts:
\\
\noindent \textbf{Time-Dependent Part}: $Z(\pm\omega_{i}\tau), i=1, 2$, which is used to correct the unmatched rotating frames of the two qubits, as analyzed above.
\\
\noindent \textbf{Time-Independent Part}: $Z(\phi_{i}), i = 1, 2, 3, 4$, which is used to compensate for the phase shift caused by the frequency change of the qubits from the idle frequency to the interaction frequency during gate operation.
\\
In the experiment, the time-dependent part $Z(\pm\omega_{i}\tau), i=1, 2$, can be easily calculated, while the time-independent part $Z(\phi_{i}), i = 1, 2, 3, 4$, can be derived by QPT. 

It is worth noting that the virtual Z gate is hardly compatible with general two-qubit gates because most of them do not propagate the phase shift cleanly, disrupting the intended computation. Hence, they cannot serve as reliable phase carriers for virtual Z operations. As a consequence, physical Z gates are typically introduced to compensate for these Z phase shifts. In this work, we use the PMW-4 scheme instead to generate these compensatory Z gates, as shown in Fig.~\ref{figszphase}b. PMW-4 scheme can generate arbitrary single-qubit gates in $\SU{2}$ with a sequence of four $\pi/2$ pulses with appropriately chosen phases. Hence, all consecutive single-qubit gates acting on the same qubit, including those compensatory Z gates, can be consolidated into a single PMW-4 sequence.

\section{\label{sec: crosstalk}  Microwave XY crosstalk}

\begin{figure}[hbt]
\includegraphics[scale=1]{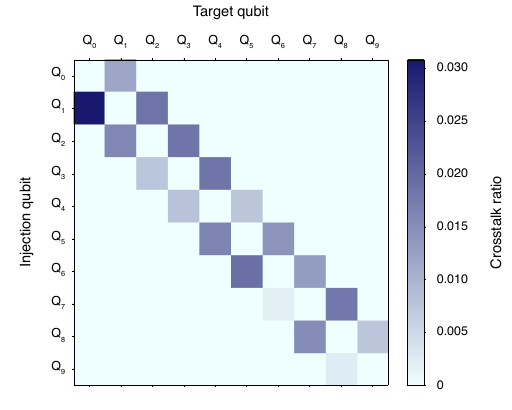}
\caption{ \textbf{The XY crosstalk ratio}. The columns are target qubits. The rows are qubits to which the control signals are sent. We only show the measured crosstalk ratio between the nearest-neighbor qubit pairs. The crosstalk ratio is defined as the measured Rabi frequency using the crosstalked drive (control qubit and target qubit are different) divided by the Rabi frequency using the dedicated drive. }
\label{figscorss}
\end{figure}

Microwave crosstalk is an important issue here, owing to its significance in performing single-qubit gates and AshN-generated two-qubit gates simultaneously. We measure the microwave crosstalk between nearest-neighbor qubits by measuring the relative Rabi frequency as shown in Fig.~\ref{figscorss}. The average crosstalk ratio is 1.3\% in our device.

\begin{figure}[hbt]
\includegraphics[scale=1]{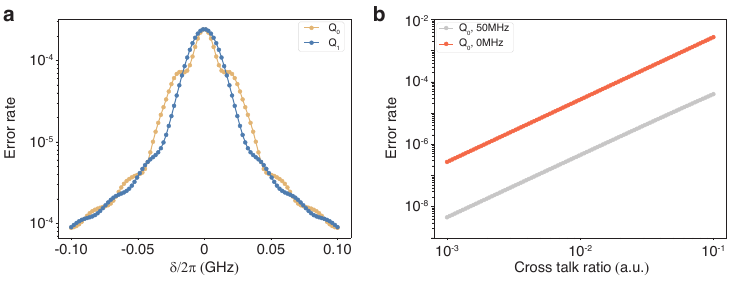}
\caption{ \textbf{Gate error due to XY crosstalk.} \textbf{a}, The simulated gate error for an AshN-generated B gate on $\mathrm{Q_0}$ and $\mathrm{Q_1}$ as a function of the detuning of a crosstalk drive $\delta$ from the resonance frequency applied to one of the qubit. The main drive for the B gate is applied to $\mathrm{Q_0}$. The crosstalk drive is assumed at a relative ratio of 3\%. \textbf{b}, The simulated gate error as a function of the crosstalk ratio.}
\label{figscorsst}
\end{figure}

Next, we evaluate the influence of this crosstalk level on gate performance. We numerically simulate an AshN-generated $\BGate$ gate and estimate the gate error in the presence of additional crosstalk signals. From Fig.~\ref{figscorsst}a, it can be seen that the error caused by the crosstalk drive, given the strongest crosstalk ratio in our device at 3\%, is still well below $10^{-3}$ even for the resonant case. The error rapidly decreases as the drive frequency is tuned further away. Moreover, we simulate the dependence of the error rate on the crosstalk ratio, which, as expected for unitary errors, follows a quadratic relationship.

\section{\label{sec: SQ} Single-qubit gate calibration}

\begin{figure}[hbt]
\includegraphics[scale=1]{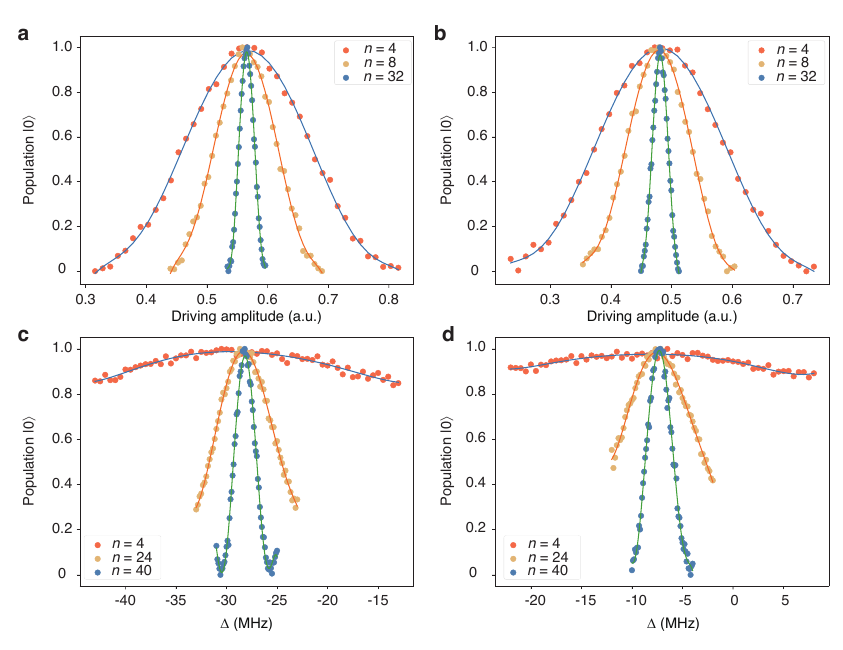}
\caption{ \textbf{Calibration of the $\pi/2$ pulse of $\mathrm{Q_{11}}$ and $\mathrm{Q_{12}}$}. $\bf{a}$, $\bf{b}$, Calibration of driving amplitude for $\mathrm{Q_{11}}$, $\mathrm{Q_{12}}$, respectively. $\bf{c}$, $\bf{d}$, Calibration of DRAG $\delta$ for $\mathrm{Q_{11}}$, $\mathrm{Q_{12}}$, respectively.}
\label{figssq}
\end{figure}

In this work, we employ the PMW-4 scheme to compile arbitrary single-qubit gates. Consequently, it is only necessary to calibrate the $\pi/2$ pulse to implement all single-qubit gates. The form of the $\pi/2$ pulse can be expressed as,
\begin{eqnarray}
\begin{cases} 
A(t)\mathrm{sin}(\omega t+\phi), &0<t<t_{g}\\
0 ,    &\mathrm{otherwise},
\end{cases}
\label{sqshape}
\end{eqnarray}
where $t_{g}$ is the time of a $\pi/2$ pulse, which is 16 ns (with 4 ns buffer and 20 ns in total) in our experiment and $\omega, \phi$ are the frequency and phase of the driving, respectively. $A(t)$ describes the shape of the pulse, which we chose cosine-shaped envelop with amplitude $A$ in our experiment:
\begin{eqnarray}
A(t) = \frac{A}{2}(1-\mathrm{cos}\frac{2\pi t}{t_{g}}), 
\label{driving1}
\end{eqnarray}
To avoid leakage to higher levels, we employ the derivative removal by adiabatic gate (DRAG) pulse method. Hence $A(t)$ becomes:
\begin{eqnarray}
A'(t) = A(t)-i\frac{\eta}{\alpha}\dot{A}(t), 
\label{driving2}
\end{eqnarray}
where $\alpha$ is the anharmonicity and $\eta$ is weighting factor, and in our experiment we keep $\eta/\alpha = 1 $ ns.

To calibrate the $\pi/2$ pulse, we apply sequences consisting of an even number $n$ of pulses, $[R_x(\pi/2)]^n$, alter the pulse amplitude, and subsequently measure the population of the state $\ket{0}$. By sweeping different numbers $n$, the driving amplitude will converge to its optimal value, as shown in Figs.~\ref{figssq}a and~\ref{figssq}b.

Due to the AC-stark shift, the driving may cause a phase error, and it can be removed by detuning the driving such that $\omega_\mathrm{e} = \omega+\Delta$. Therefore, We perform sequences with an even number $n$ of pulses and alter the parameter $\Delta$, as shown in Figs.~\ref{figssq}c and d. By calibrating the driving amplitude and $\Delta$ back and forth, the error rate of the single-qubit gate reaches its minimum.

\section{\label{sec: iswap} Two-qubit iSWAP gate calibration}

\begin{figure}[hbt]
\includegraphics[scale=1]{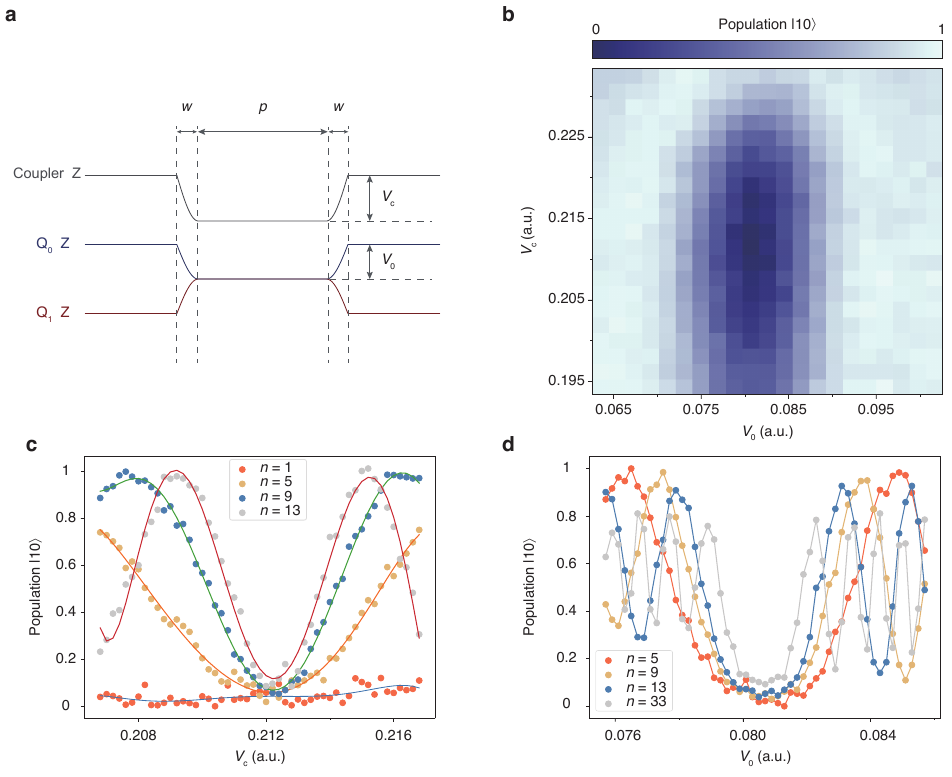}
\caption{ \textbf{Calibration of the iSWAP gate}. $\bf{a}$, pusle sequence used to generate the iSWAP gate. $\bf{b}$, measurement of $\mathrm{Q_{0}}$ Z bias $V_0$ and coupler Z bias $V_c$ to determine where the two qubit are fully swapped. $\bf{c}$, Pulse train calibration of $V_c$. $\bf{d}$,  Pulse train calibration of Z bias $V_0$.}
\label{figsiswap}
\end{figure}

The $\iSWAP$ gate is the foundation of the AshN gates because they share the same interaction frequency. To implement the $\iSWAP$ gate, the $\ket{10}$ and $\ket{01}$ states are brought into resonance and allowed to complete half of a swap oscillation, resulting in full exchange of the two states. The envelop of the gate pulse in this work contains three parts, which are rising and falling edges, each with duration $w$, a plateau with duration $p$, and has the form:
\begin{eqnarray}
\begin{cases} 
A\times\frac{\mathrm{cosh}(-e/2)-\mathrm{cosh}[e*(t-w)/2w]}{\mathrm{cosh}(-e/2)-1}, &0\leq t<w\\
A ,    &w\leq t \leq w+p\\
A\times\frac{\mathrm{cosh(-e/2)}-\mathrm{cosh}[e*(t-p-w)/2w]}{\mathrm{cosh}(-e/2)-1}, &w+p < t \leq t_{g}\\
0, &\mathrm{otherwise},
\end{cases}
\label{iswapshape}
\end{eqnarray}
where $A$ is the amplitude of the pulse and $e$ describes the pulse steepness.

As shown in Fig.~\ref{figsiswap}a, the gate duration $t_{g} = 2w+p$ we chose is 40 ns, with rising edge $w$ = 2.5 ns, falling edge $w$ = 2.5 ns, a plateau $p$ = 35 ns and steepness $e$ is 10. 

To obtain an $\iSWAP$ gate with high fidelity, we first carefully choose an interaction frequency where both qubits will not interact with any two-level systems (TLS) or neighboring qubits. Then we apply an X gate on the first qubit $\mathrm{Q_{0}}$, fix the Z bias of the second qubit $\mathrm{Q_{1}}$ and vary the coupler Z bias $V_{\mathrm{c}}$ and $\mathrm{Q_{0}}$ Z bias $V_{\mathrm{0}}$, as shown in Fig.~\ref{figsiswap}. We select the appropriate $V_{\mathrm{0}}$ and $V_{\mathrm{c}}$ where the two qubits are fully swapped. To accurately calibrate the Z bias of the qubit and coupler, we perform sequences composed of an odd number $n$ of $\iSWAP$ pulses and vary $V_{\mathrm{c}}$
 (Fig.~\ref{figsiswap}c) and $V_{\mathrm{0}}$ (Fig.~\ref{figsiswap}d). By increasing the pulse number $n$
 and alternating the calibration, both the biases of the qubit and the coupler reach their optimal values.

\section{\label{sec:level11} Cross-entropy benchmarking and precise calibration}

Most two-qubit gates are non-Clifford and can hardly be accurately characterized using the commonly employed Clifford-based randomized benchmarking (RB) method. Cross-entropy benchmarking (XEB), on the other hand, provides a robust tool for evaluating arbitrary quantum gates. Therefore, we employ XEB to benchmark the AshN gates in this experiment.

The XEB used in our experiment contains a repetitive gate sequence, and each cycle is constructed by a pair of single-qubit gates randomly chosen from the set $\{U(\pi/2, n\pi/4, -n\pi/4)\}$, $1 \leq n \leq 8, n\in \mathbb{Z}$, followed by an AshN two-qubit gate. Here, $U(\alpha, \beta, \gamma)$ represents the single-qubit gate described by three Euler angles $\alpha, \beta, \gamma$.

The XEB sequence fidelity $\mathcal{F}$ is computed using the cross-entropy $S$ between two probability distributions $P=\{p_i\}$ and $Q=\{q_i\}$ over the set of bitstrings, as $S(P, Q) = - \sum_{i}p_{i}\mathrm{ln}(q_{i})$ and has the form:
\begin{eqnarray}
\mathcal{F} = \frac{S(P_{\mathrm{incoherent}}, P_{\mathrm{expected}})-S(P_{\mathrm{measured}}, P_{\mathrm{expected}})}{S(P_{\mathrm{incoherent}}, P_{\mathrm{expected}})-S(P_{\mathrm{expected}})},
\label{XEB}
\end{eqnarray}
where $P_{\mathrm{incoherent}}, P_{\mathrm{expected}}, P_{\mathrm{measured}}$ are the incoherent, expected, and measured probability distributions, respectively. The XEB cycle error can be obtained by fitting the fidelity of the sequence $\mathcal{F}$ with an exponential decay.

As discussed in previous sections, arbitrary two-qubit gates can be generated roughly with simple steps. However, achieving high-fidelity gates requires precise calibration. In this work, we employ a Bayesian optimizer to fine-tune the calibration of the AshN gates.
\begin{figure}[hbt]
\includegraphics[scale=1]{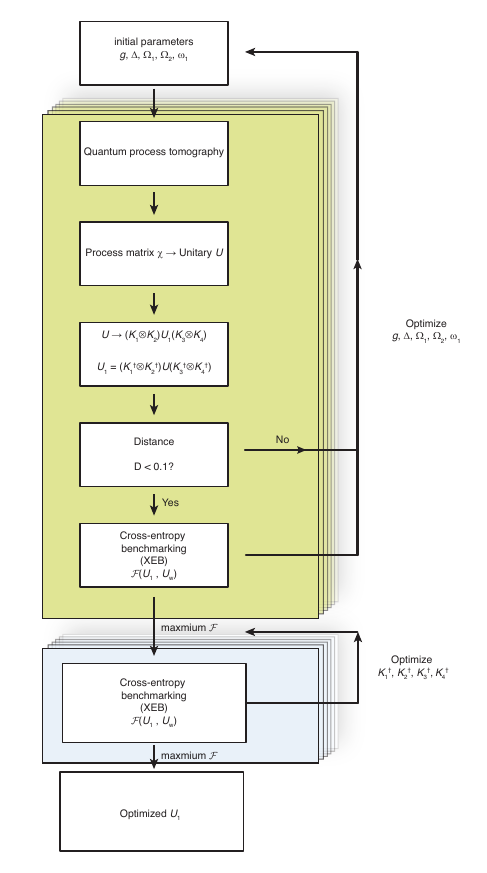}
\caption{\textbf{Optimization procedure}.}
\label{figsopt}
\end{figure}

Fig.~\ref{figsopt} shows the optimization procedure used in our work. Firstly, we prepare rough gate implementations with parameters $g, \Omega_{1}, \Omega_{2}, \Delta, \omega_{1}$. Next, we perform  QPT to obtain the process matrix $\bm{\chi}$ and the corresponding $U$. Subsequently, we decompose $U$ and acquire the Weyl chamber coordinates $(a,b,c)$ and four single-qubit correction gates $K_1, K_2, K_3, K_4$. Then we calculate the coordinate distance :

\begin{eqnarray}
D = \sqrt{(a-a_\mathrm{0})^2+(b-b_\mathrm{0})^2+(|c|-|c_\mathrm{0}|)^2},
\label{distance}
\end{eqnarray}

where $(a_\mathrm{0},b_\mathrm{0},c_\mathrm{0})$ is the coordinate of the target gate $U_{w}$. We optimize the control parameters $g, \Omega_{1}, \Omega_{2}, \Delta, \omega_{1}$ until the distance $D$ is reduced to less than 0.1. Subsequently, we apply the Hermitian conjugate of the single-qubit correction gates, $K_{1}^\dagger, K_{2}^\dagger, K_{3}^\dagger, K_{4}^\dagger$, to the generated two-qubit gate, yielding the actual gate $U_\mathrm{1}$. Next, we use the XEB fidelity $\mathcal{F}(U_{1},U_{w})$ as a cost function to further optimize $g, \Omega_{1}, \Omega_{2}, \Delta, \omega_{1}$ until $\mathcal{F}(U_{1},U_{w})$ reaches its maximum. The typical number of steps required for the first stage of optimization ranges from $50$ to $100$.

It should be noted that single-qubit correction gates $K_1, K_2, K_3, K_4$ also contain errors arising from state preparation and measurement (SPAM) imperfections and thus require further fine-tuning. However, these four correction gates introduce an additional twelve parameters, significantly increasing the optimization time, which becomes computationally prohibitive. To address this, we employ a second-stage optimization specifically designed to fine-tune the correction gates, as illustrated at the bottom of Fig.~\ref{figsopt}. After iterative rounds of optimization, the error rate is minimized.

\begin{figure}[hbt]
\includegraphics[scale=1]{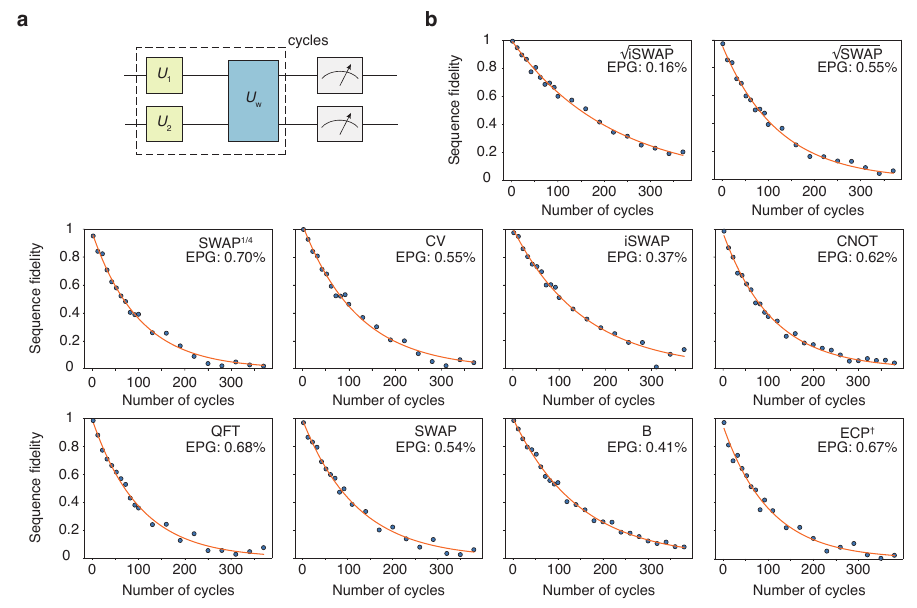}
\caption{ \textbf{Cross-entropy benchmarking (XEB)}. $\bf{a}$, XEB sequence for the AshN gates. $\bf{b}$, XEB fidelity of ten AshN gates. Each datapoint is the average of 50 random realizations.}
\label{figsxeb}
\end{figure}

\begin{figure}[hbt]
\includegraphics[scale=1.5]{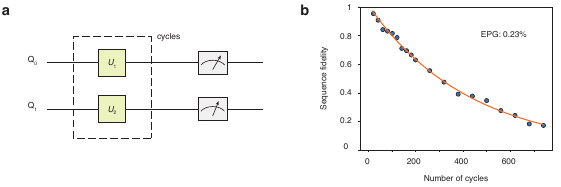}
\caption{\textbf{Single qubit gate cross-entropy benchmarking}. $\bf{a}$, XEB sequence for the single-qubit gates. $\bf{b}$, XEB fidelity of single-qubit gates with fitted error rate per gate being 0.23\%.}
\label{figssqxeb}
\end{figure}

Fig.~\ref{figsxeb} shows one of the repeated XEB measurements for each of the ten optimized AshN gates, as discussed in Fig.~2 of the main text. The errors per gate listed in Fig.~\ref{figsxeb} correspond to the measured XEB cycle error with the single-qubit gate error subtracted. For single-qubit gates, we also conduct XEB experiments in which each cycle consists solely of a random single-qubit gate $U_{1}, U_{2}$, as shown in Fig.~\ref{figssqxeb}.

\section{\label{sec: decoherence} Decoherence error of the AshN gate}

\begin{figure}[hbt]
\includegraphics[scale=1.3]{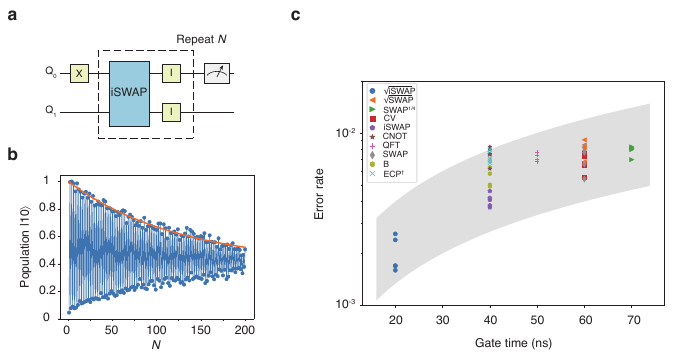}
\caption{\textbf{Decoherence errors of the AshN gates}. $\bf{a}$, Circuit of repeated iSWAP gates. We apply an X gate on $\mathrm{Q_0}$ before a train of $N$ iSWAP gates and finally measure the population of state $\ket{10}$. $\bf{b}$, The population of $\ket{10}$ as a function of the iSWAP gate count. The solid line is an exponential fit. $\bf{c}$, Error rates of the AshN gates versus their gate time. The shaded area denotes the upper and lower limits predicted from the measured iSWAP decoherence.}
\label{figserr}
\end{figure}

In the main text, we present the gate errors for commonly used two-qubit gates, which vary to a relatively large extent. To study the contribution of decoherence, we performed a simple test. Specifically, we repeated the calibrated $\iSWAP$ gate and measured the characteristic decay rate during the $\iSWAP$ pulse train, as shown in Fig.~\ref{figssqxeb}a. Note that the duration of the $\iSWAP$ pulse is 40~ns and the idling time between pulses is 80~ns, equivalent to the duration of four $\pi/2$ pulses. The extracted exponential gate count decay constant is $N=125$ (Fig.~\ref{figssqxeb}b). 
Assuming the idling periods between pulses do not contribute to the decay (i.e., decay occurs only during the 40-ns iSWAP pulse), the decoherence during $\iSWAP$ gates is lower bounded by a decay time constant of $T_\mathrm{iSWAP}=N \times 40~\mathrm{ns}=5~\mu$s. On the other hand, if we assume that the decay during idling periods is as fast as that during $\iSWAP$ gates, the time constant is then upper bounded by $T_\mathrm{iSWAP}=N \times 120~\mathrm{ns} =15~\mu$s. This analysis provides an estimated range for the decoherence during $\iSWAP$ operations.

In Fig.~\ref{figserr}c, we plot the measured AshN gate errors versus gate times. The data exhibit an increasing trend, bounded by the upper and lower limits predicted by the learned iSWAP decoherence time, following the relationship $\tau/T_\mathrm{iSWAP}$. This indicates that the decoherence error during the AshN gate is comparable to that of the $\iSWAP$ gate without additional drive. These results highlight the sensitivity of AshN gate errors to gate duration, underscoring the critical importance of time optimality in the AshN gate scheme.

\section{\label{sec: wstate} W state and dicke state}

We use quantum state tomography (QST) to restore the density matrices of the generated entangled states~\cite{song201710}. The fidelity of the generated state is calculated using its reconstructed density matrix, $\rho_\mathrm{e}$, as follows:~\cite{jozsa1994fidelity}:
\begin{eqnarray}
F(\rho_\mathrm{e},\rho) = \left(\mathrm{tr}\sqrt{\sqrt{\rho_\mathrm{e}}\rho\sqrt{\rho_\mathrm{e}}}\right)^2,
\label{fidelity}
\end{eqnarray}
where $\rho$ is the density matrix of the ideal state. The cost of QST used for a generic $N$-qubit state scales as $3^{N}\times2^{N}$. Specifically, this involves choosing and applying operators from the set $\{I, R_{x}(\pi/2), R_{y}(\pi/2)\}^{\otimes N}$. For each operator, projective measurements are conducted to obtain all diagonal terms, resulting in $2^{N}$ probabilities. To mitigate the impact of readout errors on QST, we repeat the experiment multiple times. 
To balance experimental cost and result deviation, a sample size of $2048 \times 7$ is chosen for the experiment. 

\begin{figure}[hbt]
\includegraphics[scale=1.2]{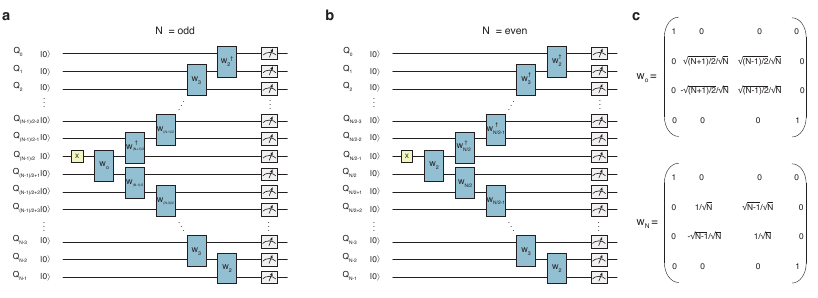}
\caption{\textbf{General Circuits to prepare $N$-qubit $W$ state}. $\bf{a}$, Circuit to prepare $N$-qubit W state, as $N$ is a odd number. $\bf{b}$, Circuit to prepare $N$-qubit W state, as $N$ is a even number. $\bf{c}$, matrixes of the gates used in $\bf{a}$ and $\bf{b}$. The phase of the $W$ state can be conveniently controlled by the single-qubit Z phase gates between the two-qubit gates, which are not shown here for simplicity.}
\label{figswcircuit}
\end{figure}

As discussed in the main text, the $N$-qubit $W$ state can be prepared with $(N-1)$ two-qubit gates. Here we propose circuits for generating the $N$-qubit $W$ state, as shown in Fig.~\ref{figswcircuit}. Generating a $W$ state with an even larger number of qubits is possible; however, we choose $N=10$ as it is sufficient to demonstrate the superiority of the AshN scheme in generating high-fidelity entangled states while keeping the experimental cost, particularly for QST, reasonable.

\begin{figure}[hbt]
\includegraphics[scale=1.2]{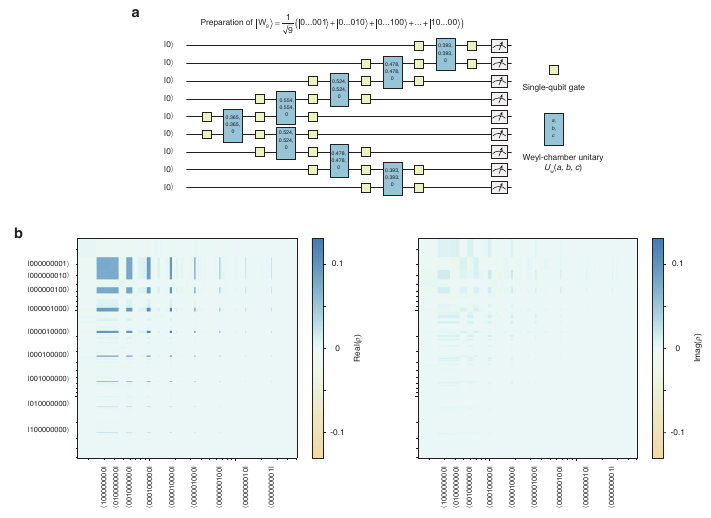}
\caption{\textbf{9-qubit W state $W_{9}$.} $\bf{a}$, Circuit to prepare the 9-qubit W state $\ket{W_{9}}$. $\bf{b}$, Quantum state tomography of the $\ket{W_{9}}$ state, and the estimated fidelity is $\mathrm{0.908 \pm 0.010 }$.}
\label{figsw9}
\end{figure}

To demonstrate the versatility of the circuit in Fig.~\ref{figswcircuit}, in addition to the $10$-qubit W state, we also perform a $9$-qubit W state, as shown in Fig.~\ref{figsw9}. The $\ket{W_9}$ state yields a fidelity of $\mathrm{0.908 \pm 0.010}$, which is slightly lower than $\ket{W_{10}}$, we attribute it to the fact that one of the qubits $Q_{1}$ in Table~\ref{tab1} was affected by a two-level system and became unstable during the experiment.

To compare with $\CNOT$-based state preparations, we use approximate synthesis to estimate the number of $\CNOT$ gates needed to generate $\ket{W_{10}}$, ensuring an error threshold set at $10^{-15}$. For $N = 8$, approximately synthesizing a single data point required roughly $15$ hours using $48$ BLAS threads, making it impractical to generate reliable results on larger scales. As such, we limit our synthesis to the cases where $N \leq 8$. By repeatedly applying approximate synthesis, we determined that the required gate counts of $\CNOT$ are $5$, $9$, and $13$ for $N = 4, 6, 8$, respectively. This leads us to conjecture that the number of $\CNOT$ gates needed scales as $2N-3$, which is significantly higher than the exact $\SU{4}$ gate count of $N-1$. These findings underscore the significant reduction in gate count achieved by employing $\SU{4}$ operations instead of $\CNOT$ gates, demonstrating the increased efficiency and scalability of the proposed approach.

Applying the same approximate synthesis protocol as
above to double-excitation Dicke states of 4- and 6-qubit results in $\CNOT$ gate counts of 6 and 15, respectively, compared to 5 and 9 $\SU{4}$ gates, further highlighting the significant advantages of $\SU{4}$ operations over $\CNOT$ gates. Approximate synthesis provides a tool for estimating the theoretical bound with a reasonable level of accuracy, but in our implementation, we adopt exact synthesis to eliminate errors arising from approximate synthesis and ensure a fair comparison with the existing literature. By utilizing the full $\SU{4}$ expressivity of the AshN gates, we implement a circuit that generates the Dicke state $\ket{D_{4}^{2}}$ with double-excitation of four qubits
with only eight $\SU{4}$ operations, illustrated as Fig. 3c in the main text, compared to the fourteen $\CNOT$ gates required otherwise.

\section{\label{sec: derror} Gate error of two-qubit gates synthesized using B gates}

In this section, we analyze the features observed in the experimental error map of all gates in the Weyl chamber through two applications of the $\BGate$ gate, as shown in Fig.~4 in the main text. We will show that dephasing noise is the primary factor behind key observation---gate errors are notably smaller for unitaries located along the line connecting $\SWAP$ and $\SWAP^\dagger$ within the tetrahedron, specifically where $a=b=\pi/4$.

The low-frequency dephasing noise can be represented by the $ZI$ and $IZ$ terms in the Hamiltonian,
\begin{align}
H'_{\mathrm{B}} = H_{\tilde{\mathrm{B}}} + \delta_{1} ZI+ \delta_{2} IZ,    
\end{align}
where $H_{\tilde{\rm{B}}}$ is the AshN Hamiltonian used to generate the $\BGate$ gate in Eq.~\ref{BAshN}, and $\delta_{1}$($\delta_{2}$) is the off-resonance or detuning noise on each qubit. The corresponding unitary is $U_{B}' = {\rm exp}(-i\frac{\pi}{2g}H'_{B})$.
During single-qubit $\pi/2$ operations, the effective Hamiltonian including the dephasing term is,
\begin{align}
H_{\pi/2}(\theta) = \frac{\pi}{2}({X\rm cos}\theta + Y{\rm sin} \theta) +  \frac{\pi}{2}\delta Z,
\label{zerr}
\end{align}
with the unitary of $U_{\pi/2}(\theta) = {\rm exp}[-\frac{i}{2}H_{\pi/2}(\theta)]$ and $\delta/2t_g =\delta_1$ or $\delta_2$, $t_g$ is the single-qubit gate time.

\begin{figure}[hbt]
\includegraphics[scale=2]{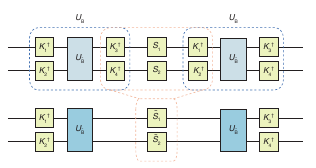}
\caption{\textbf{B-composed Weyl chamber.} The B-composed Weyl chamber contains two applications of B gate and six single-qubit gates. $U_{\tilde{B}}$ is the AshN-generated tow-qubit gate, which can be transformed into the standard B gate with four single-qubit corrections. When implementing in the experiment, we merge successive single-qubit operations into one single-qubit operation. }
\label{figswErAC}
\end{figure}

As shown in Fig.~\ref{figswErAC}, each unitary can be synthesized using two B gates combined with single-qubit operations $S_1$ and $S_2$, as expressed in Eq.~\ref{bsu4}. The B gate is implemented through the AshN-generated $U_{\rm{\tilde{B}}}$, along with four single-qubit unitary corrections $K_i^{\dagger}$ ($i=1,2,3,4$), which are obtained as:

\begin{eqnarray}
K_1^{\dagger} =& \begin{pmatrix}
0.829i & 0.559 \\ 
-0.559 & -0.829i
\end{pmatrix}, \quad 
K_2^{\dagger} = \begin{pmatrix}
 0 & -i \\
 -i & 0
\end{pmatrix}, \nonumber\\ 
K_3^{\dagger} =& \begin{pmatrix}
-0.829i & 0.559 \\
-0.559 & 0.829i
\end{pmatrix}, \quad
K_4^{\dagger} = \begin{pmatrix}
 0 & -i \\ 
 -i & 0
\end{pmatrix}.
\label{k1t4}
\end{eqnarray}

By merging the compensatory single-qubit gates required to convert the AshN-generated $\BGate$ gate into the ideal $\BGate$ gate, a mapping from the Weyl chamber coordinates to two parameterized single-qubit operations $\tilde{S}_i(\alpha,\beta,\gamma)=\begin{pmatrix}
e^{i\alpha}\cos{\gamma} & -e^{-i\beta}\sin{\gamma} \\
e^{i\beta}\sin{\gamma} & e^{-i\alpha} \cos{\gamma}
\end{pmatrix} $ $(i=1,2)$ is obtained. The single-qubit gate parameters are bounded as $-\pi/2\leq\alpha\leq\pi/2$, $-\pi\leq\beta\leq\pi$, and $ 0\leq\gamma\leq \pi/2$. For example, when the Weyl chamber coordinates satisfy $a=b=\pi/4$, the value of $\gamma$ exceeds $0.2\pi$. In practical implementation, $\tilde{S}(\alpha,\beta,\gamma)$ is further decomposed into four physical $\pi/2$ pulses according to the PMW-4 scheme,
\begin{eqnarray}
\tilde{S}(\alpha,\beta,\gamma) = X_{\pi/2}(-\alpha+\beta)X_{\pi/2}(\beta-\gamma+\pi)X_{\pi/2}(\beta-\gamma+\pi)X_{\pi/2}(\alpha+\beta).
\label{pmw4}
\end{eqnarray}
Using these relations, we can explicitly calculate the sensitivity of these single-qubit gate errors to the dephasing noise.

\begin{figure}[hbt]
\includegraphics[scale=0.8]{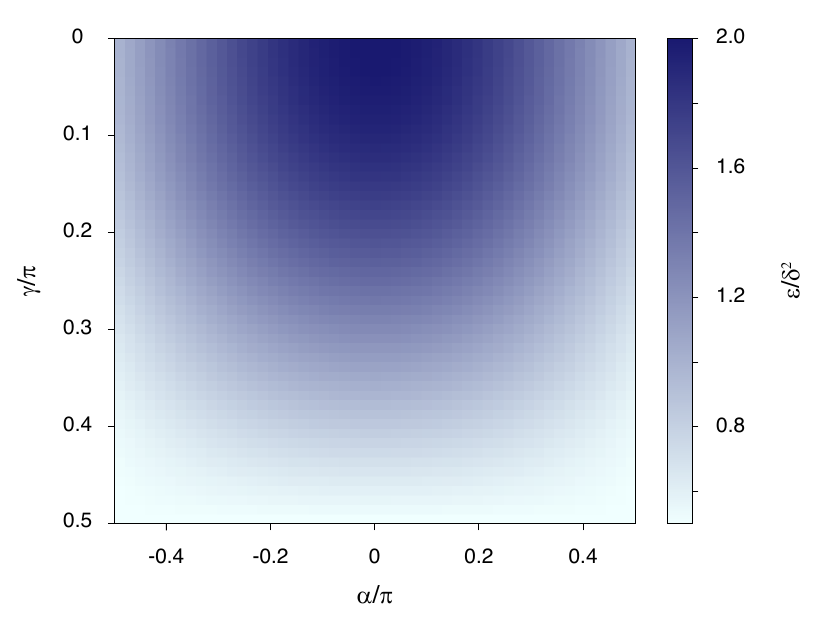}
\caption{\textbf{Simulated error map for single-qubit operations.} $\alpha$ and $\gamma$ are single-qubit gate parameters. The error of the single-qubit operation decreases as $\gamma$ increases.}
\label{figswErA1}
\end{figure}

First, $\beta$ is a common phase that appears in all four $\pi/2$ pulses, effectively a global phase shift of the combined rotation. Hence, the gate error sensitivity shall not depend on $\beta$.
The dependence of the off-resonance errors on $\alpha$ and $\gamma$ can be analytically derived and expanded up to the second order as
\begin{eqnarray}
\epsilon = \frac{1}{4}(3 + 4\cos \alpha \cos \gamma + \cos 2\gamma )\delta^2 + O(\delta^3) \;,
\label{pmw4f}
\end{eqnarray}
and is plotted in Fig. \ref{figswErA1}. It can be seen that, in the range shown, the robustness against dephasing noise for a single-qubit gate compiled by PMW-4 increases with $\gamma$.

\begin{figure}[hbt]
\includegraphics[scale=1]{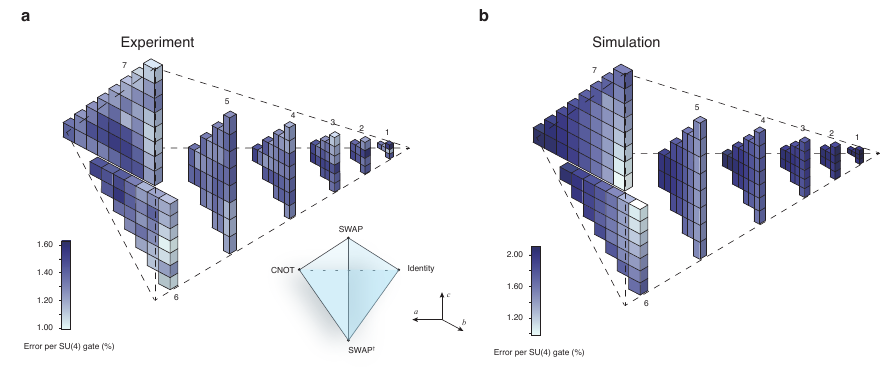}
\caption{\textbf{Gate error for B-composed Weyl chamber.} $\bf{a}$, Experimental results (same as that in Fig.~4 of the main text). $\bf{b}$, Simulated results.}
\label{figswErA2}
\end{figure}

Next, we numerically simulate the complete B-composed Weyl chamber assuming low-frequency dephasing noise only. We calculate the average gate fidelity over different $\delta$ values (as defined in Eq.~\ref{zerr}) sampled from a Gaussian distribution of standard deviations $\rm \sigma_1 = 0.765 MHz$ and $\rm \sigma_2 = 1.258 MHz$ for the two qubits respectively.
The gate fidelity is defined by the following formula:
\begin{eqnarray}
F(U_\mathrm{sim},U_\mathrm{ideal}) = \frac{|{\rm Tr}(U_\mathrm{ideal}{U_\mathrm{sim}}^{\dagger})|}{4},
\label{fid}
\end{eqnarray}
where $U_\mathrm{sim}$ and $U_\mathrm{ideal}$ represent the simulated and ideal unitary of a given two-qubit gate.
The results are shown in Fig.~\ref{figswErA2} and compared with experimental data.
It can be seen that gates near the $a=b=\pi/4$ line connecting $\SWAP$ and $\SWAP^\dagger$ are expected to be less vulnerable to such dephasing noise, which qualitatively agrees with the measured results.

\section{\label{implicat}Theoretical Implications and Challenges}
The demonstrated AshN gate scheme is closely tied to the concept of controllability, that is, the ability to achieve any element of the Lie group SU(4) by appropriately designing the time-dependent Hamiltonian. Controllability lies at the heart of quantum control, with its origins tracing back to the pioneering days of NMR research~\cite{vandersypen2004nmr}. If the given control Hamiltonians, along with their commutators, generate the complete Lie algebra $\mathfrak{su}(4)$ of the target group SU(4), then the entire group becomes accessible, allowing the implementation of any unitary operation through a sequence of exponentials of the control Hamiltonians.

In fact, most systems can be proven to be controllable. In our case, it suffices to verify that the operators $XX+YY$, $ZI+IZ$, $XI$, and $IX$ generate the Lie algebra $\mathfrak{su}(4)$ by iteratively applying the Lie bracket operation. However, while controllability can often be established in theory, constructing explicit control schemes, such as the AshN scheme, to achieve all unitary operations is highly nontrivial.

Furthermore, complex pulse engineering approaches, effective for NMR, trapped ions, and other platforms with slower operations and longer relaxation times, are much less favorable for superconducting platforms due to their significantly shorter $T_1$ relaxation times, leading to higher errors during pulse concatenations. Although there is a theoretical equivalence between single-pulse and multipulse approaches in control theory---specifically between Lie-Cartan coordinates of the first kind (expressing elements as a single exponential of a linear combination of basis elements) and the second kind (expressing elements as a product of exponentials of basis elements)~\cite{sastry2013nonlinear}---in practice, leveraging this equivalence is challenging because it requires using the Baker-Campbell-Hausdorff (BCH) formula, which involves handling an infinite series of increasingly complex nested commutators~\cite{sastry2013nonlinear}.

Representing a stronger form of the well-established controllability results, the implemented AshN scheme requires only the exchange interaction and qubit driving to generate arbitrary two-qubit gates via a single pulse.

An attentive reader might notice certain subtleties in the demonstration presented in the previous section. For example, the AshN scheme is not effective in realizing operations near the identity region. Gates in this small region typically have lower entangling capabilities and are therefore less favored; however, they remain essential for Trotterization in Hamiltonian simulations. Implementing such gates using the AshN scheme may require unbounded amplitudes, which is experimentally infeasible. Aside from slightly sacrificing the optimal duration and employing Hamiltonian engineering approaches, this issue can also be resolved at the software level. By inserting a $\SWAP$ operation, a gate near identity can be mirrored into the corner corresponding to $\SWAP$, which is easy to implement using the AshN scheme. Moreover, the effects of inserting $\SWAP$ operations can be effectively absorbed in the qubit routing stage.

We have thoroughly discussed the maximal expressivity of the AshN gate scheme, which represents just one aspect of its performance. Equally important is its accuracy, which we will now address.

Theoretical lower bound for the time required to implement two-qubit unitary operators was originally established in~\cite{hammerer2002characterization}. This foundational result defines the minimum time necessary for a sequence of time evolutions governed by a two-qubit Hamiltonian, interspersed with arbitrary single-qubit gates, to achieve a target two-qubit unitary operation. Building on this, \cite{chen2024one} demonstrated that all two-qubit unitary operators can be implemented in optimal time through the AshN gate scheme, with the exception of a small region near the identity operator. In this region, achieving the optimal duration would require physically infeasible unbounded amplitudes.

This time optimality is particularly crucial when decoherence dominates, as the ratio of gate time to coherence time becomes a critical determinant of physical gate error. By minimizing the required gate time, AshN ensures high-fidelity realization of quantum gates. Moreover, as discussed earlier in this section, AshN achieves maximal expressivity by enabling the implementation of arbitrary two-qubit gates in $\SU{4}$---a feature of significant importance given that single- and two-qubit gates are the primary constructs in quantum computation, as multi-qubit gates are hindered by severe control challenges.

By uniting these two strengths---{\em time optimality for high fidelity} and {\em maximal expressivity}---AshN delivers a gate scheme that epitomizes optimal performance for a quantum computer.

A more practical issue is the calibration cost. While we have demonstrated the capability to implement arbitrary two-qubit operations, realizing the full $\SU{4}$ remains experimentally infeasible due to the high calibration requirements. However, we believe that this challenge can also be addressed at the software level and does not constitute fundamental limitations. For instance, most quantum programs can be synthesized into structured blocks, resulting in recurring patterns throughout. This suggests that the required two-qubit operations will exhibit similarities across different building blocks, effectively bounding the overall calibration cost. In the main text, we demonstrate another approach that, while slightly sacrificing accuracy, effectively mitigates the calibration cost.

\section{AshN Algorithm}

\begin{algorithm}[ht]
\caption{AshN}\label{alg:ashn}
{\small
\KwData{$(x,y,z)\in W; g\in \mathbb{R}_+;h\in[-g,g];$\\$ r\in[0,\frac{(1-|h|)\pi}{2}]$}
\KwResult{$(\tau,\Omega_1,\Omega_2,\delta)$}
$\tau_{ND}\gets 2x$\;
$\tau_{EA+}\gets 2(x+y+z)/(2+h)$\;
$\tau_{EA-}\gets 2(x+y-z)/(2-h)$\;
$\tau'_{ND}\gets \pi-2x$\;
$\tau'_{EA+}\gets 2(\pi/2-x+y-z)/(2+h)$\;
$\tau'_{EA-}\gets 2(\pi/2-x+y+z)/(2-h)$\;
$\tau_1\gets \max\{\tau_{ND}, \tau_{EA+},\tau_{EA-}\}$\;
$\tau_2\gets \max\{\tau'_{ND}, \tau'_{EA+},\tau'_{EA-}\}$\;
\eIf{$\min\{\tau_1,\tau_2\}\leq r$} {
\Return{AshN-ND-EXT($x,y,z,g,h$)};
}
{
\If{$\tau_2< \tau_1$}{
$x\gets \pi/2 - x$\;
$z\gets - z$\;
$\tau_{ND}\gets\tau'_{ND}$\;
$\tau_{EA+}\gets\tau'_{EA+}$\;
$\tau_{EA-}\gets\tau'_{EA-}$\;
}
\eIf{$\tau_{ND}\geq \max\{\tau_{EA+},\tau_{EA-}\}$}
{\Return{AshN-ND($x,y,z,g,h$)};}
{\eIf{$\tau_{EA+}\geq\tau_{EA-}$}
{\Return{AshN-EA+($x,y,z,g,h$)};}
{\Return{AshN-EA-($x,y,z,g,h$)};}
}
}
}
\end{algorithm}

\begin{algorithm}[ht]
\caption{AshN-ND}\label{alg:ashna0}
\KwData{$(x,y,z)\in W;g\in\mathbb{R}_+;h\in[-g,g]$}
\KwResult{$(\tau,\Omega_1,\Omega_2,\delta)$}
$\tau\gets 2x$\;
$r_1\gets 2\sinc^{-1}(\frac{2\sin(y+z)}{(1+h)\tau})/\tau$ \Comment{$\sinc(x):=\sin(x)/x$}\;
$r_2\gets 2\sinc^{-1}(\frac{2\sin(y-z)}{(1-h)\tau})/\tau$
\Comment{$\sinc^{-1}:[0,1]\rightarrow [0,\pi]$}\;
$\gamma_1\gets \sqrt{r_1^2-1}/4$\;
$\gamma_2\gets \sqrt{r_2^2-1}/4$\;
\Return{$\tau/g, 2(\gamma_1+\gamma_2)g,2(\gamma_1-\gamma_2)g,0$};
\end{algorithm}

\begin{algorithm}[ht]
\caption{AshN-ND-EXT}\label{alg:ashna1}
\KwData{$(x,y,z)\in W;g\in\mathbb{R}_+;h\in[-g,g]$}
\KwResult{$(\tau,\Omega_1,\Omega_2,\delta)$}
$\tau\gets \pi-2x$\;
$r_1\gets 2\sinc^{-1}(\frac{2\sin(y-z)}{(1+h)\tau})/\tau$\;
$r_2\gets 2\sinc^{-1}(\frac{2\sin(y+z)}{(1-h)\tau})/\tau$\;
$\gamma_1\gets \sqrt{r_1^2-1}/4$\;
$\gamma_2\gets \sqrt{r_2^2-1}/4$\;
\Return{$\tau/g, 2(\gamma_1+\gamma_2)g,2(\gamma_1-\gamma_2)g,0$};
\end{algorithm}

\begin{algorithm}[ht]
\caption{AshN-EA+}\label{alg:ashnb}
\KwData{$(x,y,z)\in W;g\in\mathbb{R}_+;h\in[-g,g]$}
\KwResult{$(\tau,\Omega_1,\Omega_2,\delta)$}
$\tau\gets 2(x+y+z)/(2+h)$\;
$\tau'\gets (1-h)\tau$\;
$(x',y',z')\gets (x,y,z)-(h\tau/2,h\tau/2,h\tau/2)$\;
$S\gets e^{i(y'-x'-z')}-e^{i(x'-y'-z')}-e^{i(z'-x'-y')}$\;
Find a pair $(\alpha,\beta)\in[0,1]\times [0, 2\pi/\tau]$ such that $\frac{(1-\alpha)\beta e^{i\tau'(\alpha+\beta)}}{(2\alpha+\beta)(1+\alpha+2\beta)}-\frac{(1-\alpha)(1+\alpha+\beta)e^{-i\tau'(1+\beta)}}{(1-\alpha+\beta)(1+\alpha+2\beta)}-\frac{\beta(1+\alpha+\beta) e^{-i\tau'\alpha}}{(1-\alpha+\beta)(2\alpha+\beta)}=S$\;
$\gamma\gets\sqrt{(1+\alpha+\beta)(1-\alpha)\beta}/2$\;
$d\gets\sqrt{(\alpha+\beta)\alpha(1+\beta)}/2$\;

\Return{$\tau/g, 2(1-h)\gamma g,2(1-h)\gamma g,2(1-h)d g$};
\end{algorithm}

\begin{algorithm}[ht]
\caption{AshN-EA-}\label{alg:ashnc}
\KwData{$(x,y,z)\in W;g\in\mathbb{R}_+;h\in[-g,g]$}
\KwResult{$(\tau,\Omega_1,\Omega_2,\delta)$}
$T',\Omega'_1,\Omega'_2,\delta'\gets \text{AshN-EA+}(x,y,-z,g,-h)$\;

\Return{$T', 0,\Omega'_1,-\delta'$};
\end{algorithm}

\bibliography{AshN}